\documentclass[fleqn,usenatbib]{mnras}

\usepackage{newtxtext,newtxmath}

\usepackage[T1]{fontenc}
\usepackage{ae,aecompl}
\usepackage{natbib}

\newcommand\nata{Nature Astronomy}
\newcommand{\angstrom}{\mbox{\normalfont\AA}}

\def\fesc{\ifmmode f_{\rm esc} \else $f_{\rm esc}$\fi}

\usepackage{graphicx}	
\usepackage{amsmath}	

\title[Ly$\alpha$ and LyC emission in Mg~{\sc ii}-selected SFGs]{Lyman alpha and Lyman continuum emission of Mg~{\sc ii}-selected star-forming galaxies}

\author[Y. I. Izotov et al.]{
Y. I. Izotov$^{1}$\thanks{E-mail: yizotov@bitp.kiev.ua},
J. Chisholm$^{2}$, 
G. Worseck$^{3}$, 
N. G. Guseva$^{1}$,
D. Schaerer$^{4,5}$, 
\newauthor 
J. X. Prochaska$^{6}$
\\
$^{1}$Bogolyubov Institute for Theoretical Physics,
National Academy of Sciences of Ukraine, 14-b Metrolohichna str., Kyiv,
03143, Ukraine,\\
$^{2}$Astronomy Department, University of Texas at Austin,
2515 Speedway, Stop C1400 Austin, TX 78712-1205, USA, \\
$^{3}$ Institut f\"ur Physik und Astronomie, Universit\"at Potsdam, Karl-Liebknecht-Str. 24/25, D-14476 Potsdam, Germany,\\
$^{4}$Observatoire de Gen\`eve, Universit\'e de Gen\`eve, 
51 Ch. des Maillettes, 1290, Versoix, Switzerland,\\
$^{5}$IRAP/CNRS, 14, Av. E. Belin, 31400 Toulouse, France,\\
$^{6}$University  of  California  Observatories-Lick  Observatory,  University  
of  California,  1156 High Street, Santa Cruz, CA 95064, USA\\
}

\date{Accepted XXX. Received YYY; in original form ZZZ}

\pubyear{2022}

\begin{document}
\label{firstpage}
\pagerange{\pageref{firstpage}--\pageref{lastpage}}
\maketitle

\begin{abstract}
  We present observations with the Cosmic Origins Spectrograph onboard the
  Hubble Space Telescope of seven compact low-mass star-forming galaxies at
  redshifts, $z$, in the range 0.3161 -- 0.4276, with various O$_3$Mg$_2$ =
  [O~{\sc iii}] $\lambda$5007/Mg~{\sc ii} $\lambda$2796+2803 and
  Mg$_2$ = Mg~{\sc ii} $\lambda$2796/Mg~{\sc ii} $\lambda$2803
  emission-line ratios. We aim to study the dependence of leaking Lyman
  continuum (LyC) emission on the characteristics of Mg~{\sc ii} emission
  together with the dependencies on other indirect indicators of escaping
  ionizing radiation. LyC emission with escape fractions
  $f_{\rm esc}$(LyC) = 3.1 -- 4.6 per cent is detected in four galaxies,
  whereas only 1$\sigma$ upper limits of $f_{\rm esc}$(LyC) in the
  remaining three galaxies were derived. A strong narrow Ly$\alpha$ emission
  line with two peaks separated by $V_{\rm sep}$ $\sim$ 298 -- 592 km s$^{-1}$
  was observed in four galaxies with detected LyC emission and very
  weak Ly$\alpha$ emission is observed in galaxies with LyC non-detections.
  Our new data confirm the tight anti-correlation between $f_{\rm esc}$(LyC) and
  $V_{\rm sep}$ found for previous low-redshift galaxy samples. $V_{\rm sep}$
  remains the best indirect indicator of LyC leakage among all considered
  indicators. It is found that escaping LyC
  emission is detected predominantly in galaxies with Mg$_2$ $\ga$ 1.3.
  A tendency of an increase of $f_{\rm esc}$(LyC) with increasing of both the
  O$_3$Mg$_2$ and Mg$_2$ is possibly present. However, there is substantial
  scatter in these
relations not allowing their use for reliable prediction of $f_{\rm esc}$(LyC).
\end{abstract}




\section{Introduction}\label{intro}

It was established during last decade that Lyman continuum (LyC)
emission, which is produced in copious amount in both the high redshift
star-forming galaxies (SFGs) at $z$\,$\sim$\,2\,-\,4
\citep{Va15,B16,Sh16,B17,Va18,RT19,Sa20,Mes20,Vi20,Fl19,Ma17,Ma18,St18}
and the low-redshift SFGs at $z$ $\la$ 0.4
\citep{L13,B14,L16,C17,I16a,I16b,I18a,I18b,I21a,F22a,F22b,Xu22}, can escape from
the galaxies resulting in ionization of the intergalactic medium (IGM).
These galaxies are considered as analogues of the galaxies at redshifts
6 - 8, which are presumably the main sources of the reionization
of the Universe \citep*{O09,WC09,Y11,M13,B15a,Fi19,Le20,N20,Me20}.

It was also found that $f_{\rm esc}$(LyC) 
in many discovered galaxies is of the order of 10 - 20 per
cent or higher. This could be sufficient for efficient reionization of the IGM
at $z$ $\ga$ 6 \citep[e.g. ][]{O09,R13,Robertson15,D15,K16}.

Direct LyC observations of high-redshift galaxies are difficult because
of their faintness, the increasing of IGM opacity, and contamination
by lower-redshift interlopers \citep[e.g. ][]{V10,V12,Inoue14,Gr16}.
Furthermore, the knowledge of the galaxy H$\beta$ or H$\alpha$ luminosity is
needed to derive the production
rate of ionizing photons and thus the $f_{\rm esc}$(LyC). This is not possible
yet for most of high-$z$ LyC emitters. Low-redshift galaxies are
brighter, but observations
from space, with the aid of {\sl Hubble Space Telescope}
({\sl HST}), are needed for the detection of LyC emission in $z$ $\ga$ 0.3
galaxies. This can be
done only for limited samples of low-$z$ galaxies. On the other hand,
the H$\beta$ and H$\alpha$ emission lines can easily be observed in low-$z$
galaxies from the ground. In fact, many such
galaxies were observed in the course of the Sloan Digital Sky Survey
(SDSS). This survey was succesfully used to select promising
LyC leaking candidates and their subsequent observations with the
{\sl HST} \citep[this paper, ][]{I16a,I16b,I18a,I18b,I21a,W21,F22a,Xu22}.

Due to difficulties of direct detection of LyC emission in both the high- and
low-redshift SFGs indirect indicators for the determination of the
$f_{\rm esc}$(LyC) can be used. However, at present, it cannot be very reliably
determined from most indicators due to the large scatter in their
correlations with $f_{\rm esc}$(LyC).

The shape of the Ly$\alpha$ line can be considered as the prime indicator of the
$f_{\rm esc}$(LyC) value, since it depends on the distribution of the neutral
hydrogen around the galaxy, which also determines the escape of ionizing
radiation \citep[e.g. ][]{V15}. In most galaxies with the Ly$\alpha$ emission
line it has a two-peak shape due to scattering in the neutral gas with a
  relatively high column density of H~{\sc i}, with a weaker blue peak and a
stronger red peak. The offset of the peaks from the line centre serves
as a measure of the neutral hydrogen optical depth along the line
of sight \citep[e.g. ][]{V15}. In particular, a tight correlation
between the Ly$\alpha$ blue and red peak separation and the escape
fraction of ionizing radiation was found \citep{I18b}.
More complex Ly$\alpha$ profiles with three or
  more peaks are rarely observed \citep{Va18,I18b,RT17,RT19}. They show
  significant central line emission, an indication of direct escape through
  porous channels in addition to escape via scattering. In these cases
the separation of the Ly$\alpha$ emisson peaks is a poor tracer of
$f_{\rm esc}$(LyC) because of
  a combination of two distinct modes of Ly$\alpha$ escape \citep{N22}.
We also note that
at redshifts $z$ $\ga$ 6 the detection of Ly$\alpha$ is difficult because
of declining Ly$\alpha$ transmission with redshift \citep{Gr21}.
This decline with redshift
is sharper on the blue side of Ly$\alpha$ making it more difficult to detect
the blue peak.

  \begin{table*}
  \caption{Coordinates, redshifts, distances, oxygen abundances, EW(H$\beta$),
    O$_{32}$, O$_3$Mg$_2$ and Mg$_2$ ratios of selected galaxies
\label{tab1}}
\begin{tabular}{lrrccrccrcc} \hline
  Name&R.A.(2000.0)&Dec.(2000.0)&$z$&$D_L$$^{\rm a}$&\multicolumn{1}{c}{$D_A$$^{\rm b}$}&12+logO/H$^{\rm c}$&EW(H$\beta$)$^{\rm d}$&O$_{32}$$^{\rm e}$&O$_3$Mg$_2$$^{\rm f}$&Mg$_2$$^{\rm g}$ \\ \hline
J0130$-$0014&01:30:32.37&$-$00:14:32.52&0.31606&1664& 961&7.97$\pm$0.02&200& 7.4$\pm$0.4& $\ga$100\,~~~~~& ... \\ 
J0141$-$0304&01:41:42.85&$-$03:04:51.12&0.38161&2075&1087&8.06$\pm$0.02&220& 5.6$\pm$0.2&     15$\pm$1&1.62$\pm$0.31\\ 
J0844$+$5312&08:44:57.90&$+$53:12:30.11&0.42764&2374&1165&8.02$\pm$0.02&196& 4.9$\pm$0.2&     10$\pm$1&2.38$\pm$0.43\\ 
J1014$+$5501&10:14:23.78&$+$55:01:43.82&0.37297&2019&1071&7.96$\pm$0.02&240& 6.8$\pm$0.4&     14$\pm$1&1.19$\pm$0.28\\ 
J1137$+$3605&11:37:47.77&$+$36:05:04.62&0.34387&1836&1017&7.81$\pm$0.01&280& 7.4$\pm$0.3&     22$\pm$2&1.53$\pm$0.44\\ 
J1157$+$5801&11:57:44.80&$+$58:01:42.69&0.35210&1887&1032&7.81$\pm$0.01&263& 9.0$\pm$0.5&     35$\pm$7&0.97$\pm$0.52\\ 
J1352$+$5617&13:52:35.80&$+$56:17:01.41&0.38818&2117&1099&8.05$\pm$0.03&172& 3.8$\pm$0.2&     10$\pm$1&1.50$\pm$0.26\\ 
\hline
\end{tabular}

\hbox{$^{\rm a}$Luminosity distance in Mpc \citep[NED, ][]{W06}.}

\hbox{$^{\rm b}$Angular size distance in Mpc \citep[NED, ][]{W06}.}

\hbox{$^{\rm c}$Oxygen abundance derived by the direct $T_{\rm e}$ method.}

\hbox{$^{\rm d}$Equivalent width of the H$\beta$ emission line in \AA.}

\hbox{$^{\rm e}$O$_{32}$ is the extinction-corrected 
[O~{\sc iii}]$\lambda$5007/[O~{\sc ii}]$\lambda$3727 flux
ratio derived in this paper from the SDSS spectrum.}

\hbox{$^{\rm f}$O$_3$Mg$_2$ is the extinction-corrected
  [O~{\sc iii}]$\lambda$5007/Mg~{\sc ii} $\lambda$2796+2803 flux ratio derived
  in this paper from the SDSS spectrum.}

\hbox{$^{\rm g}$Mg$_2$ is the Mg~{\sc ii} $\lambda$2796/Mg~{\sc ii} $\lambda$2803
  flux ratio derived in this paper from the SDSS spectrum.}

  \end{table*}

Therefore, other indirect indicators are needed, for example,
those, which use strong emission lines in the rest-frame optical and
UV ranges, or UV absorption lines, including hydrogen lines of the
Lyman series and heavy element lines, such as Si~{\sc ii} $\lambda$1260 that can
measure the Lyman continuum escape fraction
\citep[e.g. ][]{Ga18,Ga20,Ch18,F22a,F22b,SL22}.

\citet{JO13} and \citet{NO14} proposed to use the
O$_{32}$ = [O~{\sc iii}]$\lambda$5007/[O~{\sc ii}]$\lambda$3727 flux ratio
arguing that its high values of up to $\sim$ 60 in some low-$z$ galaxies
\citep*{S15,I21b} may indicate that the ISM is
predominantly ionized, allowing the escape of Lyman continuum
photons. Indeed, \citet{I16a,I16b,I18a,I18b,I21a} obtained
{\sl HST}/COS observations of compact SFGs at redshifts $z$ $\sim$ 0.3 - 0.4
with O$_{32}$ = 5 - 28 and an escape fraction in the range of 2 - 72
per cent. Although they did find some trend of increasing $f_{\rm esc}$(LyC)
with increasing O$_{32}$, the dependence is weak, with a large scatter.

It has also been suggested that $f_{\rm esc}$(LyC) tends to be higher in
low-mass galaxies \citep{W14,T17}. However, \citet{I18b,I21a}
added low-mass LyC leakers and found rather a relatively weak anti-correlation
between $f_{\rm esc}$(LyC) and stellar mass $M_\star$ in a wide range between
10$^7$ - 10$^{10}$ M$_\odot$. A similar correlation is also found in the
Low-$z$ Lyman Continuum Survey (LzLCS) in \citet{F22b}.

\begin{figure*}
\hbox{
\includegraphics[angle=-90,width=0.32\linewidth]{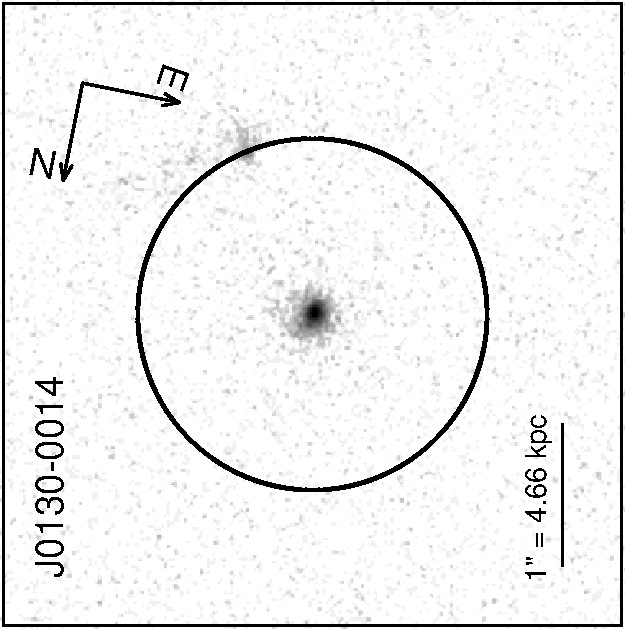}
\includegraphics[angle=-90,width=0.32\linewidth]{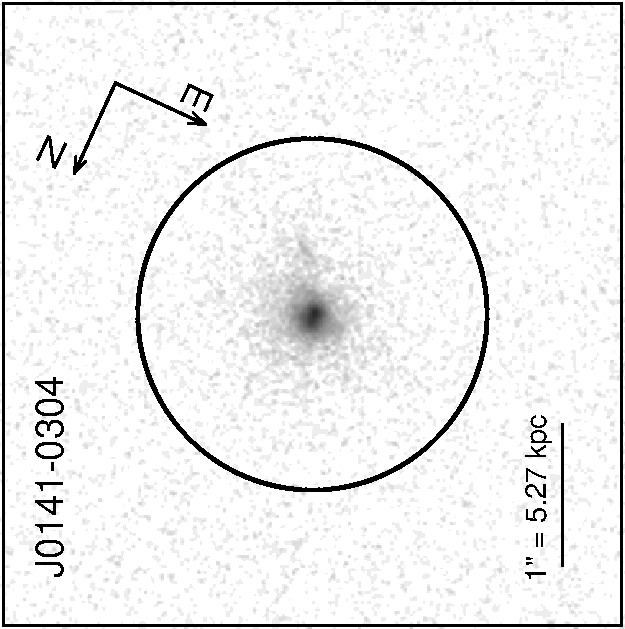}
\includegraphics[angle=-90,width=0.32\linewidth]{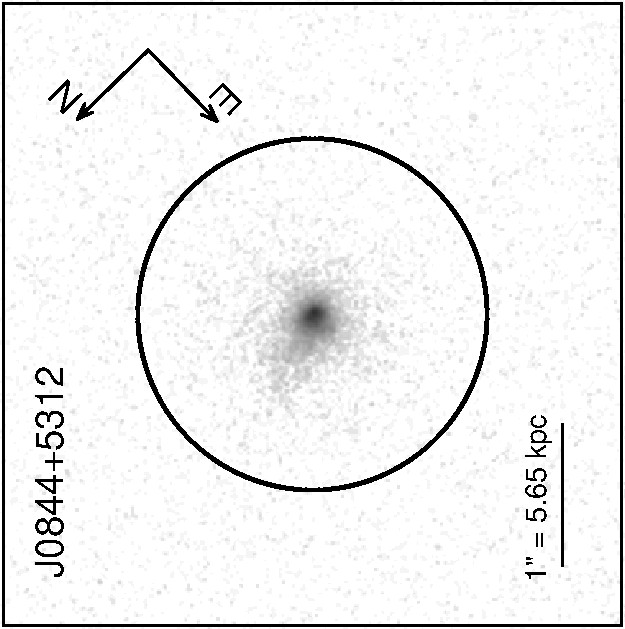}
}
\centering
\includegraphics[angle=-90,width=0.32\linewidth]{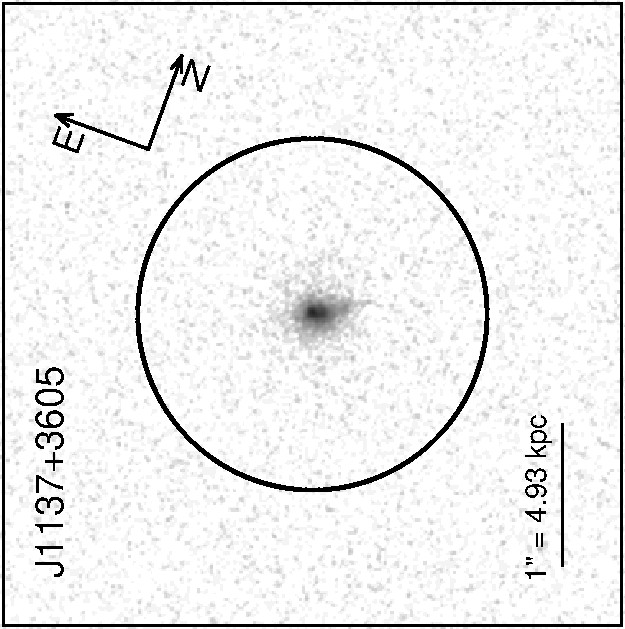}
\includegraphics[angle=-90,width=0.32\linewidth]{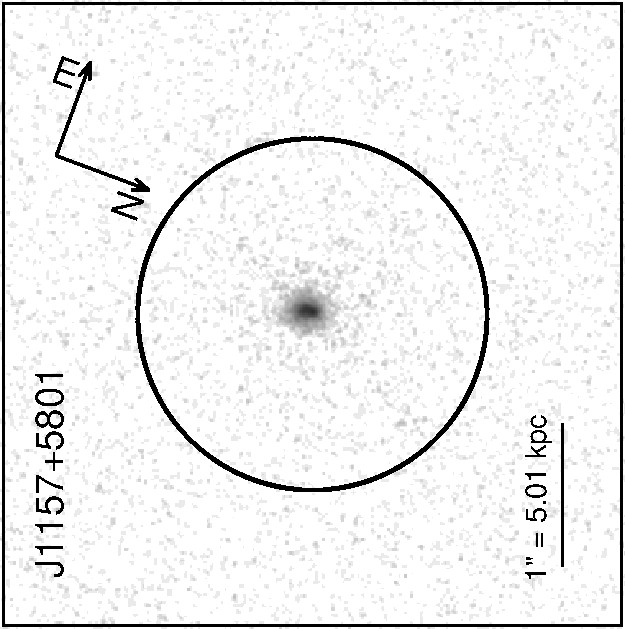}
\caption{The {\sl HST}/COS NUV acquisition images of the candidate LyC leaking
galaxies in a log surface brightness scale. The COS spectroscopic aperture
with a diameter of 2.5 arcsec is shown in all panels by a circle.
The linear scale in each panel is derived adopting an angular size distance.
\label{fig1}}
\end{figure*}

  \begin{table*}
  \caption{Integrated characteristics \label{tab2}}
  \begin{tabular}{lccccccccc} \hline
Name&$M_{\rm FUV}$$^{\rm a}$&log $M_\star$$^{\rm b}$ &log $L$(H$\beta$)$^{\rm c}$&$t_b$$^{\rm d}$&SFR$^{\rm e}$&$\alpha$$^{\rm f}$&$r_{50}$$^{\rm g}$&$\Sigma$$_1^{\rm h}$&$\Sigma$$_2^{\rm i}$\\
\hline   
J0130$-$0014&$-$18.24& 8.63    &41.22&0.4& 3.7    &0.20&0.10&27&118\\
J0141$-$0304&$-$19.47& 9.99    &42.23&2.0&36.0\,~~&0.63&0.21&30&269\\
J0844$+$5312&$-$20.79& 8.18    &42.07&3.8&25.3\,~~&0.54&0.19&28&228\\
J1014$+$5501&$-$19.32& 7.48    &41.60&3.0& 8.6    & ...$^{\rm j}$& ...$^{\rm j}$& ...$^{\rm j}$&...$^{\rm j}$\\
J1137$+$3605&$-$18.46& 9.17    &41.99&0.3&21.1\,~~&0.46&0.19&33&191\\
J1157$+$5801&$-$18.49& 9.28    &41.77&3.0&13.0\,~~&0.21&0.16&95&163\\
J1352$+$5617&$-$19.48& 9.40    &41.76&2.7&12.6\,~~& ...$^{\rm j}$& ...$^{\rm j}$& ...$^{\rm j}$&...$^{\rm j}$\\
\hline
  \end{tabular}

\hbox{$^{\rm a}$Absolute FUV magnitude derived from the intrinsic rest-frame SED in mag.}


\hbox{$^{\rm b}$$M_\star$ = $M_{\rm y}$ $+$ $M_{\rm o}$ is the total stellar mass in
  M$_\odot$, where $M_{\rm y}$ and $M_{\rm o}$ are masses of the young and old
  stellar population, respectively.}

\hbox{$^{\rm c}$$L$(H$\beta$) is the H$\beta$ luminosity corrected for the Milky Way and internal extinction
in erg s$^{-1}$.}

\hbox{$^{\rm d}$$t_b$ is the starburst age in Myr.}

\hbox{$^{\rm e}$Star-formation rate corrected for the Milky Way and internal extinction, and escaping LyC radiation in M$_\odot$ yr$^{-1}$.}


\hbox{$^{\rm f}$Exponential disc scale length in kpc.}

\hbox{$^{\rm g}$Galaxy radius, at which NUV intensity is equal to half of maximal intensity, in kpc.}

\hbox{$^{\rm h}$Star-formation rate surface density assuming galaxy radius is equal to $\alpha$ 
in M$_\odot$ yr$^{-1}$kpc$^{-2}$.}

\hbox{$^{\rm i}$Star-formation rate surface density assuming galaxy radius is equal to $r_{50}$ 
in M$_\odot$ yr$^{-1}$kpc$^{-2}$.}

\hbox{$^{\rm j}$Acquisition image not obtained.}
  \end{table*}

Mg~{\sc ii} $\lambda$2796, 2803 emission may also provide a constraint of the
LyC escape and its doublet ratio can be used to infer the neutral gas
column density \citep{Hen18,Ch20,Xu22,N22,Ka22}. These
two lines in emission are commonly seen in the spectra of local
compact star-forming galaxies \citep{G13,G19} including
LyC leaking galaxies \citep{Ch20,G20} and might be more likely to leak
  LyC than similar galaxies without strong Mg~{\sc ii} \citep{Xu22}.
They are also detected in $z$ $\sim$ 1 - 2 galaxies \citep{W09,E12,F17,N22}
and in a $z$ $\sim$ 5 star-forming galaxy \citep{Wi21}. \citet{Hen18} found that
the Mg~{\sc ii} escape fraction correlates with the Ly$\alpha$ escape
fraction, and that the Mg~{\sc ii} emission line profiles are broader and
more redshifted in galaxies with low escape fractions. They and
\citet{Ch20} pointed out that the link between Ly$\alpha$ and
Mg~{\sc ii} can be used for a LyC diagnostic at high redshifts, where Ly$\alpha$
and LyC are difficult to observe. However, \citet{Ka22} pointed out from
  the numerical simulations that Mg~{\sc ii} is a useful diagnostic of escaping
ionizing radiation only in the optically thin regime.

The goal of this paper is to determine $f_{\rm esc}$(LyC)
for seven low-mass galaxies with various
Mg$_2$ = Mg~{\sc ii} $\lambda$2796/Mg~{\sc ii} $\lambda$2803
flux ratios and various O$_3$Mg$_2$ =
[O~{\sc iii}]$\lambda$5007/Mg~{\sc ii} $\lambda$2796+2803 flux ratios.
The O$_3$Mg$_2$ flux ratios range
from 10 to 35 in six galaxies and $\ga$ 100 in one galaxy, where Mg~{\sc ii}
emission is almost undetected. We aim to study the dependence of leaking LyC
emission on the characteristics of Mg~{\sc ii} emission. We also wish to
enlarge the known sample of low-redshift LyC leakers, to search
for and to improve reliable diagnostics for the indirect estimation of
$f_{\rm esc}$(LyC).
The properties of the selected SFGs derived from
observations in the optical range are presented in Section~\ref{sec:integr}. The
{\sl HST} observations and data reduction are described in Section~\ref{sec:obs}.
The surface brightness profiles in the UV range are discussed in
Section~\ref{sec:sbp}. In Section \ref{sec:global}, we compare the {\sl HST}/COS spectra with the
extrapolation of the SEDs modelled with the SDSS spectra to the
UV range. Ly$\alpha$ emission and escaping Lyman continuum emission
are discussed in Section~\ref{sec:lya} together with the corresponding escape
fractions. The indirect indicators of escaping LyC emission are
considered in Section~\ref{sec:ind}. Mg~{\sc ii} diagnostics are discussed in
Section~\ref{sec:MgII}. We summarize our findings in
Section~\ref{sec:summary}.

  \begin{table}
  \caption{{\sl HST}/COS observations \label{tab3}}
  \begin{tabular}{lcccc} \hline
\multicolumn{1}{c}{}&\multicolumn{1}{c}{}&\multicolumn{3}{c}{Exposure time (s)} \\ 
\multicolumn{1}{c}{Name}&\multicolumn{1}{c}{Date}&\multicolumn{3}{c}{(Central wavelength (\AA))} \\ 
    &    &MIRRORA&G140L&G160M \\ \hline
J0130$-$0014&2020-11-20&2$\times$700     &  5321& 3550\\
            &          &         &(800)&(1533)\\
J0141$-$0304&2021-01-18&2$\times$700     &  5321& 3560\\
            &          &         &(800)&(1589)\\
J0844$+$5312&2021-05-18&2$\times$700     &  5516& 3932\\
            &          &         &(800)&(1623)\\
J1014$+$5501&2021-05-14&2$\times$0$^{\rm a}$     &  5711& 4054\\
            &          &         &(800)&(1589)\\
J1137$+$3605&2021-02-04&2$\times$700     &  5437   & 3678\\
            &          &         &(800)&(1577)\\ 
J1157$+$5801&2020-10-18&2$\times$700     &  5705& 4054\\
            &          &         &(800)&(1589)\\ 
J1352$+$5617&2021-02-01&2$\times$0$^{\rm a}$     &  5815& 4044\\
            &          &         &(800)&(1623)\\ 
\hline
\end{tabular}

\hbox{$^{\rm a}$Failed exposure.}


  \end{table}

\section{Integrated properties of selected galaxies}\label{sec:integr}

We selected a sample of local compact low-mass SFGs from the
SDSS in the redshift range $z$ = 0.32 - 0.43 with O$_3$Mg$_2$ in a wide
range to observe their Ly$\alpha$ and LyC emission with {\sl HST}/COS.
These galaxies are chosen to be sufficiently bright,
to have high O$_{32}$ ratios and high equivalent widths EW(H$\beta$) of the
H$\beta$ emission line. This ensures that a galaxy can be acquired and observed
with low- and medium-resolution gratings in one visit, consisting
of 4 orbits. Finally we selected a total sample of 7 galaxies
with EW(H$\beta$) $>$ 170\,\AA\ and O$_{32}$\,$\ga$\,4. They
are listed in Table\,\ref{tab1}. All galaxies are nearly unresolved by the SDSS
5-band images and have FWHMs of $\sim$\,1.0 arcsec, so that all the
galaxy’s light falls within the 2.5 arcsec diameter COS aperture and
within the 2 arcsec diameter SDSS aperture. This ensures that global
quantities can be derived from both the UV and optical spectra.
We note, however, that Mg~{\sc ii} lines are located in the noisy parts of SDSS
spectra and detected with a low signal-to-noise ratio, at least in some
galaxies. As such, their fluxes, and especially the Mg~{\sc ii} flux ratio
Mg$_2$ = Mg~{\sc ii}\,$\lambda$2796/Mg~{\sc ii}\,$\lambda$2803 should only be
considered tentatively.
We note that follow up spectroscopy of these galaxies
with high signal-to-noise ratio covering the wavelength range with Mg~{\sc ii}
emission will be presented in King et al., in preparation.

The SDSS, {\sl GALEX} and {\sl WISE} apparent magnitudes of the
selected galaxies are shown in Table \ref{taba1}, indicating that these SFGs
are among the faintest low-redshift LyC leaker candidates selected
so far for {\sl HST} observations.

To derive absolute magnitudes and other integrated parameters
we adopted luminosity and angular size distances \citep[NASA Extragalactic
Database (NED),][]{W06}
with the cosmological parameters $H_0$ = 67.1 km s$^{-1}$ Mpc$^{-1}$,
$\Omega_\Lambda$ = 0.682, $\Omega_m$ = 0.318 \citep{P14}.
These distances are presented in Table \ref{tab1}.

Internal interstellar extinction $A$($V$)$_{\rm int}$ has been derived from
the observed decrement of hydrogen emission lines in the SDSS
spectra after correction for the Milky Way extinction with $A$($V$)$_{\rm MW}$
from the NED, adopting the \citet*{C89} reddening
law and $R$($V$)$_{\rm int}$ = 2.7 and $R$($V$)$_{\rm MW}$ = 3.1.
The motivation of the adopted $R$($V$)$_{\rm int}$ value is following.
\citet{I17} modelled UV FUV and NUV magnitudes of the large sample of
SDSS galaxies and found that the FUV magnitudes of galaxies
better match the observed magnitudes with $R$($V$)$_{\rm int}$ = 2.7 if
EW(H$\beta$) $>$ 150\AA, which is the case for our galaxies, whereas
$R$($V$)$_{\rm int}$ = 3.1 is more appropriate for galaxies with lower
EW(H$\beta$)s. However, we note that in the optical range, which is used for
SED fitting, the determination of intrinsic fluxes of the Lyman continuum and
of the elemental abundances, extinction does only slightly depend on
$R$($V$)$_{\rm int}$.

The extinction-corrected
emission lines are used to derive ionic and total element abundances
following the methods described in \citet{I06} and \citet{G13}.

The emission-line fluxes $I$($\lambda$) relative to the H$\beta$ flux corrected
for both the Milky Way and internal extinctions, the restframe
equivalent widths, the Milky Way ($C$(H$\beta$)$_{\rm MW}$) and internal
($C$(H$\beta$)$_{\rm int}$) extinction coefficients, and extinction-corrected
H$\beta$ fluxes are shown in Table \ref{taba2}. It is seen in the Table that
the extinction-corrected fluxes of the H$\delta$, H$\gamma$ and H$\alpha$
emission lines in all galaxies are consistent within the errors with
theoretical recombination values indicating that $C$(H$\beta$)$_{\rm int}$
is derived correctly.

\begin{figure*}
\hbox{
\includegraphics[angle=-90,width=0.32\linewidth]{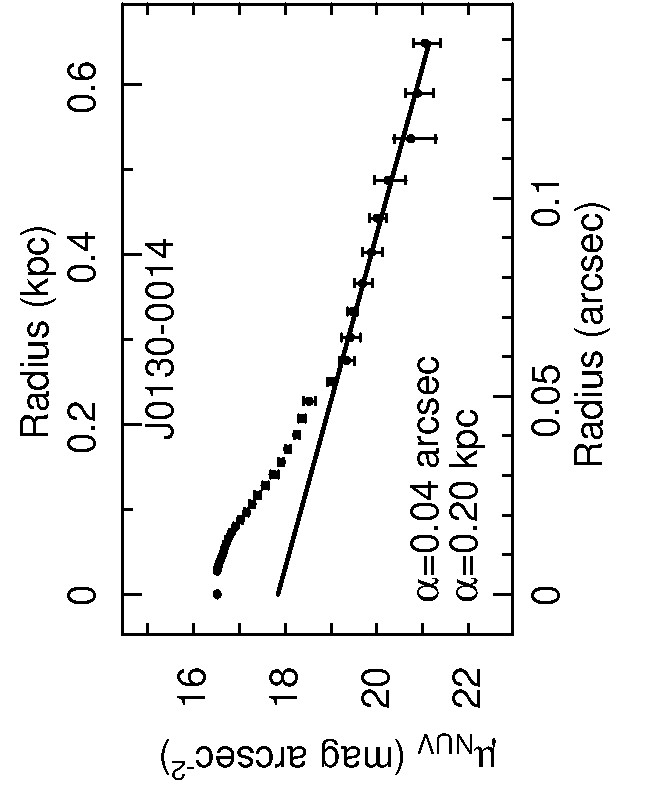}
\hspace{0.2cm}\includegraphics[angle=-90,width=0.32\linewidth]{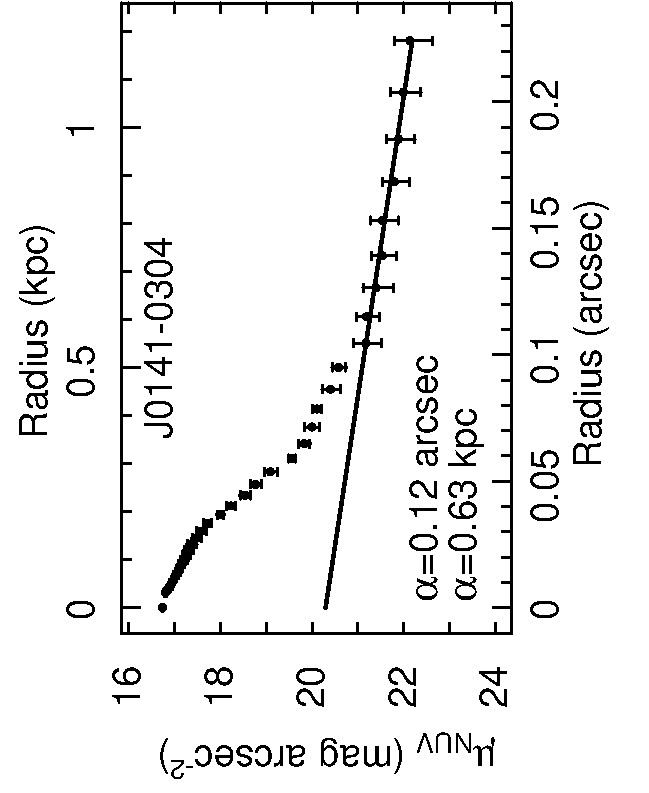}
\hspace{0.2cm}\includegraphics[angle=-90,width=0.32\linewidth]{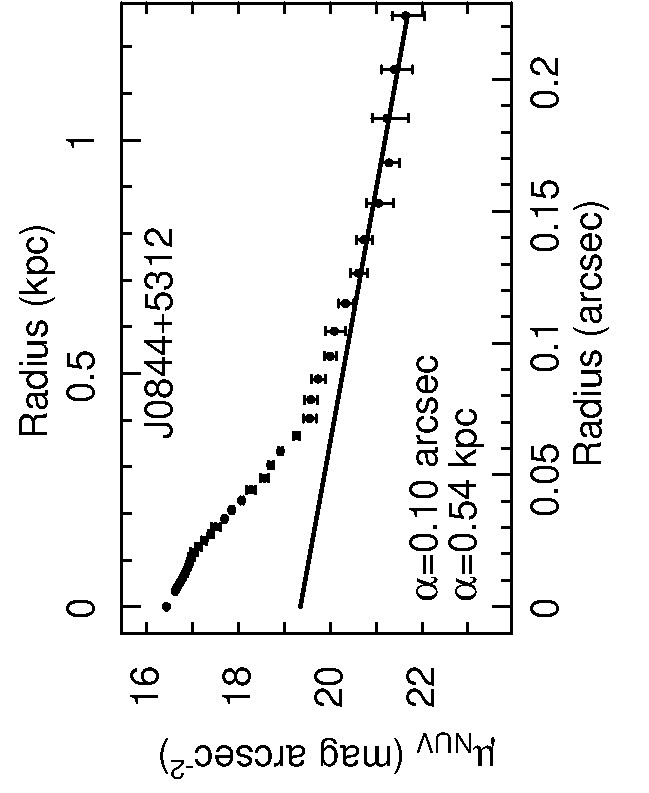}
}
\centering
\includegraphics[angle=-90,width=0.32\linewidth]{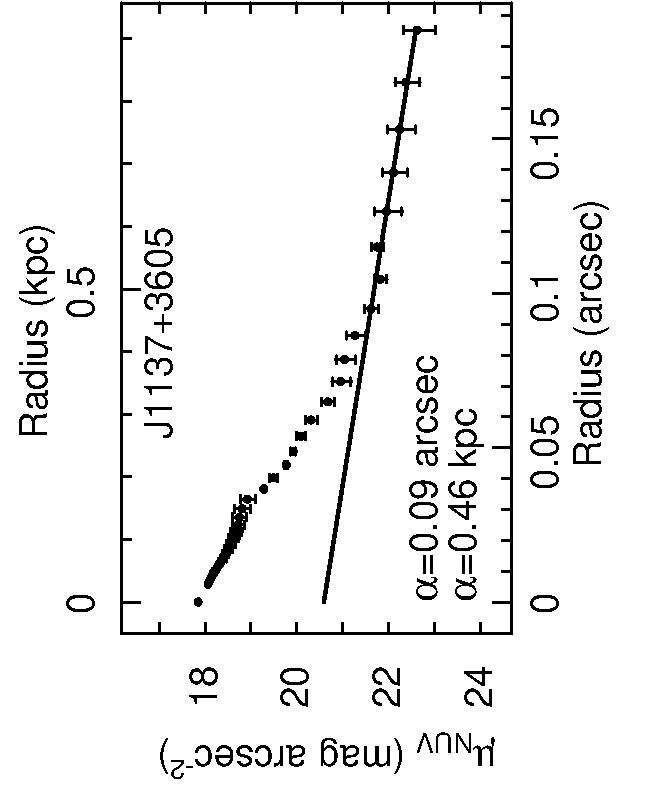}
\hspace{0.2cm}\includegraphics[angle=-90,width=0.32\linewidth]{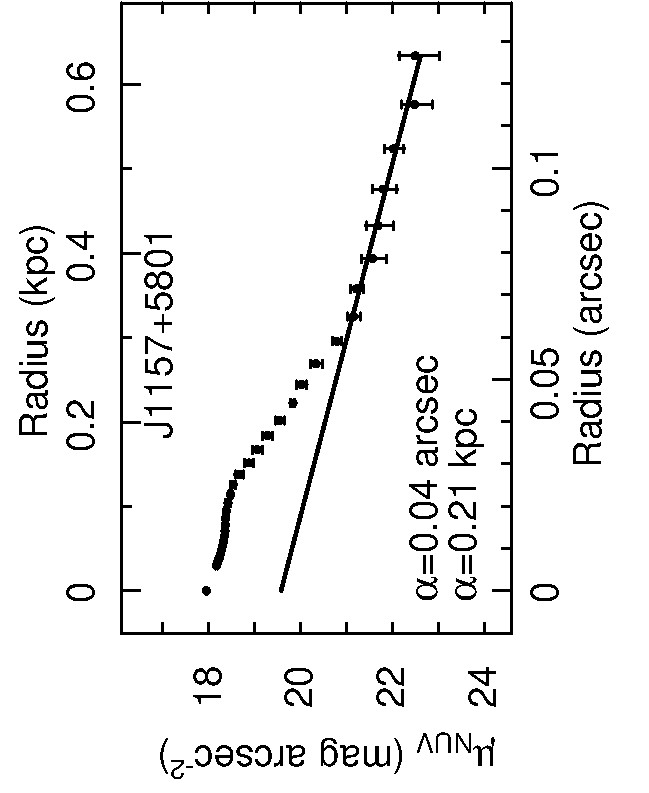}
\caption{NUV surface brightness profiles of galaxies indicated by the dots
  with error bars. Straight lines are linear fits of surface brightness
  $\mu_{\rm NUV}$ in outer galaxy regions, and $\alpha$s are exponential
  scale lengths in arcsec and kpc.
\label{fig2}}
\end{figure*}

The fluxes and the direct
$T_{\rm e}$ method are used to derive the physical conditions (electron
temperature and electron number density) and the element abundances in the
H~{\sc ii} regions. These quantities are shown in Table \ref{taba3}. The derived
oxygen abundances are comparable to those in known low-redshift
LyC leakers by \citet{I16a,I16b,I18a,I18b,I21a}. The ratios
of the $\alpha$-element (neon and magnesium) abundances to oxygen
abundance are similar to those in dwarf emission-line galaxies
\citep[e.g. ][]{I06,G13}. On the other hand, the
nitrogen-to-oxygen abundance ratios in some galaxies are somewhat
elevated, similar to those in other LyC leakers at $z$ $\ga$ 0.3.

We determine absolute FUV magnitudes from the fluxes of the intrinsic
(i.e. extinction-corrected)
SEDs at the rest-frame wavelength $\lambda$ = 1500\,\AA, which are reddened
adopting extinction derived from the observed decrement of hydrogen Balmer
lines. 
The attenuations are, on average, similar to the ones for other
$z$ $\sim$ 0.3 - 0.4 LyC
leakers and the $M_{\rm FUV}$ are similar as observed at high-redshift.

The H$\beta$ luminosities $L$(H$\beta$) and the corresponding star-formation
rates, SFR, were obtained from the extinction-corrected
H$\beta$ fluxes, using the relation from \citet{K98} for the SFR and adopting
$I$(H$\alpha$)/$I$(H$\beta$) from Table\,\ref{taba2}. SFRs
are increased by a factor 1/[1~$-$~$f_{\rm esc}$(LyC)] to take into account
the escaping ionizing radiation which is discussed later. The SFRs
corrected for escaping LyC radiation are shown in Table \ref{tab2}. They are
somewhat below the range of 14 - 80 M$_\odot$ yr$^{-1}$ for the other LyC
leakers studied by \citet{I16a,I16b,I18a,I18b,I21a}.

We use the SDSS spectra of our LyC leakers to fit the SED
in the optical range and derive their stellar masses. The fitting method,
using a two-component model with a young burst and older continuosly formed
stellar population, is described for example in \citet{I18a,I18b}.
Spectral energy distributions of instantaneous bursts
in the range between 0 and 10 Gyr with evolutionary tracks
of non-rotating stars by \citet{G00} and a combination
of stellar atmosphere models \citep*{L97,S92} are used to produce the integrated
SED for each galaxy. The star formation history is approximated
by a young burst with a randomly varying age $t_b$ in the range
$<$ 10 Myr, and a continuous star formation for older ages between
times $t_1$ and $t_2$, randomly varying in the range 10 Myr - 10 Gyr, and
adopting a constant SFR. The contribution of the two components
is determined by randomly varying the ratio of their stellar masses,
$b$ = $M_{\rm o}$/$M_{\rm y}$, in the range 0.1 - 1000, where $M_{\rm o}$ and
$M_{\rm y}$ are the masses of the old and young stellar populations.

The nebular continuum emission, including free-free and free-bound
hydrogen and helium emission, and two-photon emission, is
also taken into account using the observed H$\beta$ flux (i.e. not corrected
for escaping LyC emission), the ISM temperature,
and density. The fraction of nebular continuum emission in
the observed spectrum near H$\beta$ is determined by the ratio of the observed
H$\beta$ equivalent width EW(H$\beta$)$_{\rm obs}$, shifted to the rest frame,
to the equivalent width EW(H$\beta$)$_{\rm rec}$ for pure nebular emission.
EW(H$\beta$)$_{\rm rec}$
varies from $\sim$\,900\,\AA\ to $\sim$\,1100\,\AA, for electron temperatures in the
range $T_{\rm e}$\,=\,10000\,-\,20000\,K. We note that non-negligible nebular
emission in the continuum is produced only by the young burst with ages
of a few Myr.

The \citet{S55} initial mass function (IMF) is adopted, with a
slope of $-$2.35, upper and lower mass limits $M_{\rm up}$ and $M_{\rm low}$ of
100\,M$_\odot$ and 0.1\,M$_\odot$, respectively. \citet{I16a} compared
  differences in SEDs obtained with two different IMFs, by \citet{S55} and
  \citet{K01}. They concluded that the effect is minor.
 A $\chi$$^2$ minimization technique
was used 1) to fit the continuum in such parts of the restframe wavelength
range 3600 - 6500\,\AA, where the SDSS spectrum is least noisy and
free of nebular emission lines, and 2) to reproduce the observed H$\beta$
and H$\alpha$ equivalent widths.

\begin{figure*}
\hbox{
\includegraphics[angle=-90,width=0.32\linewidth]{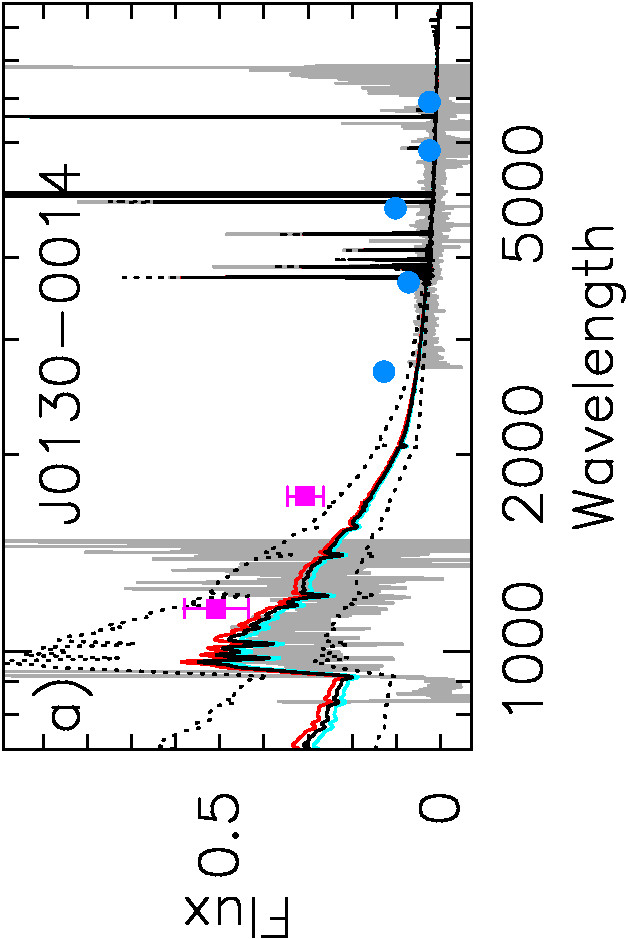}
\hspace{0.2cm}\includegraphics[angle=-90,width=0.32\linewidth]{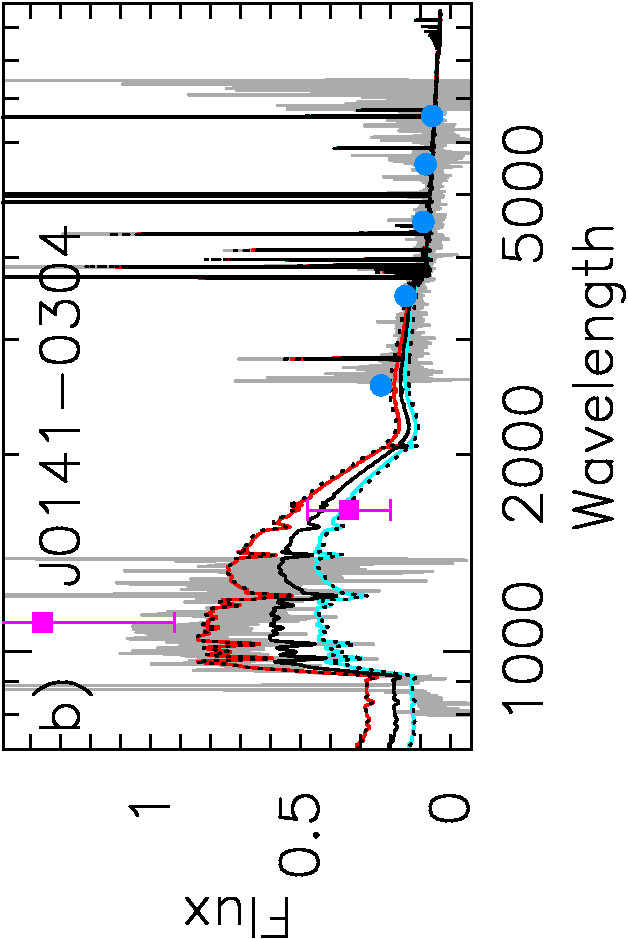}
\hspace{0.2cm}\includegraphics[angle=-90,width=0.32\linewidth]{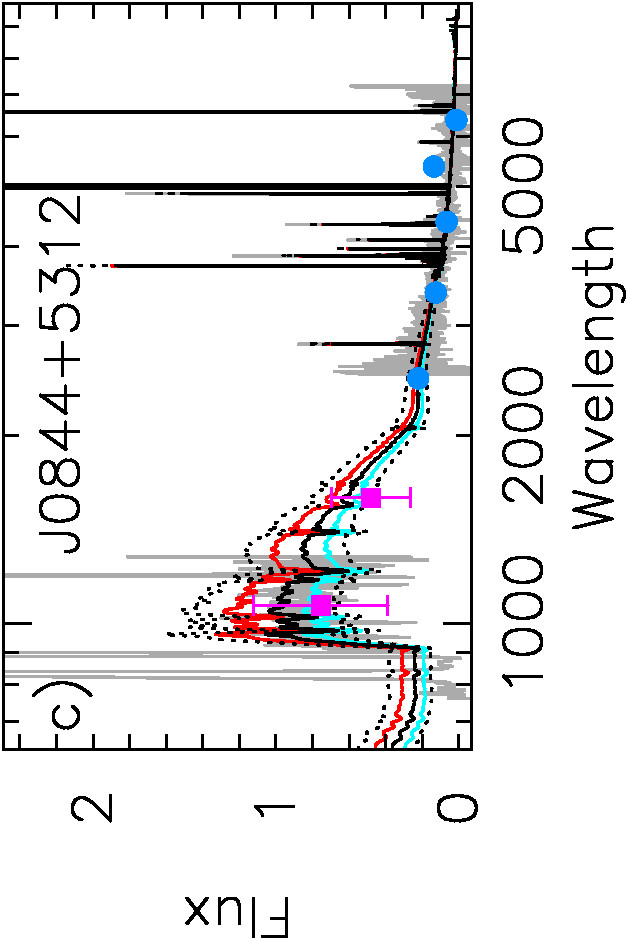}
}
\hbox{
\includegraphics[angle=-90,width=0.32\linewidth]{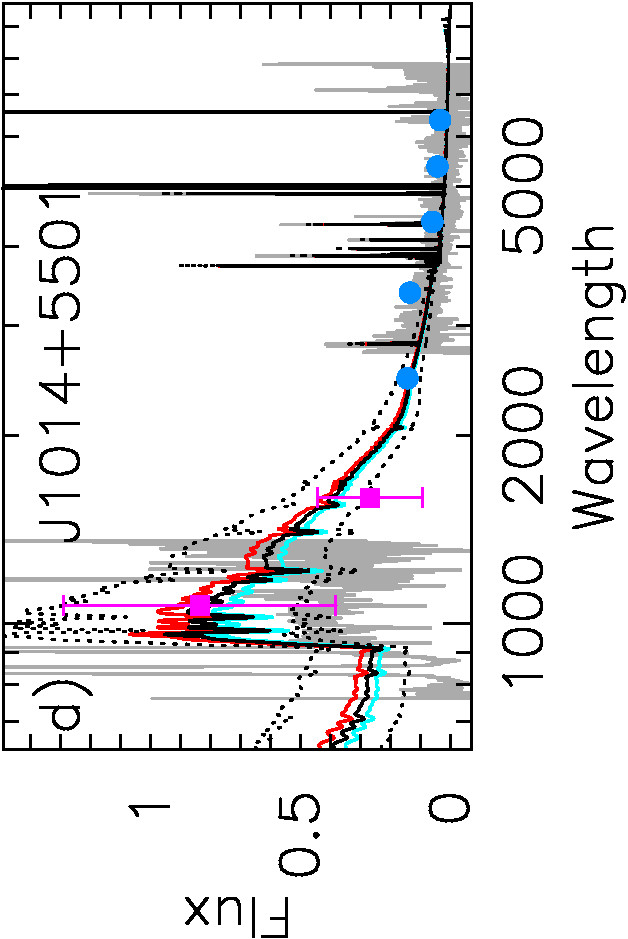}
\hspace{0.2cm}\includegraphics[angle=-90,width=0.32\linewidth]{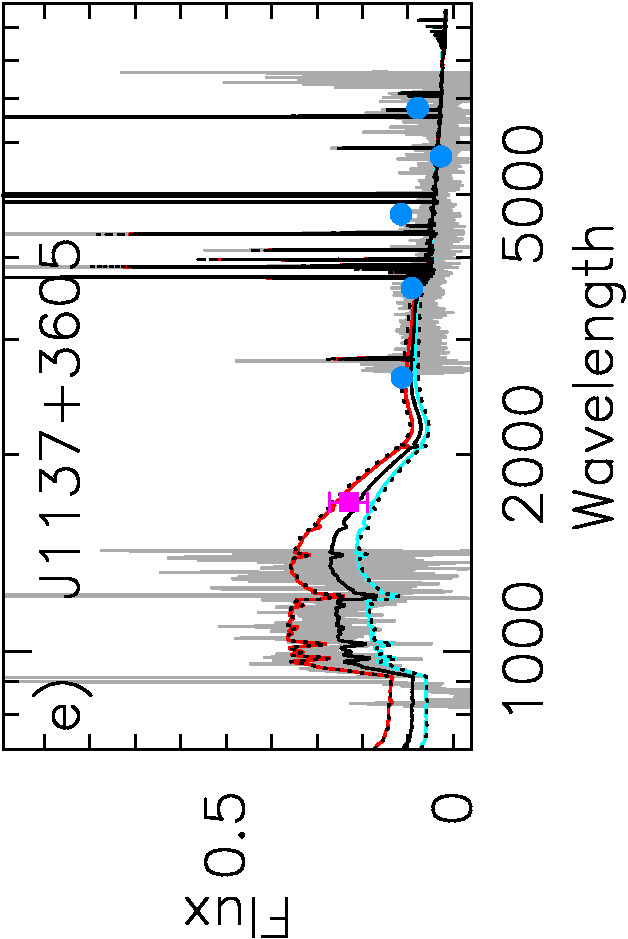}
\hspace{0.2cm}\includegraphics[angle=-90,width=0.32\linewidth]{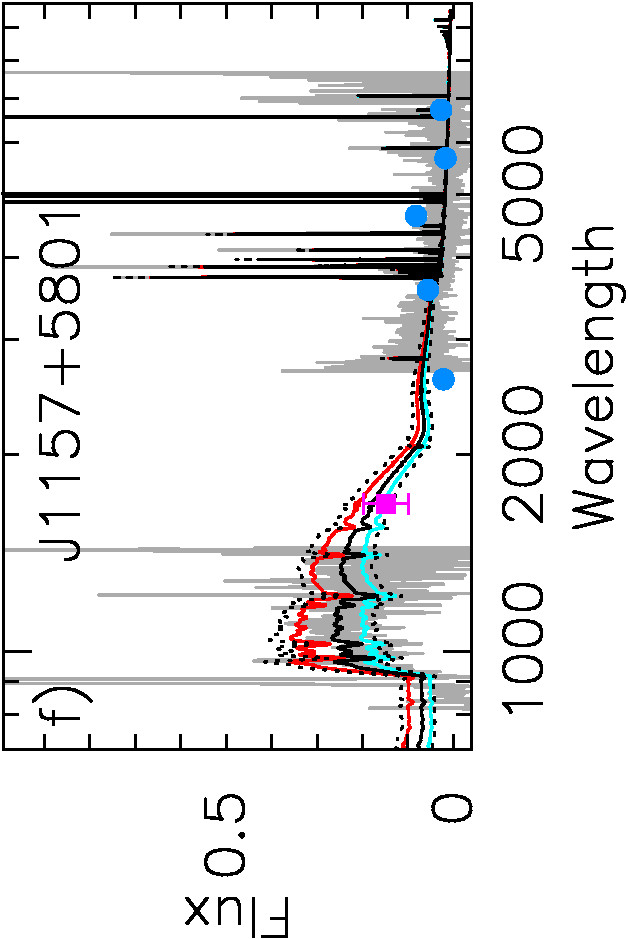}
}
\centering
\includegraphics[angle=-90,width=0.32\linewidth]{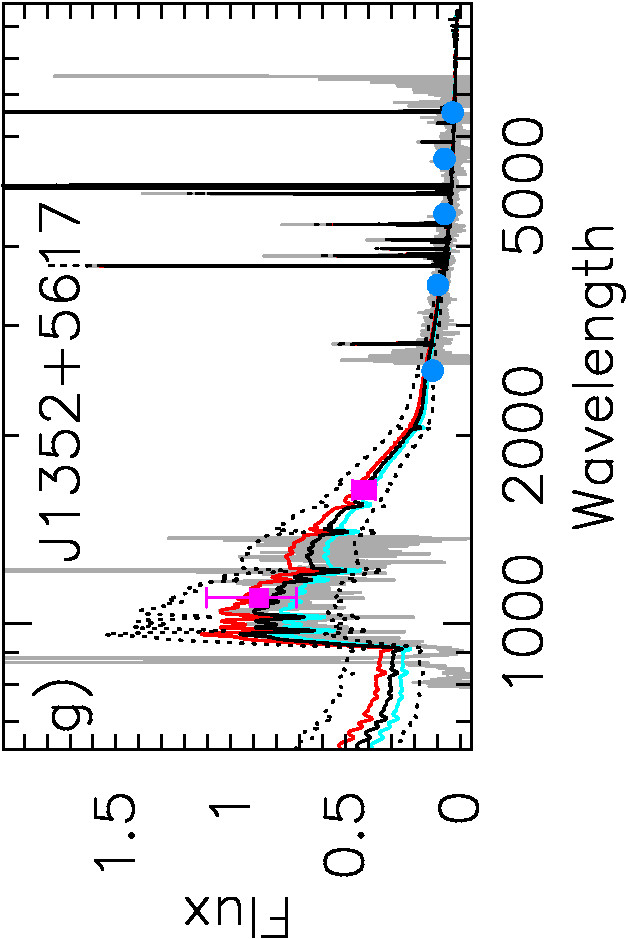}
\caption{A comparison of the COS G140L and SDSS spectra (grey lines), and
  photometric data together with the modelled SEDs of the optical
  spectra and their extrapolation to the UV range in the restframe wavelength
  scale. {\sl GALEX} FUV and NUV fluxes and SDSS fluxes 
in $u,g,r,i,z$ bands are shown by magenta-filled squares with 1$\sigma$
deviations and blue-filled 
circles, respectively. Modelled intrinsic SEDs and their extrapolation to
the UV range, which are reddened by the Milky 
Way extinction with $R(V)_{\rm MW}$ = 3.1 and
internal extinction with $R(V)_{\rm int}$ = 3.1, 2.7 and 2.4, are shown by red,
black and cyan solid lines, respectively. The black dotted lines show the
1$\sigma$ spread of the SED fit reddened with the $C$(H$\beta$)$_{\rm int}$
values and $R(V)_{\rm int}$ = 2.7. Fluxes are in 
10$^{-16}$ erg s$^{-1}$ cm$^{-2}$\AA$^{-1}$, wavelengths are in \AA. \label{fig3}}
\end{figure*}

The total stellar masses ($M_\star$ = $M_{\rm y}$ + $M_{\rm o}$) of our LyC leakers
derived from SED fitting are presented in Table \ref{tab2}. They are derived
in exactly the same way as the stellar masses of the other LyC leakers studied
by \citet{I16a,I16b,I18a,I18b,I21a}, permitting a direct comparison.

\section{{\sl HST}/COS observations and data 
reduction}\label{sec:obs}

{\sl HST}/COS spectroscopy of the seven selected galaxies was obtained
in program GO~15845 (PI: Y.\ I.\ Izotov) during the period October 2020 --
May 2021. The observational details
are presented in Table \ref{tab3}. As in our previous programs
\citep{I16a,I16b,I18a,I18b,I21a}, the galaxies were directly acquired by COS
near ultraviolet (NUV) imaging. All these galaxies are compact (as compact
  as all the other targets from our previous programs) and they have accurate
  SDSS astrometry for direct imaging acquisition.
The NUV-brightest region of each target was centered in the
2.5\,arcsec diameter spectroscopic aperture (Fig.~\ref{fig1}).
We note, however, that the acquisition exposure failed for J1014$+$5501
and J1352$+$5617 due to guide star acquisition failure in both cases
  because the acquisition of the guide stars was delayed. This is a frequent {
    \sl HST} gyro issue. For safety reasons,
  the shutter remained closed and no acquisition image was obtained.
  Therefore, both galaxies were blindly acquired. The blind acquisition
  accuracy is $\sim$ 0.3 arcsec, which will result in very modest vignetting
  for a compact galaxy, possibly
introducing uncertainties in the wavelength and flux
calibration in the partially vignetted COS aperture. For J1352$+$5617 the
vignetting is negligible, because the COS spectrophotometric magnitude
(FUV$=21.90$ mag) agrees well with the {\sl GALEX} FUV$=21.83\pm 0.17$ mag. For
J1014$+$5501 the spectrophotometry (FUV$=22.69$ mag) is still consistent with
the {\sl GALEX} magnitude (FUV$=21.88\pm 0.59$ mag), considering the significant
Eddington bias for the latter. The wavelength calibration was confirmed with
Lyman series absorption lines of the galaxies.

The spectra were obtained with the 
low-resolution grating G140L and medium-resolution grating G160M, applying all 
four focal-plane offset positions.
The 800\,\AA\ setup was used for the G140L grating 
(sensitive wavelength range 1100--1950\,\AA, resolving
power $R\simeq 1050$ at 1150\,\AA) to include the redshifted LyC
emission for all targets. We obtained resolved spectra of the galaxies'
Ly$\alpha$ emission lines with the G160M grating ($R\sim 16000$ at
1600\,\AA), varying the G160M central wavelength with galaxy redshift
to cover the emission line and the nearby continuum on a single detector 
segment. 

The individual exposures were reduced with the \textsc{calcos} pipeline v3.3.10,
followed by accurate background subtraction and co-addition as required for our
Poisson-limited data with \textsc{FaintCOS} v1.09 \citep{Ma21}.
We used the same methods and extraction
aperture sizes as in \citet{I18a,I18b,I21a} to achieve a homogeneous
reduction of the galaxy sample observed in multiple programmes.We corrected
  for scattered geocoronal Ly$\alpha$ according to \citet{W16}. The accuracy of
  our custom correction for scattered light in COS G140L data was checked
by comparing the LyC fluxes obtained in the total exposure and in 
orbital night, respectively. We find that the differences in LyC fluxes
for five galaxies are less or similar to the 1$\sigma$ errors.
Due to insufficient time spent in orbital night,
this check was not possible for J1157$+$5801 and J1352$+$5617. However, we
verified that the detected LyC flux of J1352$+$5617 (Section~\ref{sec:lya}) is
insignificantly affected by residual uncertainties in the G140L scattered
light model.

\section{Acquisition images and surface brightness profiles in the NUV range}\label{sec:sbp}

The acquisition images of five galaxies in the NUV range are shown
in Fig.\,\ref{fig1}. All galaxies are very compact with angular diameters
considerably smaller than the COS spectroscopic aperture (the circles
in Fig.\,\ref{fig1}) and linear diameters of $\sim$ 1 -- 4 kpc. However, two
of the most compact galaxies, J0130$-$0014 and J1157$+$5801, appear to
be non-leaking LyC galaxies, whereas LyC emission is detected in the remaining
three galaxies with more extended envelopes (see Section\,\ref{sec:lya}).
We use these images to derive the surface brightess (SB) profiles
of our galaxies, in accordance with previous studies by
\citet{I16b,I18a,I18b,I21a}. No SB profiles have been derived for
galaxies J1014$+$5501 and J1352$+$5617 because their acquisition exposures
failed, as noted before.
In accordance with \citet{I16b,I18a,I18b,I21a} we have
found that the outer parts of our galaxies are characterised by a linear
decrease in SB (in mag per square arcsec scale), characteristic of a
disc structure, and by a sharp increase in the central part due to the
bright star-forming region (Fig.\,\ref{fig2}).
The scale lengths $\alpha$ of our galaxies, defined in Eq. 1 of
\citet{I16b}, are in the range $\sim$\,0.2\,--\,0.6\,kpc (Fig.\,\ref{fig2}), lower
than $\alpha$\,=\,0.6\,--\,1.8\,kpc in other LyC leakers \citep{I16b,I18a,I18b},
but similar to scale lengths of low-mass LyC leakers with
masses $<$\,10$^8$ M$_\odot$ \citep{I21a}. The corresponding surface
densities of star-formation rate in the studied galaxies, $\Sigma$ =
SFR/($\pi\alpha^2$), are similar to those of other LyC leakers. The half-light
radii $r_{50}$ of our galaxies in the NUV are considerably smaller than
$\alpha$ because of the bright compact star-forming regions
in the galaxy centres (see Table\,\ref{tab2}).

\section{Modelled spectral energy distributions in the UV range}\label{sec:global}

To derive the fraction of the escaping ionizing radiation we use the two
methods \citep[e.g. ][]{I18a} based on the comparison
between the observed flux in the Lyman continuum and its
intrinsic flux in the same wavelength range.
The intrinsic LyC flux is obtained 1) from SED fitting of
the SDSS spectra simultaneously with reproducing the observed H$\beta$ and
H$\alpha$ equivalent widths (and thus corresponding observed H$\beta$ and
H$\alpha$ fluxes) or 2) from the flux of the
H$\beta$ emission line. The attenuated extrapolations
of SEDs to the UV range along with the observed COS spectra
are shown in Fig.\,\ref{fig3}. For comparison, we also show the {\sl GALEX} FUV
and NUV fluxes with magenta filled squares and the fluxes in the SDSS
$u, g, r, i, z$ filters with blue filled circles. We find that
the spectroscopic and photometric data in the optical range are
consistent, indicating that almost all the emission of our galaxies
is inside the SDSS spectroscopic aperture. Therefore, aperture corrections
are not needed.

The attenuated modelled intrinsic SEDs in the optical range and
  their extrapolations
  to the UV range (Fig.\,\ref{fig3}) are obtained by assuming
that extinctions for stellar and nebular emission are equal and adopting
the extinction coefficients $C$(H$\beta$)$_{\rm MW}$ from the NED and
$C$(H$\beta$)$_{\rm int}$ derived from the hydrogen Balmer decrement
(Table\,\ref{taba2}), and the reddening law by \citet{C89} at
  $\lambda$ $\geq$ 1250\AA\ and its extension to shorter wavelengths by
  \citet{Ma90} with $R(V)_{\rm int}$ = 3.1
(red solid lines), $R(V)_{\rm int}$ = 2.7 (black solid lines) and $R(V)_{\rm int}$
= 2.4 (cyan solid lines). \citet{Ma90} presents the data only for
$R(V)$ = 3.1. For practical use, we fit them with polynomials and adjusted in
such a way to have the same values at $\lambda$=1250\AA\ and for a variety of
$R(V)$s as the values from the \citet{C89} reddening law at the same
  wavelength and same $R(V)$. The dotted lines indicate the range of attenuated
SEDs adopting $R(V)_{\rm int}$ = 2.7 and varying $C$(H$\beta$) within 1$\sigma$
errors of its nominal value.

  \begin{table*}
  \caption{Parameters for the Ly$\alpha$ emission line \label{tab4}}
  \begin{tabular}{rcrcrrcc} \hline
Name&$A$(Ly$\alpha$)$_{\rm MW}$$^{\rm a}$&\multicolumn{1}{c}{$I$$^{\rm b}$}&log $L$$^{\rm c}$&\multicolumn{1}{c}{EW$^{\rm d}$}&\multicolumn{1}{c}{$V_{\rm sep}$$^{\rm e}$}&\multicolumn{1}{c}{blue/red$^{\rm f}$}&\multicolumn{1}{c}{$f_{\rm esc}$(Ly$\alpha$)$^{\rm g}$} \\ 
\hline
J0130$-$0014&0.211&  1.1$\pm$0.5&40.56&  4.2$\pm$2.4&\multicolumn{1}{c}{...}&\multicolumn{1}{c}{...}& ~1.0$\pm$0.6 \\
J0141$-$0304&0.138&131.9$\pm$3.9&42.83&153.2$\pm$4.9& 308.7$\pm$51.8&28.8  &16.6$\pm$5.3 \\
J0844$+$5312&0.164& 83.4$\pm$7.8&42.75& 61.8$\pm$6.1& 298.5$\pm$41.0&23.8  &20.2$\pm$5.9 \\
J1014$+$5501&0.084& 15.8$\pm$2.8&41.89& 35.6$\pm$6.6& 591.8$\pm$61.2&247.2\,~ & ~8.2$\pm$3.3 \\
J1137$+$3605&0.107& 63.0$\pm$8.6&42.40&224.5$\pm$31.& 328.2$\pm$79.5&49.0  &11.2$\pm$5.2 \\
J1157$+$5801&0.150&  3.9$\pm$1.0&41.22& 15.7$\pm$4.5&\multicolumn{1}{c}{...}&\multicolumn{1}{c}{...}& ~1.1$\pm$0.3 \\
J1352$+$5617&0.054&104.8$\pm$9.7&42.75&111.0$\pm$12.& 394.4$\pm$27.6&12.7  &41.5$\pm$5.9 \\ \hline
  \end{tabular}

\hbox{$^{\rm a}$Milky Way extinction at the observed wavelength of the Ly$\alpha$
emission line in mags adopting \citet{C89} reddening law}

\hbox{\, with $R(V)$=3.1.}

\hbox{$^{\rm b}$Flux in 10$^{-16}$ erg s$^{-1}$ cm$^{-2}$ measured in
the COS spectrum and corrected for the Milky Way extinction.}

\hbox{$^{\rm c}$$L$ is Ly$\alpha$ luminosity in erg s$^{-1}$ corrected for the
Milky Way extinction.}

\hbox{$^{\rm d}$Rest-frame equivalent width in \AA.}

\hbox{$^{\rm e}$Ly$\alpha$ peak separation in km s$^{-1}$.}

\hbox{$^{\rm f}$Flux ratio of blue-to-red peaks in per cent.}

\hbox{$^{\rm g}$Ly$\alpha$ escape fraction in per cent.}

  \end{table*}

\begin{figure*}
\hbox{
\includegraphics[angle=-90,width=0.32\linewidth]{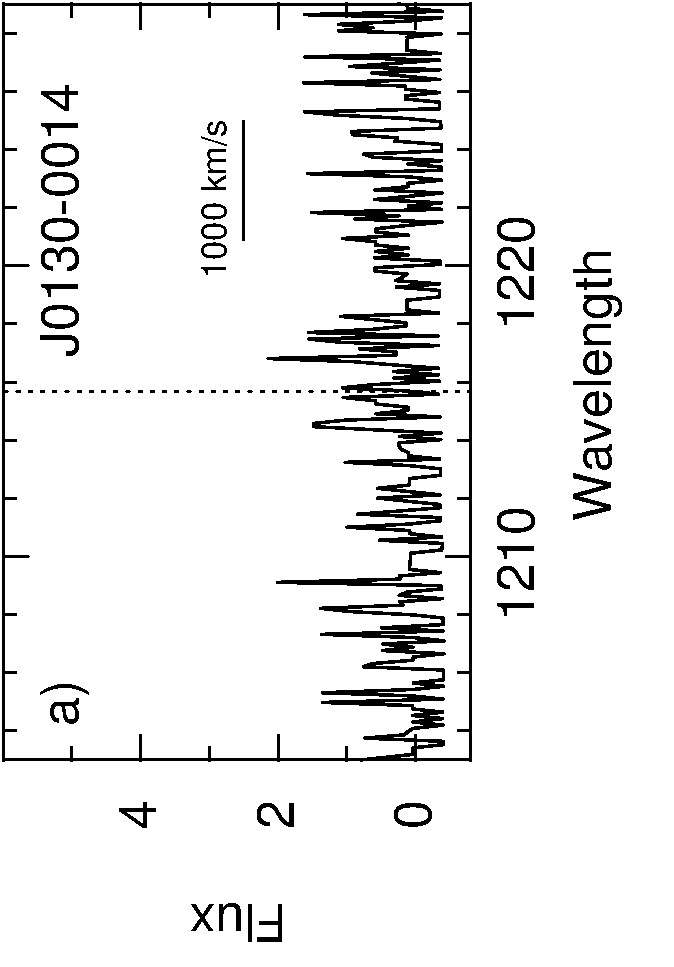}
\hspace{0.2cm}\includegraphics[angle=-90,width=0.32\linewidth]{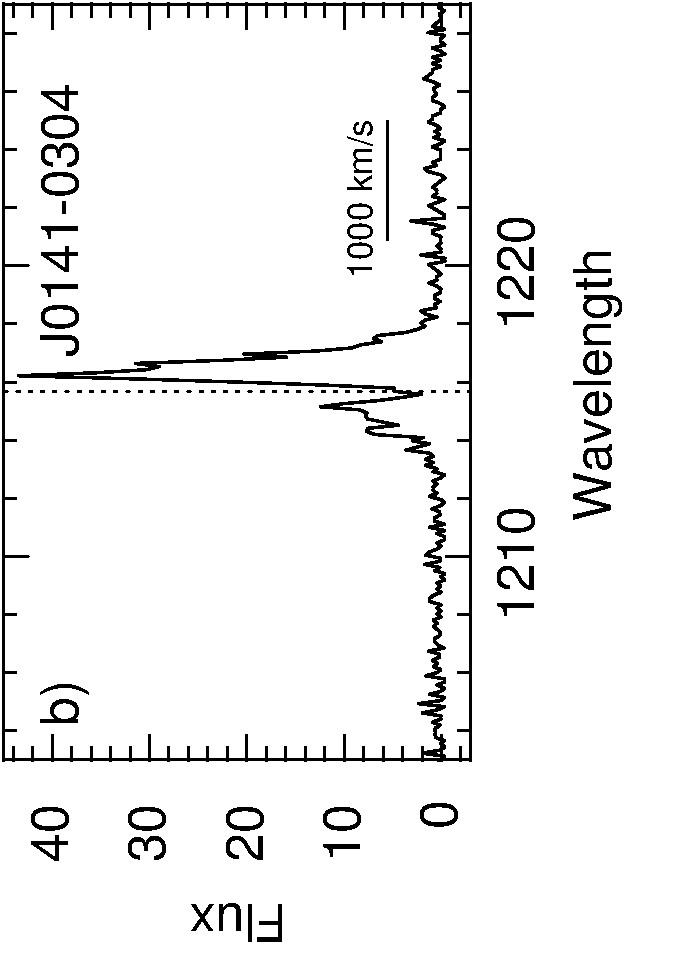}
\hspace{0.2cm}\includegraphics[angle=-90,width=0.32\linewidth]{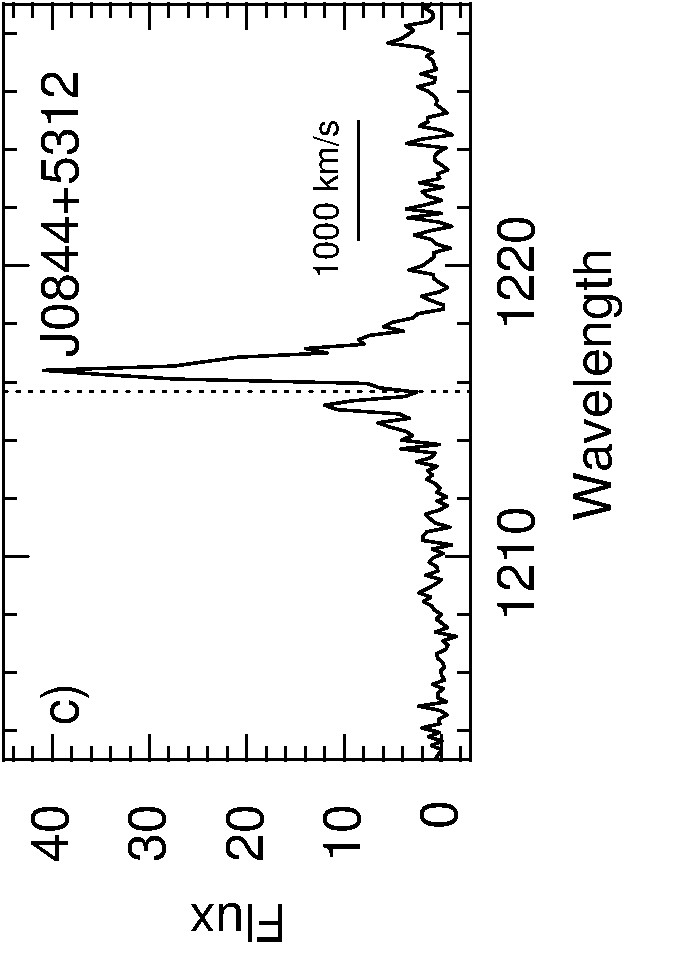}
}
\hbox{
\includegraphics[angle=-90,width=0.32\linewidth]{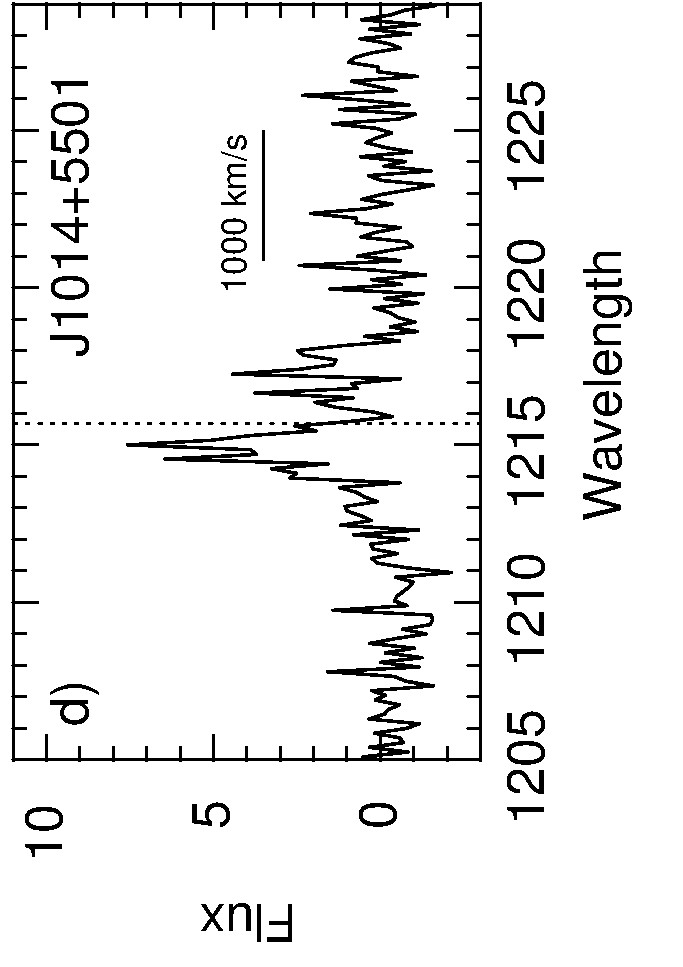}
\hspace{0.2cm}\includegraphics[angle=-90,width=0.32\linewidth]{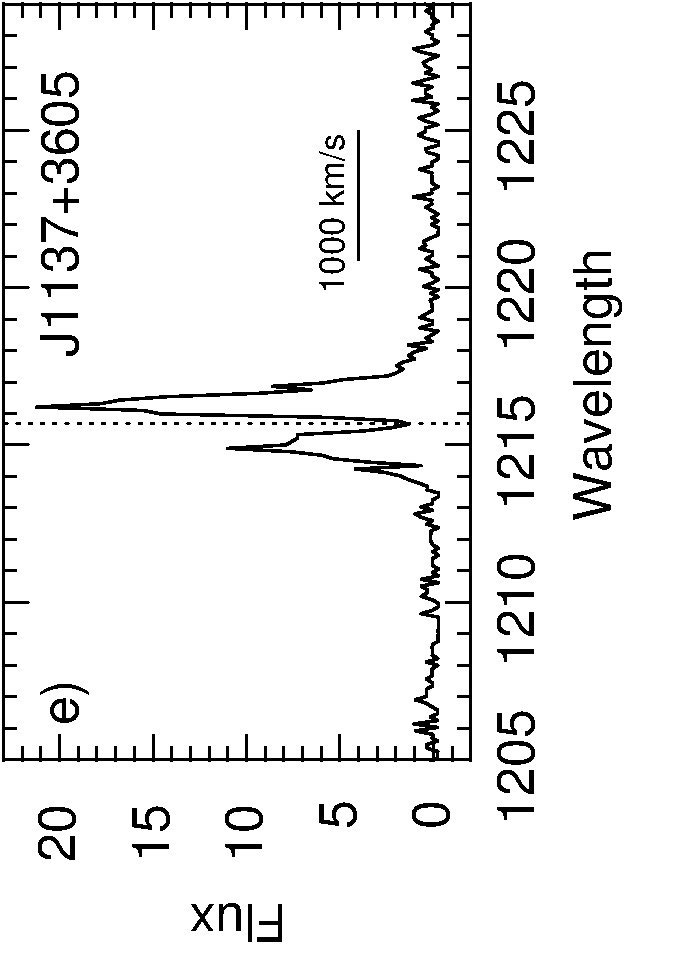}
\hspace{0.2cm}\includegraphics[angle=-90,width=0.32\linewidth]{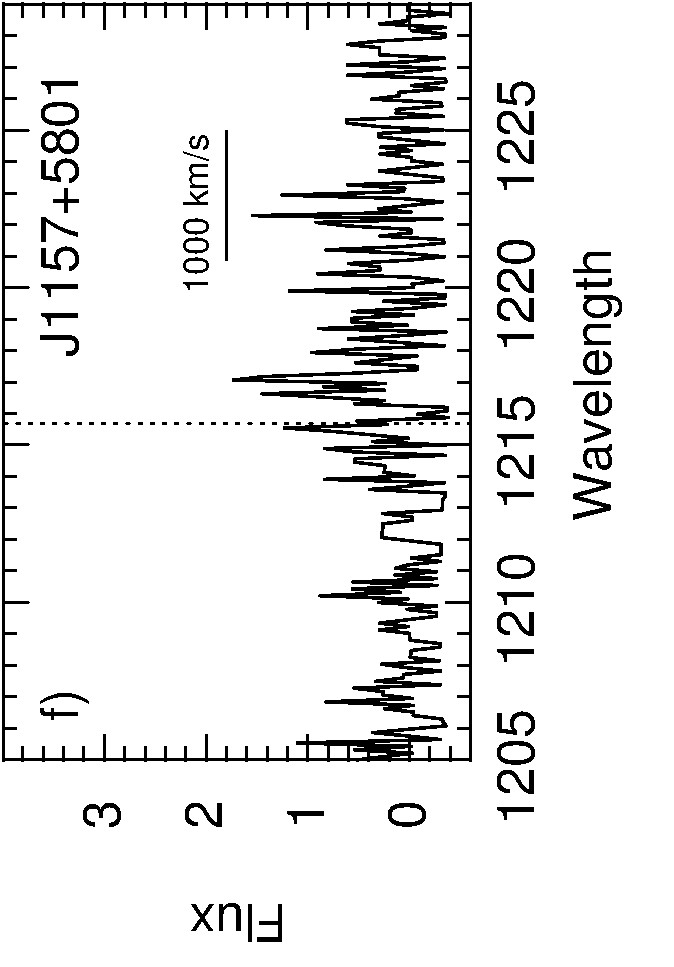}
}
\centering
\includegraphics[angle=-90,width=0.32\linewidth]{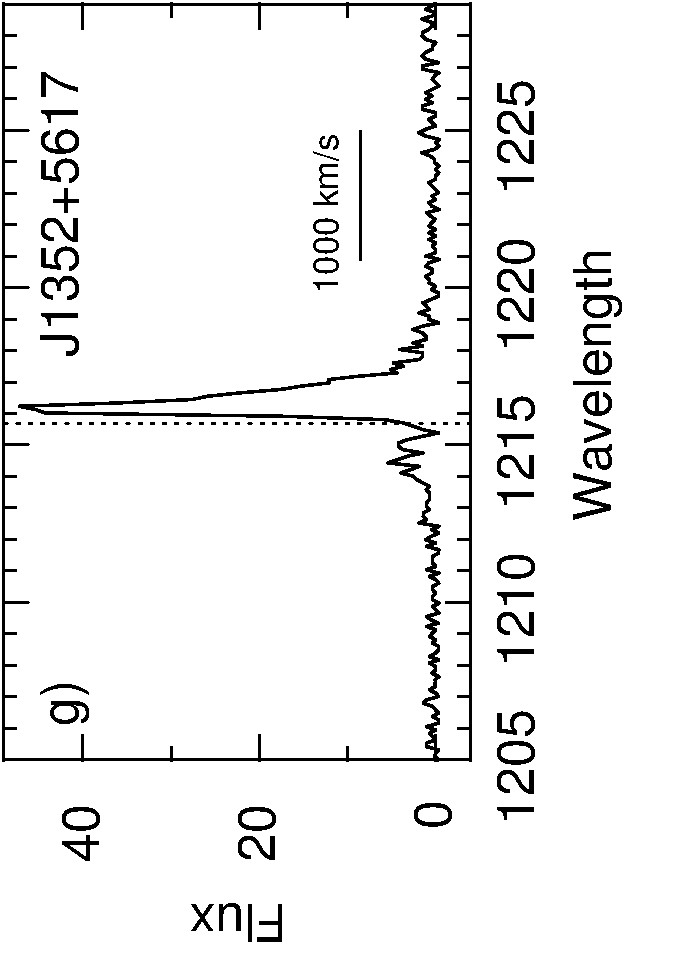}
\caption{Ly$\alpha$ profiles. Vertical dashed lines indicate the restframe 
wavelength of 1215.67\AA\ for Ly$\alpha$. 
Fluxes are in 10$^{-16}$ erg s$^{-1}$ cm$^{-2}$\AA$^{-1}$ and restframe 
wavelengths are in \AA. \label{fig4}}
\end{figure*}

\begin{figure*}
\hbox{
\includegraphics[angle=-90,width=0.32\linewidth]{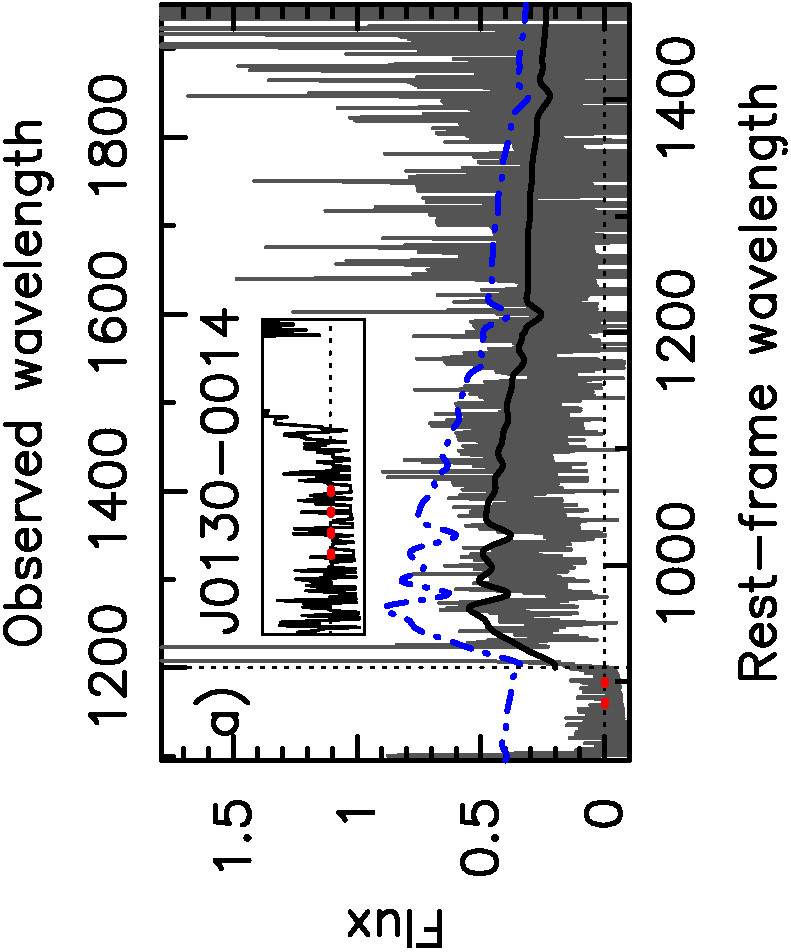}
\hspace{0.2cm}\includegraphics[angle=-90,width=0.32\linewidth]{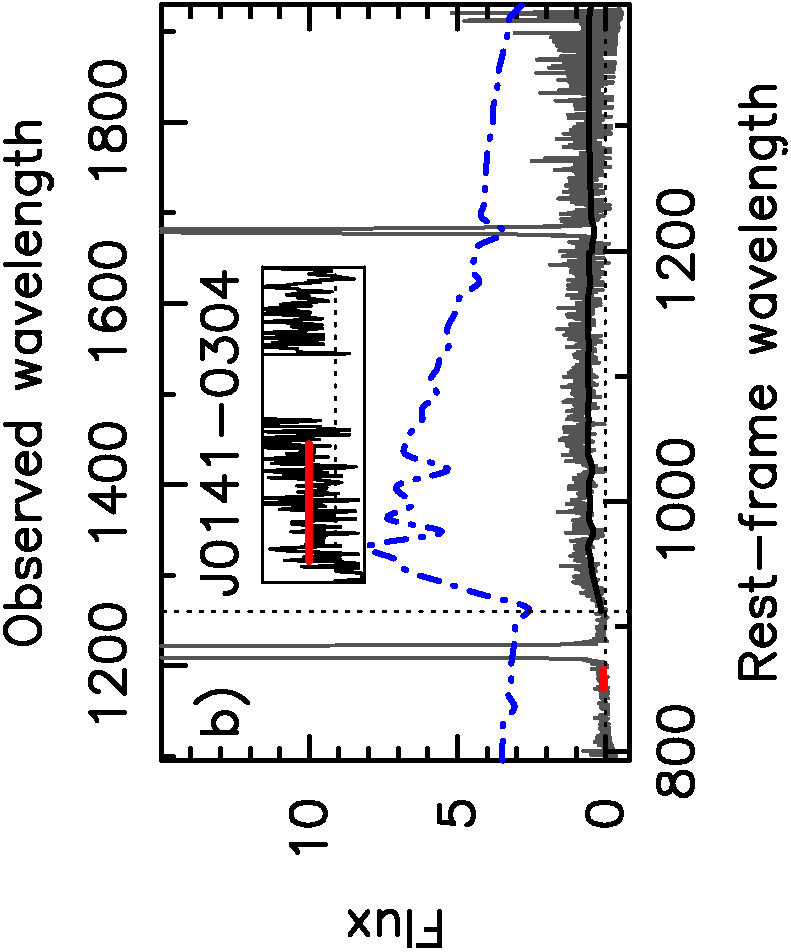}
\hspace{0.2cm}\includegraphics[angle=-90,width=0.32\linewidth]{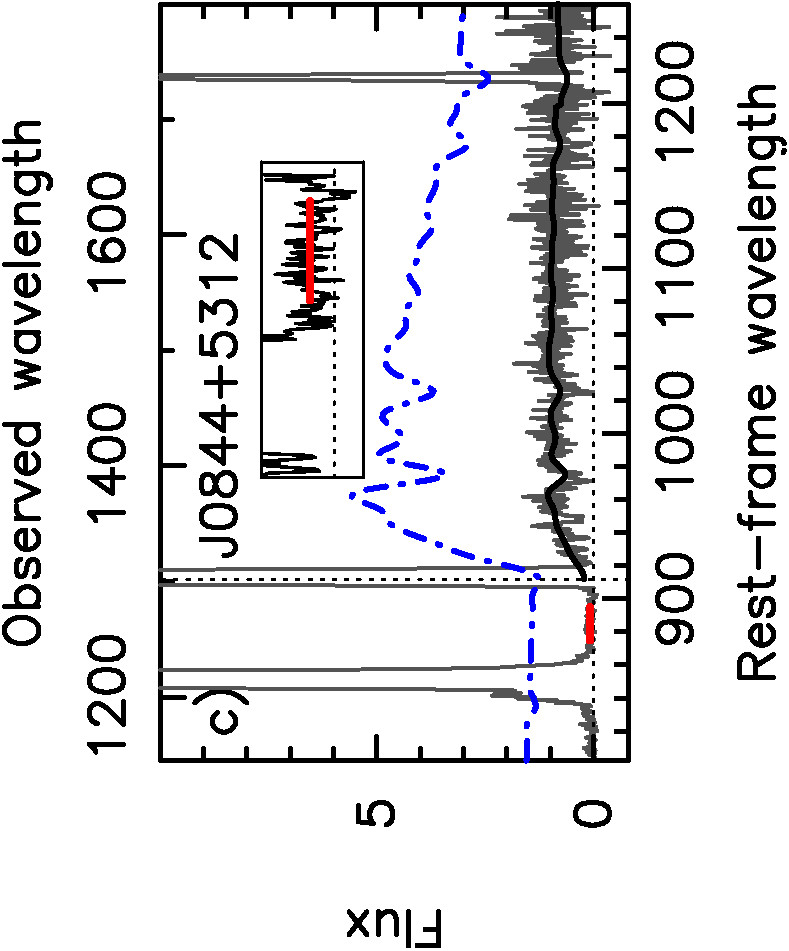}
}
\hbox{
\includegraphics[angle=-90,width=0.32\linewidth]{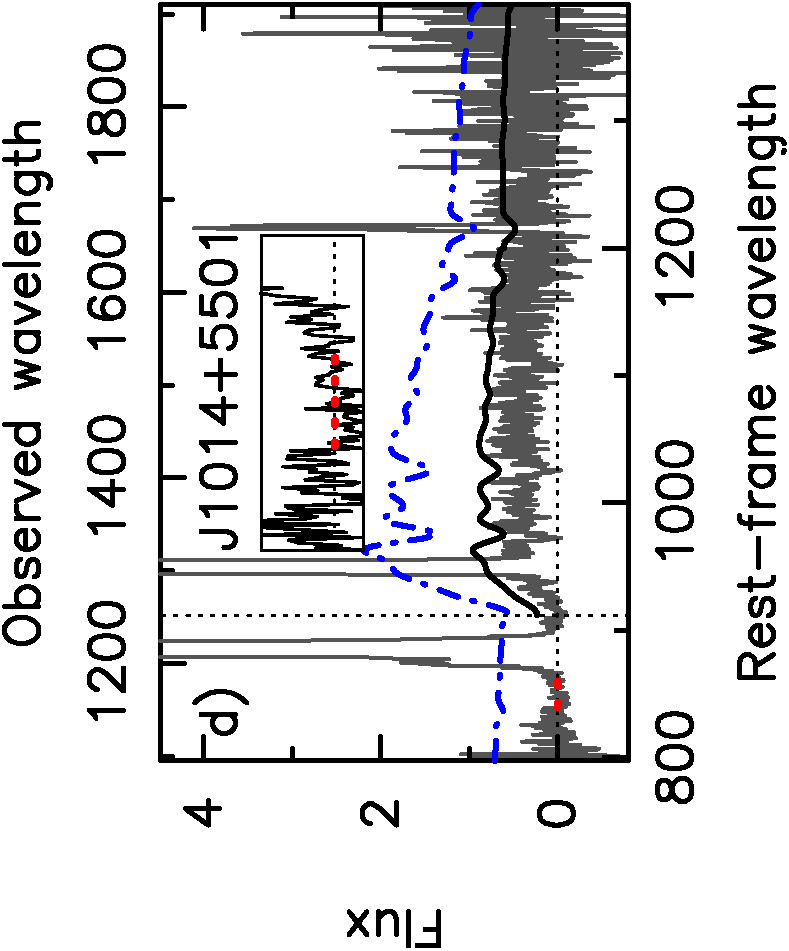}
\hspace{0.2cm}\includegraphics[angle=-90,width=0.32\linewidth]{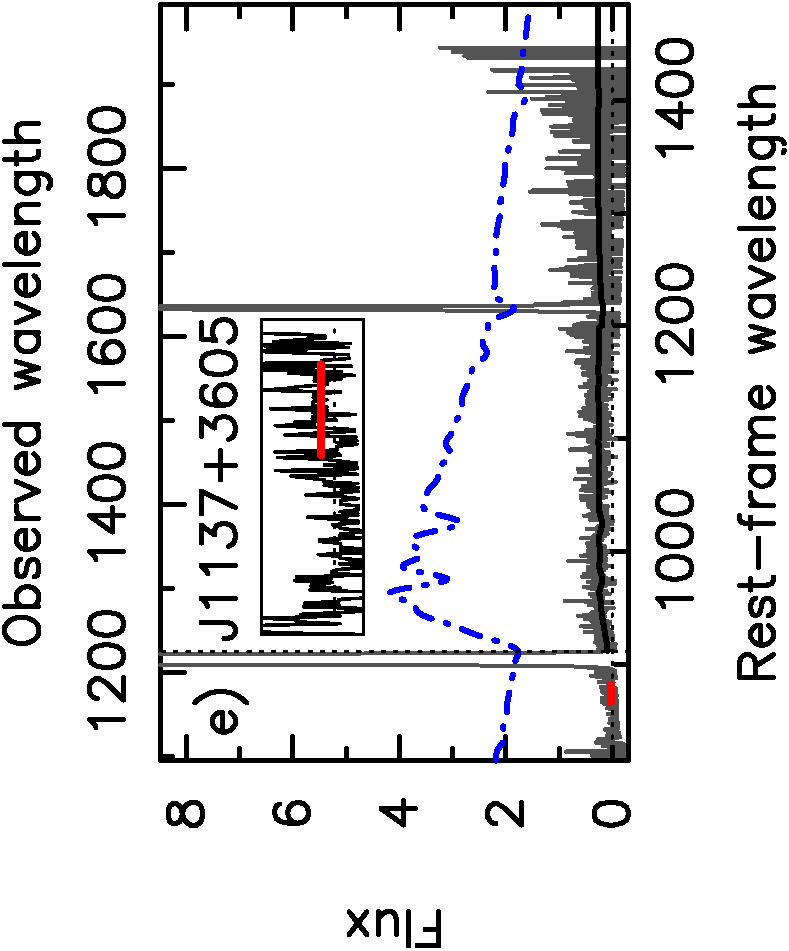}
\hspace{0.2cm}\includegraphics[angle=-90,width=0.32\linewidth]{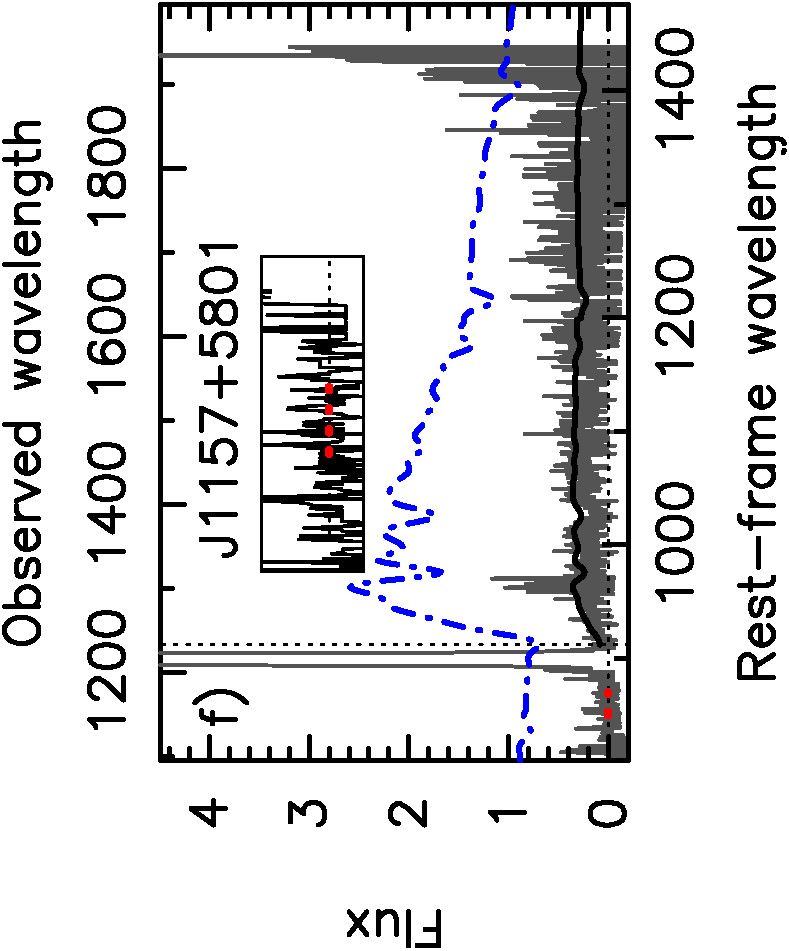}
}
\centering
\includegraphics[angle=-90,width=0.32\linewidth]{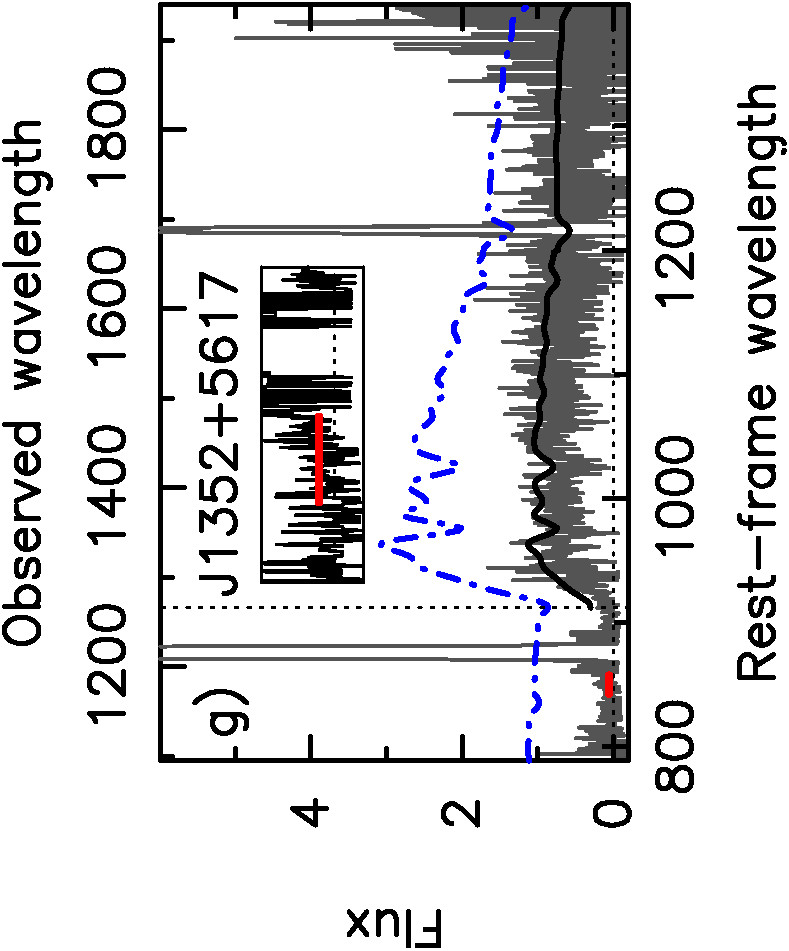}
\caption{COS G140L spectra of our sources (grey lines).
  The LyC fluxes shown by the solid (detections) and dotted (upper limits of
  non-detections) red horizontal lines are measured in the wavelength ranges
  determined by their location. The extrapolations to the UV range of
  intrinsic SEDs and of attenuated SEDs in the optical range
  adopting a $R(V)$ = 2.7, are represented by blue dash-dotted lines
  and black solid
  lines, respectively. The Lyman limit at the restframe wavelength
  912\,\AA\ is indicated by dotted vertical lines. Strong emission lines at
  the observed wavelengths 1216\,\AA\ (all panels) and 1303\,\AA\ (panels
  {\bf c)} and {\bf d)}) are geocoronal Ly$\alpha$ and [O~{\sc i}] lines.
  Zero flux is represented by dotted horizontal lines. Insets in all panels
  show expanded parts of spectra with LyC emission.
Fluxes are in 10$^{-16}$ erg s$^{-1}$ cm$^{-2}$\AA$^{-1}$, wavelengths 
are in \AA. \label{fig5}}
\end{figure*}

  \begin{table*}
  \caption{LyC escape fraction \label{tab5}}
\begin{tabular}{lccrrrrr} \hline
Name&$\lambda_0$$^{\rm a}$&$A$(LyC)$_{\rm MW}$$^{\rm b}$&$I_{\rm mod}$$^{\rm c,d}$&$I_{\rm obs}$(total)$^{\rm c,e}$&$I_{\rm esc}$(total)$^{\rm c,f}$&\multicolumn{1}{c}{$f_{\rm esc}$$^{\rm g}$}&\multicolumn{1}{c}{$f_{\rm esc}$$^{\rm h}$} \\
    &(\AA)&(mag)&&&&\multicolumn{1}{c}{(per cent)}&\multicolumn{1}{c}{(per cent)} \\
\hline
J0130$-$0014&880-900&0.287& 37.85$\pm$1.92&    $<$0.88$^{\rm i}$&     $<$1.16        &   $<$3.1          & $<$2.1 \\
J0141$-$0304&890-910&0.185&306.19$\pm$3.70&11.96$^{+1.34}_{-1.31}$&14.18$^{+1.63}_{-1.53}$& 4.6$^{+0.6}_{-0.6}$&4.3$^{+0.5}_{-0.5}$ \\
J0844$+$5312&880-900&0.208&221.40$\pm$4.87& 5.68$^{+1.13}_{-1.07}$& 6.87$^{+1.17}_{-1.11}$& 3.1$^{+0.6}_{-0.6}$&3.6$^{+0.7}_{-0.7}$ \\
J1014$+$5501&840-860&0.142& 85.68$\pm$5.30&    $<$1.09$^{\rm i}$&     $<$1.25        &    $<$1.4       &      $<$1.4      \\
J1137$+$3605&870-890&0.150&191.58$\pm$5.85& 5.11$^{+1.24}_{-1.15}$& 5.86$^{+1.30}_{-1.22}$& 3.1$^{+0.8}_{-0.9}$& 2.2$^{+0.5}_{-0.5}$ \\ 
J1157$+$5801&850-870&0.190& 80.29$\pm$6.51&    $<$1.10$^{\rm i}$&     $<$1.30        &    $<$1.7        &      $<$0.8      \\ 
J1352$+$5617&895-910&0.072&100.24$\pm$1.87& 4.17$^{+1.05}_{-0.98}$& 4.45$^{+1.09}_{-1.05}$& 4.5$^{+1.1}_{-1.1}$& 3.8$^{+0.9}_{-0.9}$ \\
\hline
  \end{tabular}

\hbox{$^{\rm a}$Restframe wavelength range in \AA\ used to determine the LyC flux.}

\hbox{$^{\rm b}$Milky Way extinction at the mean observed wavelengths of the 
range used to determine the LyC flux.} 

\hbox{\, The \citet{C89} reddening law with $R(V)$ = 3.1 is adopted.} 

\hbox{$^{\rm c}$in units of 10$^{-18}$ erg s$^{-1}$cm$^{-2}$\AA$^{-1}$.}

\hbox{$^{\rm d}$LyC flux derived from extrapolation of the modelled SED in
  the optical range to the UV range.}

\hbox{$^{\rm e}$Observed LyC flux.}


\hbox{$^{\rm f}$LyC flux which is corrected for the Milky Way extinction.}

\hbox{$^{\rm g}$$f_{\rm esc}$(LyC) = $I_{\rm esc}$(total)/$I_{\rm mod}$, where $I_{\rm mod}$ is derived from SED (first method).}

\hbox{$^{\rm h}$$f_{\rm esc}$(LyC) = $I_{\rm esc}$(total)/$I_{\rm mod}$, where $I_{\rm mod}$ is derived from H$\beta$ flux (second method).}

\hbox{$^{\rm i}$1$\sigma$ confidence upper limit.}

  \end{table*}

It is seen in Fig.\,\ref{fig3} that the SDSS spectra are reproduced by the
models quite well. On average, extrapolations of the attenuated SEDs
to the UV range with $R(V)_{\rm int}$\,=\,2.7
reproduce the observed COS spectra somewhat better with flux deviations not
exceeding $\sim$ 10 per cent for most galaxies. An exception is J1014$+$5501,
for which the difference in fluxes is as high as $\sim$ 50 per cent. This
difference can possibly be caused in part by the uncertain location of the galaxy
within the COS spectroscopic aperture as the acquisition exposure was failed.
It could also be caused by the underestimation of interstellar
extinction, which is derived from the hydrogen Balmer decrement in the SDSS
spectrum. The observed FUV shape could be fit by increasing $C$(H$\beta$)
by 0.065 from the value in Table \ref{taba2}.  This would increase the
H$\beta$ fluxes by $\sim15$\% and decrease the Ly$\alpha$ and LyC escape
fractions by a similar amount. However, in this case the extinction-corrected
fluxes of H$\delta$, H$\gamma$ and H$\alpha$ emission lines are considerably
off from their theoretical recombination values. Furthermore, the difference
between the models and observations can be caused by the non-perfect
absolute flux calibration of the SDSS spectrum.

However, we note that
Fig.\,\ref{fig3} is used only for the sake of illustration to check whether
extrapolation of the SED in the optical range reproduces the observed COS
spectrum. But it is not used for the determination of the escaping LyC fraction.
Instead the observed LyC flux is measured in COS spectra and the intrinsic LyC
flux is determined by two methods mentioned above: from the extinction-corrected
flux of the H$\beta$
emission line $I$(H$\beta$) and from simultaneous fitting of the SED in the
optical range and of observed equivalent widths of the H$\beta$ and H$\alpha$
emission lines. The fluxes of latter lines are also iteratively corrected for
the escaping ionizing radiation \citep[e.g. ][]{I18b} and they determine the
level of the intrinsic LyC emission.
It is seen in Fig.\,\ref{fig3} that the SED in the optical range is
almost independent on $R(V)_{\rm int}$. Consequently, the LyC escape fraction
$f_{\rm esc}$(LyC) is also almost independent of
$R(V)_{\rm int}$. This is because $f_{\rm esc}$(LyC) is derived
from the ratio of the observed to modelled intrinsic LyC fluxes with the latter
fluxes being derived from data in the optical range.

The relation between $I$(H$\beta$) and the intrinsic LyC flux at
900\,\AA\ $I$(900\,\AA), assuming the instantaneous burst model, takes a form
\citep{I16b}
\begin{equation}
  \frac{I({\rm H}\beta)}{I(900~\angstrom)} = 2.99 \times
       {\rm EW}({\rm H}\beta)^{0.228}~\angstrom, \label{eq:inst}
\end{equation}
where EW(H$\beta$) is in \AA, $I$(H$\beta$) and $I$(900\,\AA) are in
erg s$^{-1}$ and erg\,s$^{-1}$\,\AA$^{-1}$, respectively.
The term with EW(H$\beta$) in  Eq.~\ref{eq:inst} takes into account the weak
dependence on the starburst age. According to this equation, uncertainties on
$I$(900\,\AA) are due to small uncertainties of $C$(H$\beta$) (Table~\ref{taba2})
and, thus, on $I$(H$\beta$) are unlikely to be greater than $\sim$\,15 - 20 per
cent.

\section{Ly$\alpha$ and LyC emission}\label{sec:lya}

A resolved Ly$\alpha$ $\lambda$1216\,\AA\ emission line is detected
in the G160M medium-resolution spectra of five out of seven galaxies
(Fig.\,\ref{fig4}). Its shape is similar to that observed in most known
LyC leakers \citep{I16a,I16b,I18a,I18b,I21a} and in some other
galaxies at lower redshift \citep{JO14,H15,Y17,I20}.
Profiles with two peaks are detected in the spectra of four galaxies from the
present sample with detected LyC emission,
J0141$-$0304, J0844$+$5312, J1137$+$3605, J1352$+$5617, and in
one galaxy with non-detected LyC emission, J1014$+$5501. The blue
Ly$\alpha$ component in the latter galaxy is $\sim$ 2.5 times brighter than
the red component (Fig.\,\ref{fig4}d). This fact is at variance with that for
other galaxies, where the blue component is considerably weaker than the
red component, and may be indicative of a gas inflow.
The Ly$\alpha$ emission line is very weak in the spectra of
two galaxies, J0130$-$0014 and J1157$+$5801. The parameters of Ly$\alpha$
emission are shown in Table\,\ref{tab4}.

The observed G140L total-exposure spectra with the LyC spectral
region (grey lines) and extrapolations to the UV range of predicted
intrinsic SEDs in the optical range (blue dash-dotted
lines) are shown in Fig.\,\ref{fig5}. Additionally, we include the attenuated
extrapolations of the intrinsic SEDs (black solid lines),
the same as those with $R(V)$ = 2.7 that are shown in Fig.\,\ref{fig3}
but with different flux and wavelength scales.

The level of the observed LyC continuum is indicated by horizontal
red lines. The vertical dotted lines show the Lyman limit.
The Lyman continuum emission is detected in the spectra of four
galaxies, J0141$-$0304, J0844$+$5312, J1137$+$3605 and J1352$+$5617
(solid red lines), and only 1$\sigma$ upper limits are derived in the spectra
of the remaining three galaxies (dotted red lines). The measurements
are summarised in Table\,\ref{tab5}.

\citet{I16a,I16b,I18a,I18b} used the ratio of the escaping
fluxes $I_{\rm esc}$ to the intrinsic fluxes $I_{\rm mod}$
of the Lyman continuum to derive $f_{\rm esc}$(LyC):
\begin{equation}
f_{\rm esc}({\rm LyC}) =\frac{I_{\rm esc}(\lambda)}{I_{\rm mod}(\lambda)}, 
\label{eq:fesc}
\end{equation}
where $\lambda$ is the mean wavelength of the range near 900\,\AA\ used for
averaging the LyC flux density (see Table~\ref{tab5}).
\citet{I16b} proposed two methods to iteratively derive the intrinsic fluxes 
$I_{\rm mod}$ and, correspondingly, the LyC escape fractions $f_{\rm esc}$(LyC): 
1) from simultaneous fitting of the SED in the optical range together with
observed equivalent widths of the H$\beta$ and H$\alpha$ emission lines
and 2) from the equivalent width of the H$\beta$ 
emission line, its extinction-corrected flux
and adopting relations between $I$(H$\beta$)/$I_{\rm mod}$ 
and EW(H$\beta$) from the models of photoionized H~{\sc ii} regions
\citep[Eq.\,\ref{eq:inst}, ][]{I16b}. The extinction-corrected
flux of the H$\beta$ emission line in both methods determines the intrinsic LyC
flux at 900\,\AA\ by taking into account the starburst age which mainly
depends on the H$\beta$ and/or H$\alpha$ equivalent widths. 
We use both methods in this paper.

The escape fraction $f_{\rm esc}$(LyC) ranges between 3.1 and
4.6 per cent in four out of the seven galaxies and the 1$\sigma$ upper
limits of $f_{\rm esc}$(LyC) for the remaining galaxies are shown in
Table\,\ref{tab5}. We find that $f_{\rm esc}$(LyC) obtained by the two
methods are similar.

\begin{figure*}
\hbox{
\includegraphics[angle=-90,width=0.45\linewidth]{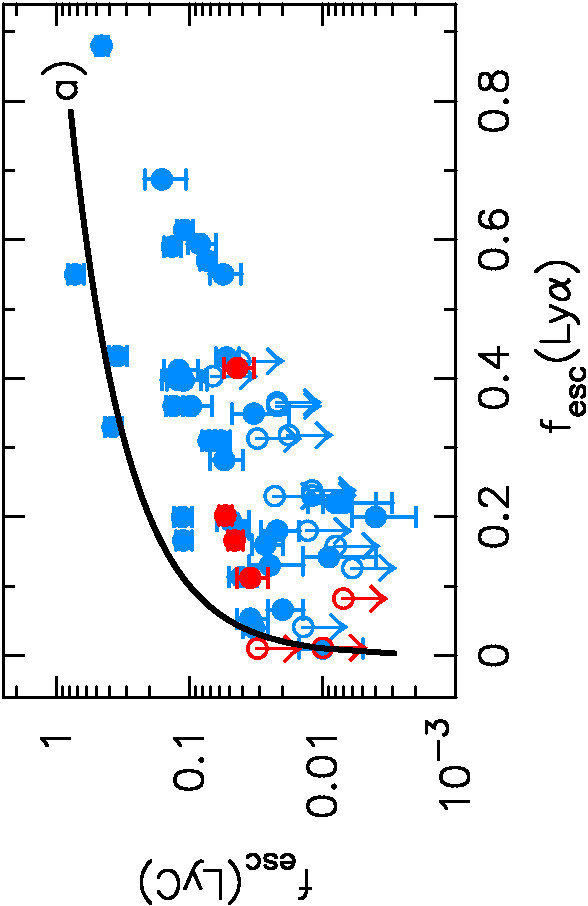}
\hspace{0.2cm}\includegraphics[angle=-90,width=0.45\linewidth]{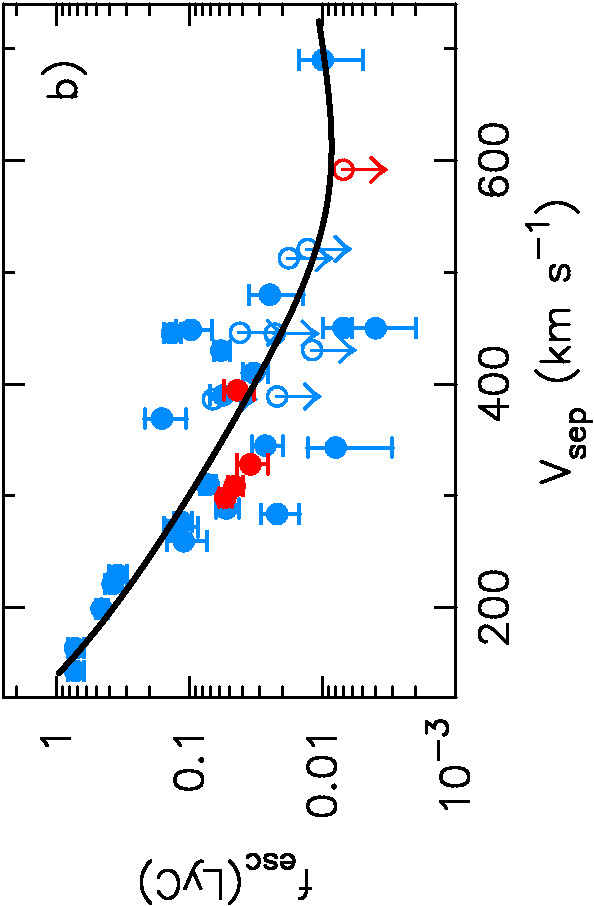}
}
\hbox{
\includegraphics[angle=-90,width=0.45\linewidth]{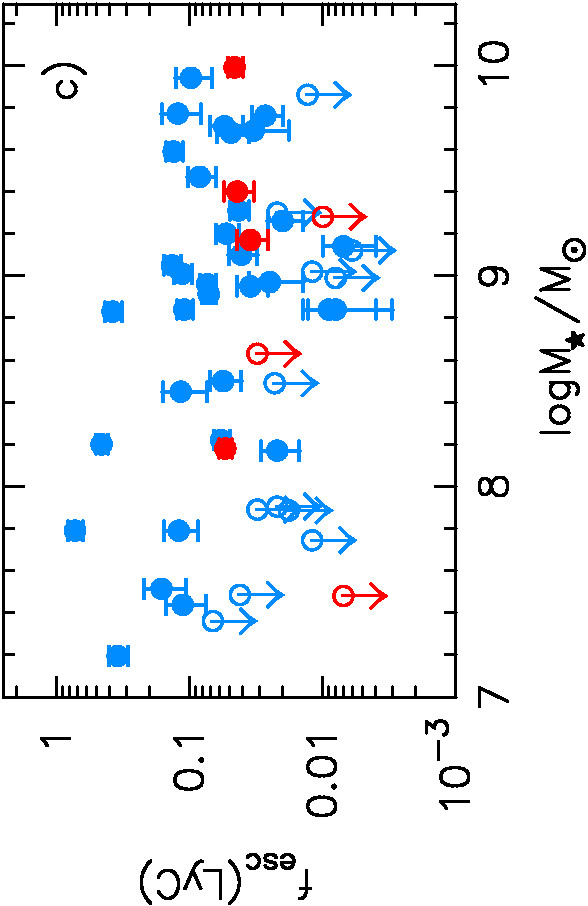}
\hspace{0.2cm}\includegraphics[angle=-90,width=0.45\linewidth]{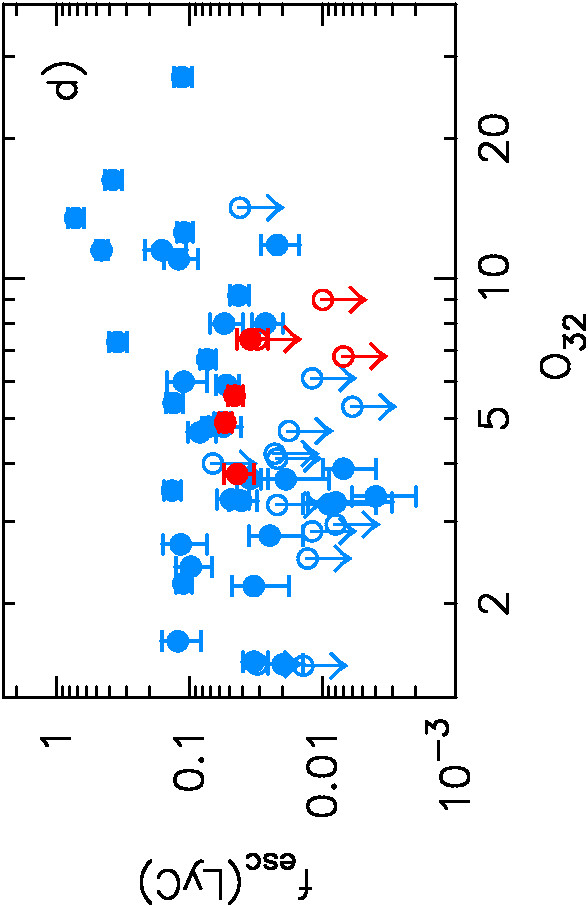}
}
\caption{Relations between the Lyman continuum escape fraction
  $f_{\rm esc}$(LyC) in low-redshift LyC leaking galaxies derived by the method
  with the use of SED fits and constraints from the observed H$\beta$ and
  H$\alpha$ equivalent widths, and {\bf a)} the 
Ly$\alpha$ escape fraction $f_{\rm esc}$(Ly$\alpha$), {\bf b)} the separation $V_{\rm sep}$ between the Ly$\alpha$ profile peaks, {\bf c)} the stellar mass
$M_\star$, {\bf d)} the [O~{\sc iii}]$\lambda$5007/[O~{\sc ii}]$\lambda$3727 
emission-line flux ratios. In all panels, the galaxies from this paper
are shown by red symbols and from 
\citet{I16a,I16b,I18a,I18b,I21a}, \citet{B14}, \citet{C17}, \citet{F22a},
  \citet{Xu22} are represented by blue symbols.
LyC leakers and galaxies with upper limits of LyC emission are shown by
filled circles and open circles with downward arrows, respectively.
The solid line in {\bf a)} is the equality line 
and the solid line in {\bf b)} represents the relation from \citet{I18b}.
\label{fig6}}
\end{figure*}

\begin{figure*}
\includegraphics[angle=-90,width=0.88\linewidth]{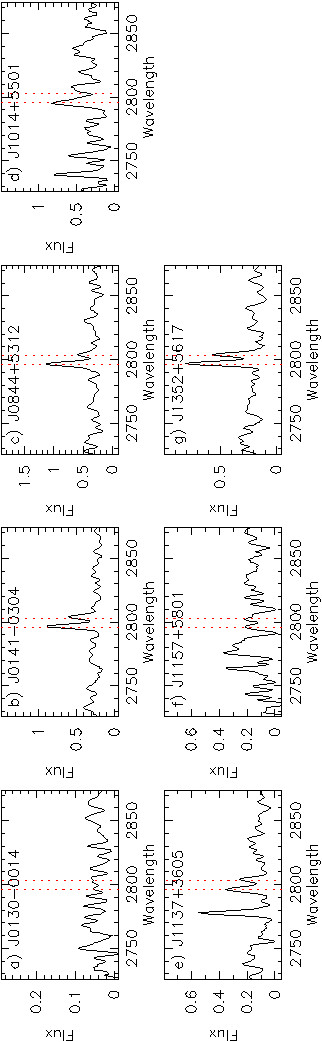}
\caption{Mg~{\sc ii} $\lambda$2796, 2803 emission lines in SDSS spectra
  of new galaxies discussed in this paper. Red dotted lines indicate
  the restframe centres of the lines.
  \label{fig7}}
\vspace{0.35cm}
\includegraphics[angle=-90,width=0.88\linewidth]{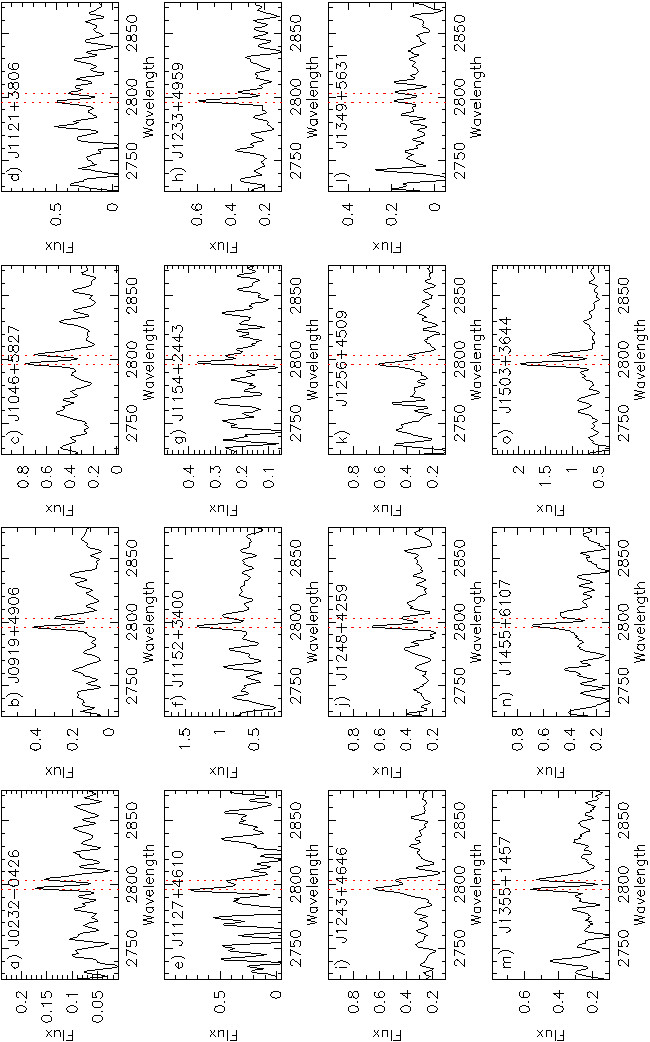}
\caption{Mg~{\sc ii} $\lambda$2796, 2803 emission lines in SDSS spectra
  of LyC leaking galaxies from \citet{I16a,I16b,I18b,I18c,I21a}.
  Red dotted lines indicate the restframe centres of the lines.
\label{fig8}}
\vspace{0.35cm}
\includegraphics[angle=-90,width=0.88\linewidth]{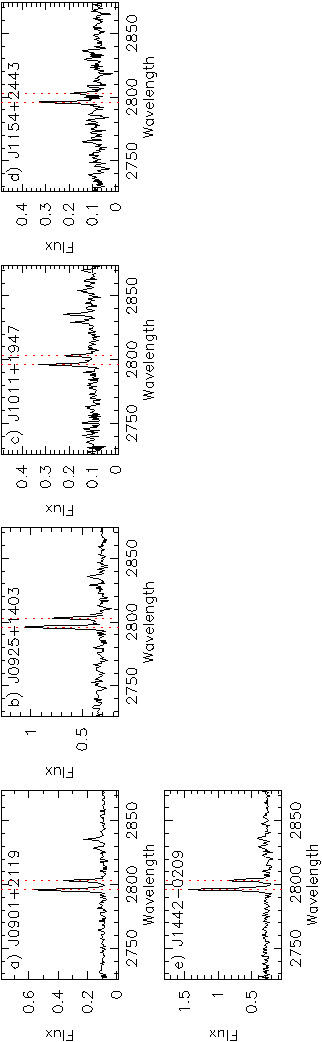}
\caption{Mg~{\sc ii} $\lambda$2796, 2803 emission lines in XShooter spectra
  of LyC leaking galaxies from \citet{G20}. Red dotted lines indicate
  the restframe centres of the lines.
\label{fig9}}
\end{figure*}

\begin{figure*}
\hbox{
\includegraphics[angle=-90,width=0.45\linewidth]{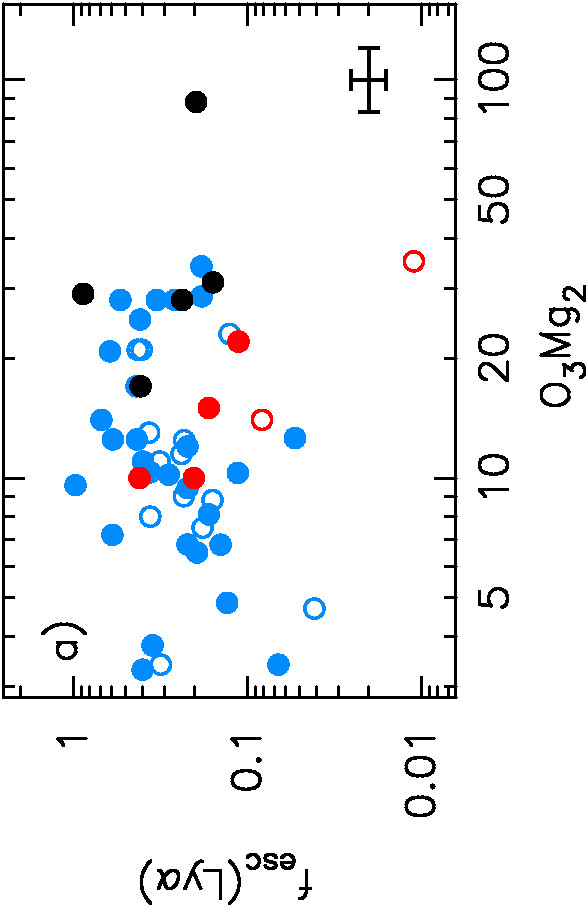}
\hspace{0.2cm}\includegraphics[angle=-90,width=0.45\linewidth]{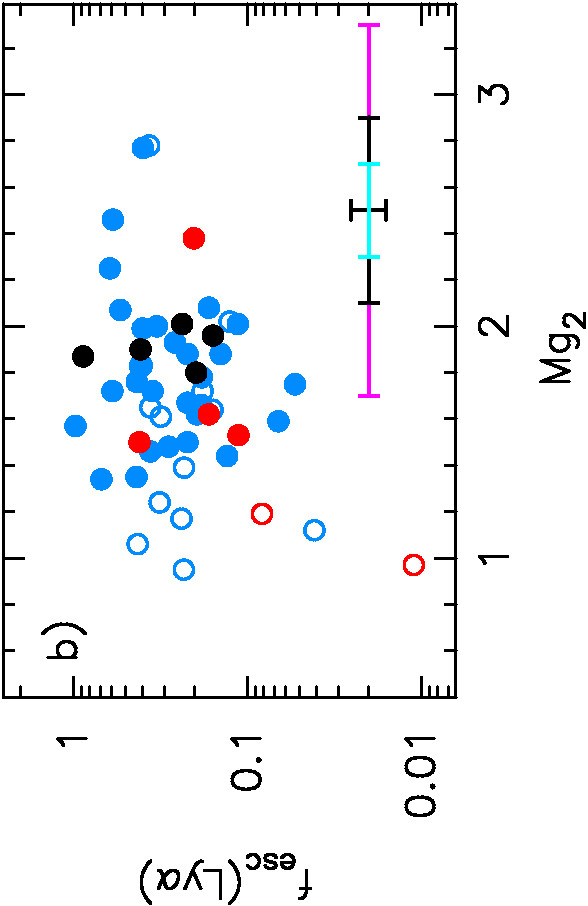}
}
\vspace{0.3cm}
\hbox{
\includegraphics[angle=-90,width=0.45\linewidth]{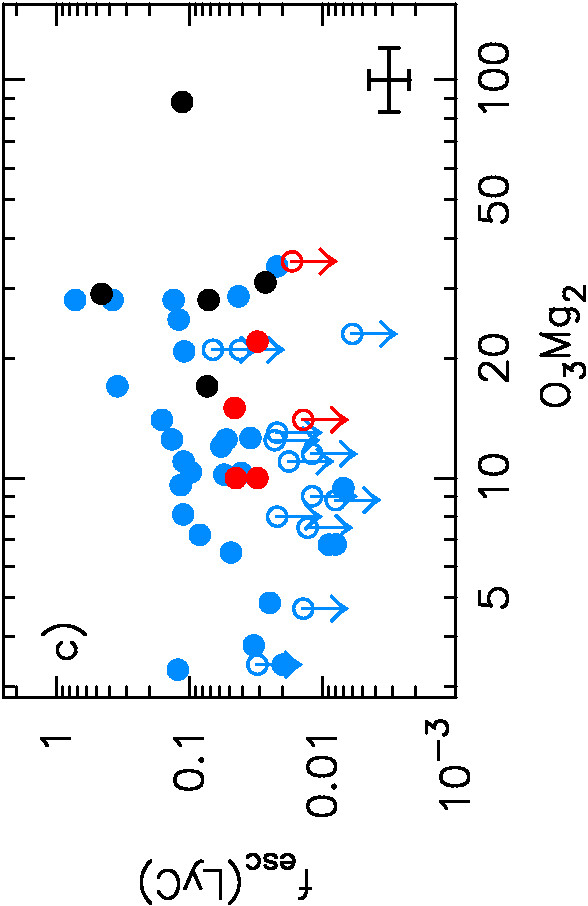}
\hspace{0.2cm}\includegraphics[angle=-90,width=0.45\linewidth]{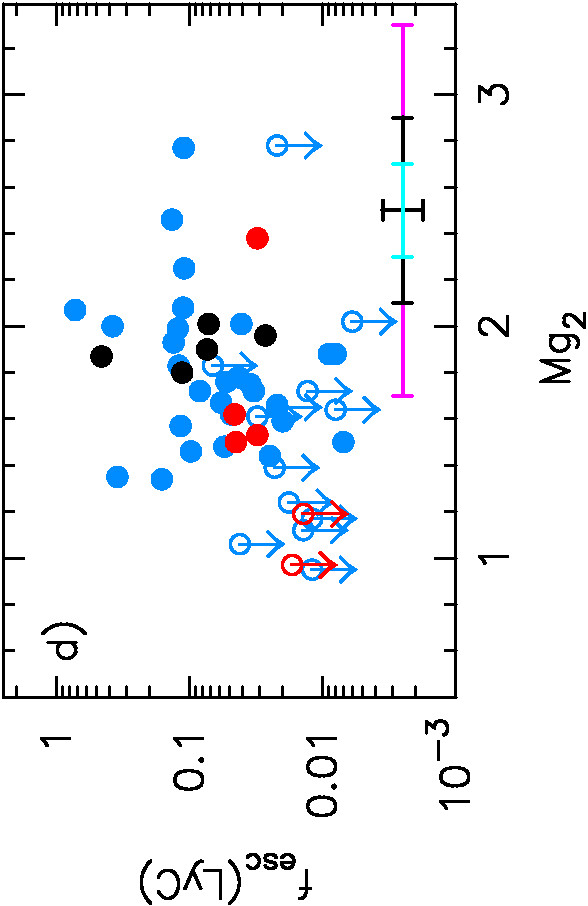}
}
\caption{{\bf a)} and {\bf b)} Relations between the Ly$\alpha$
  escape fraction
  $f_{\rm esc}$(Ly$\alpha$) in low-redshift LyC leaking galaxies and the
  O$_3$Mg$_2$ = [O~{\sc iii}]$\lambda$5007/Mg~{\sc ii} $\lambda$2796+2803 ratio
  and the Mg$_2$ = Mg~{\sc ii} $\lambda$2796/Mg~{\sc ii} $\lambda$2803 ratio,
  respectively.
  {\bf c)} and {\bf d)} Relations between the Lyman continuum
  escape fraction
$f_{\rm esc}$(LyC) in low-redshift LyC leaking galaxies derived from the SED
fits and the O$_3$Mg$_2$ and Mg$_2$,
respectively. The galaxies from this paper and from
\citet{I16a,I16b,I18a,I18b,I21a}, \citet{B14}, \citet{C17}, \citet{F22a},
  \citet{Xu22} are
represented by red and blue symbols, excluding objects observed with the
XShooter \citep[Fig.\,\ref{fig9},][]{G20}, which are shown by black symbols.
LyC leakers and galaxies with upper limits of LyC emission
are shown by filled circles and open circles with
downward arrows, respectively. Error bars in black (all panels) represent
average 1$\sigma$ deviations, whereas error bars in cyan and magenta (panels
{\bf b)} and {\bf d)}) are minimal and maximal 1$\sigma$ errors of Mg$_2$ for
the SDSS+XShooter sample, respectively. The values of O$_3$Mg$_2$ and
  Mg$_2$ for all galaxies are calculated in this paper.
\label{fig10}}
\end{figure*}

\section{Indirect determination of the LyC escape fraction} \label{sec:ind}

The direct measurement of LyC emission is the best way to derive
the LyC escape fraction. However, LyC emission in most cases is
weak and it difficult to detect in both the high-$z$ and low-$z$ galaxies.
Furthermore, only {\sl HST} can be used for the observation of the
LyC wavelength range in galaxies with $z$ $\sim$ 0.3 -- 1.0. Therefore,
reasonable indirect indicators of LyC leakage at low and high redshift
are needed, namely those which can more easily be derived from
observations, to build a larger sample for statistical studies. Several
possible indicators have been proposed, which are based mainly on
observations of strong emission lines in the UV and optical ranges.
For the analysis of possible indirect indicators we use a sample
of $\sim$ 30 -- 50 galaxies with Mg~{\sc ii} emission in their SDSS spectra
from \citet{I16a,I16b,I18a,I18b,I21a}, \citet{B14}, \citet{C17}, \citet{F22a},
\citet{Xu22} and this paper. The number of galaxies varies for different
indicators because not all indicators are determined for all galaxies in the
sample.

The Ly$\alpha$ escape fraction $f_{\rm esc}$(Ly$\alpha$), which is derived from
the observed Ly$\alpha$/H$\beta$ emission line ratio, can potentially be linked
with the LyC escape fraction. However, there are differences between mechanisms
controlling the escape of LyC and Ly$\alpha$. The LyC photons
can efficiently be absorbed by neutral hydrogen
and/or dust. On the other hand, Ly$\alpha$ photons can be ceased only via
absorption by dust and via inefficient two-photon transitions. Thus,
the fraction of escaping Ly$\alpha$ photons is expected to be higher than
that of escaping LyC photons, in agreement with theoretical predictions
\citep*{D15,JO13,NO14}. This is seen in Fig.\,\ref{fig6}a, where almost
all LyC leaking galaxies are located below the line of equal
escape fractions (black solid line). There is a tendency for $f_{\rm esc}$(LyC)
to increase with increasing $f_{\rm esc}$(Ly$\alpha$) but with a large spread
\citep[see also e.g. ][]{I18b,I21a,F22b}.
New data do not contradict with this conclusion. 

The shape of the Ly$\alpha$ profile provides the best indirect indicator
of the LyC leakage due to the fact that it depends on the column density
of the neutral hydrogen along the line of sight, which determines
the optical depth in both the Ly$\alpha$ emission line and the LyC continuum.
In particular, a non-zero intensity at the center of Ly$\alpha$ or a
small offset of its brighter red component from the center of the line
indicate low optical depth in the H~{\sc i} cloud. However, these indicators
may be influenced by insufficient spectral resolution and uncertainties
in the wavelength calibration. On the other hand, the separation
between its blue and red components in medium-resolution COS spectra is less
subject to these limitations. Previously \cite{V17} and \citet{I18b} found
a tight dependence of $f_{\rm esc}$(LyC) on the separation $V_{\rm sep}$ between the
peaks of the Ly$\alpha$ emission line in LyC leakers. This dependence
has been updated in the later paper by \citet{I21a} and
in this paper. The new data also follow the relation discussed by
\citet{I18b} (see the solid line in Fig.\,\ref{fig6}b). There is no new
galaxy in our present sample having a peak separation less than
$\sim$300 km s$^{-1}$, which is considerably higher compared to the lowest
peak separation of $\sim$ 150 km s$^{-1}$ in the sample of low-$z$ leakers
shown in Fig.\,\ref{fig6}b. The relation by \citet{I18b} is likely not
applicable for complex Ly$\alpha$ profiles with three or more peaks, indicating
  considerable direct Ly$\alpha$ escape, in addition to escape through 
  scattering in the neutral gas \citep{N22}. The Ly$\alpha$ profile
in only one galaxy, J1243$+$4646, from the \citet{I18b} sample consists of
three peaks with the peak separations of 143 and 164 km s$^{-1}$.
This galaxy does not change significantly the shape of the relation shown in
Fig.~\ref{fig6}b by the solid line because most of the galaxies in the sample
have two Ly$\alpha$ peaks. 

The new observations of Mg~{\sc ii}-selected galaxies (red symbols) support
previous findings on the existence of the tight relation between
$f_{\rm esc}$(LyC) and $V_{\rm sep}$. However, the application of this relation
for galaxies observed during epoch of reionization is limited because
of incomplete ionization of the intergalactic medium and thus high
optical depth for Ly$\alpha$ emission.

Low galaxy stellar masses are also considered as a possible
indicator of high $f_{\rm esc}$(LyC) \citep{W14,T17}.
Indeed, there is a trend of decreasing $f_{\rm esc}$(LyC) with increasing
stellar mass in galaxies with the detected LyC continuum (filled
circles in Fig.\,\ref{fig6}c). However, \citet{I21a} found several
strongly star-forming galaxies with $M_\star$ $<$ 10$^8$M$_\odot$ and non-detected
LyC (blue open circles in Fig.\,\ref{fig6}c), considerably weaking
the anti-correlation between $f_{\rm esc}$(LyC) and $M_\star$. New data in the
present paper are in agreement with the conclusion of no or only a weak
correlation between $f_{\rm esc}$(LyC) and $M_\star$.  

\citet{JO13} and \citet{NO14} proposed
a high O$_{32}$ ratio as an indication of escaping ionizing radiation.
However, the increase of this ratio is caused not only by decreasing
optical depth of the neutral hydrogen around the H~{\sc ii} region, but
also by increasing ionization parameter and/or decreasing metallicity.
These effects are difficult
to separate. O$_{32}$ in low-redshift galaxies can easily be
derived from their spectra in the optical range.
This quantity is known for all low-$z$ LyC leakers. 

The relation between $f_{\rm esc}$(LyC) and O$_{32}$ has been discussed
by \citet{F16}, \citet{I18b,I21a} and \citet{F22b}. Its updated version
from \citet{I21a} is presented in
Fig.\,\ref{fig6}d, which shows a trend of increasing $f_{\rm esc}$(LyC)
with increasing of O$_{32}$, but with a substantial scatter. This scatter,
in part, can be caused by a variety of scenarios with leakage through channels
with low optical depth and their orientation relative to the
observer. Similar conclusion can be drawn from the \citet{F22b} data.
Therefore, a high O$_{32}$ can be used for selection of the LyC
leaking candidates, but it is not a very certain indicator of
high $f_{\rm esc}$(LyC) \citep{I18b,Na20}.

\begin{figure*}
\hbox{
\includegraphics[angle=-90,width=0.45\linewidth]{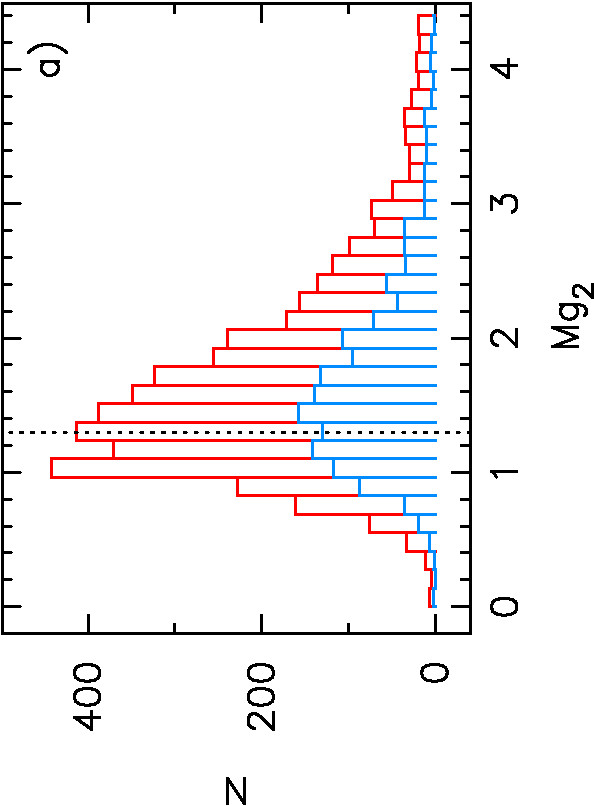}
\hspace{0.2cm}\includegraphics[angle=-90,width=0.45\linewidth]{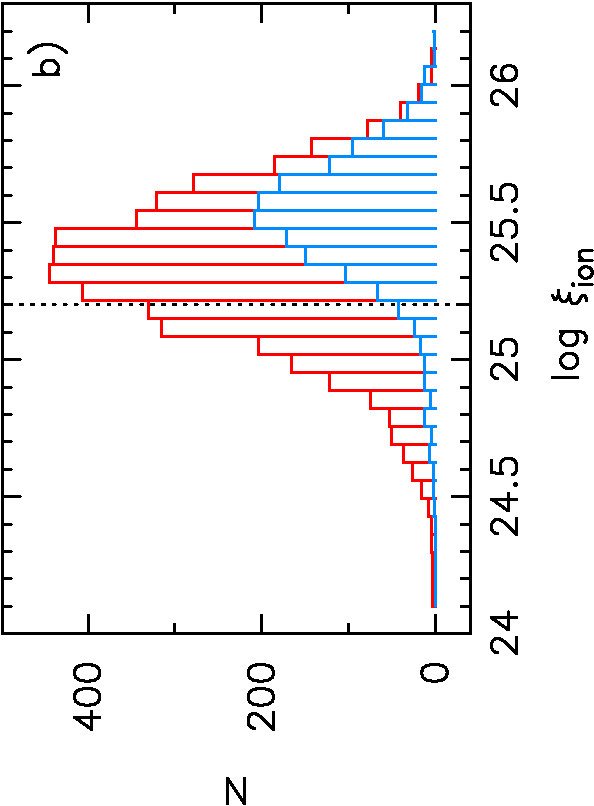}
}
\vspace{0.3cm}
\hbox{
\includegraphics[angle=-90,width=0.45\linewidth]{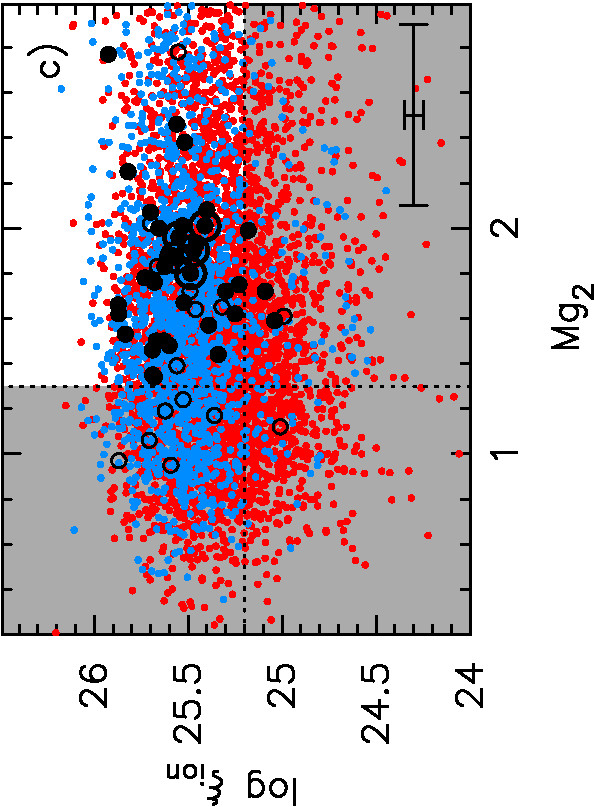}
\hspace{0.2cm}\includegraphics[angle=-90,width=0.45\linewidth]{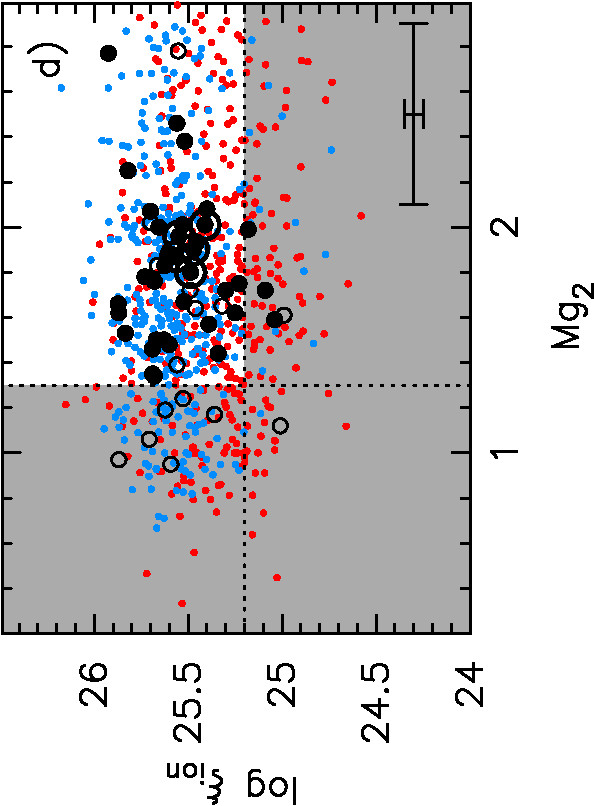}
}
\caption{{\bf a)} Distribution of the flux ratio
  Mg$_2$ = Mg~{\sc ii} $\lambda$2796/$\lambda$2803 in the $\sim$ 6000 SDSS
  compact star-forming galaxies with redshifts $z$ $\geq$ 0.3.
  {\bf b)} Distribution of ionizing photon production efficiency $\xi_{\rm ion}$
  for the
  same sample as in {\bf a)}. Dotted line in {\bf a)} separates expected LyC
  leakers (Mg$_2$ $\geq$ 1.3) from non-LyC leakers (Mg$_2$ $<$ 1.3), whereas
  the dotted line in {\bf b)} is the canonical value of ionizing photon
  production efficiency commonly adopted to complete reionization.
  {\bf c)} Relation between Mg$_2$ and $\xi_{\rm ion}$. The LyC leaking galaxies with
  detected LyC emission and its upper limits, the same as in
  Figs.~\ref{fig6}, \ref{fig10}, are represented by black filled
  and open circles, respectively. Objects with the fluxes of
Mg~{\sc ii}\,$\lambda$2796, 2803 emission lines obtained from the XShooter
spectra \citep{G20} are encircled. {\bf d)} Same as in {\bf c)}, but are shown
only galaxies from the SDSS with H$\beta$ fluxes above
5$\times$10$^{-16}$ erg s$^{-1}$cm$^{-2}$ and equivalent widths of the
Mg~{\sc ii} $\lambda$2796 emission line above 10\,\AA. In all panels SDSS
compact star-forming galaxies with equivalent widths
EW(H$\beta$) $\geq$ 100\,\AA\ and $<$ 100\,\AA\ are shown in blue and red,
respectively. Error bars in {\bf c)} represent average 1$\sigma$ errors.
\label{fig11}}
\end{figure*}

\section{Mg~{\sc ii} diagnostics} \label{sec:MgII}

\citet{Hen18} and \citet{Ch20} have proposed
to use the double resonance line of Mg~{\sc ii} $\lambda$2796, 2803 in emission
as an indicator of escaping LyC emission based on the fact that
its escape fraction correlates with the Ly$\alpha$ escape fraction.
Later, \citet{Xu22} also proposed Mg~{\sc ii} as low-$z$ tracer of Ly$\alpha$ and
  LyC, \citet{N22} pointed out that Mg~{\sc ii}$\lambda$2796/$\lambda$2803 line
  ratio is higher in $z$ $\sim$ 2 galaxies with higher $f_{\rm esc}$(LyC).
Following these papers we consider the properties of Mg~{\sc ii} emission
and their relations with the Ly$\alpha$ and LyC escape fractions. For many
low-redshift LyC leaking galaxies
\citep[][this paper]{I16a,I16b,I18a,I18b,I21a,F22a,Xu22} the wavelength range with
the redshifted Mg~{\sc ii} $\lambda$2796, 2803 emission lines is covered by the
SDSS spectra
(Fig.\,\ref{fig7}\,--\,\ref{fig8}). However, these redshifted lines are outside the
wavelength range of SDSS spectra from the releases earlier than DR10 of some
LyC leakers with lowest redshifts of
$z$\,$\approx$\,0.3 (for example, J0925$+$1403, J1011$+$1947, J1442$-$0209).
XShooter spectra covering the Mg~{\sc ii} emission (Fig.\,\ref{fig9}) are also
available for some LyC galaxies \citep{G20}, including those with
$z$\,$\approx$\,0.3.

We note that Mg~{\sc ii} emission
is located in the noisy parts of the SDSS spectra. Because of weakness of
these lines they cannot be measured with high accuracy. The spectral
resolution of SDSS spectra is insufficient to determine the
Mg~{\sc ii} emission line profiles. On the other hand,
the accuracy of measurements and spectral resolution are better for
XShooter spectra. Because of the limitations for the SDSS sample, we consider
only two characteristics for the entire SDSS+XShooter sample,
the extinction-corrected
O$_3$Mg$_2$ = [O~{\sc iii}]$\lambda$5007/Mg~{\sc ii} $\lambda$2796+2803 and
Mg$_2$ = Mg~{\sc ii} $\lambda$2796/Mg~{\sc ii} $\lambda$2803 flux ratios,
which are less subject to the uncertainties compared to those in fitting of
the Mg~{\sc ii} emission line profiles.

Mg~{\sc ii} emission is detected in most LyC leaking galaxies if it falls
in the wavelength range of SDSS spectra,
as expected in the case of low neutral gas column densities. The two galaxies
with very little (or no) Mg~{\sc ii} detections in Fig.~\ref{fig7}
(J0130$-$0014 and J1157$+$5801) also do not have LyC detections, illustrating
how a non-detection of Mg~{\sc ii} can also lead to a non-detection of LyC.
However, there is one possible exception. The galaxy J1121$+$3806 has
$f_{\rm esc}$(LyC) $\sim$ 35 per cent and strong and narrow Ly$\alpha$ emission
line \citep{I21a}. On the other hand, Mg~{\sc ii} emission in this galaxy is
barely seen (Fig.\,\ref{fig8}d). Thus, the high LyC leakage is possibly not
always associated with the presence of strong Mg~{\sc ii} emission. However,
the SNR of SDSS spectrum is low and this galaxy merits deeper observations
(King et al. in preparation).
 
Fig.\,\ref{fig10}a and \ref{fig10}b show the dependencies of the Ly$\alpha$
escape fraction $f_{\rm esc}$(Ly$\alpha$) on the O$_3$Mg$_2$ and Mg$_2$,
respectively.
It is seen that $f_{\rm esc}$(Ly$\alpha$) is almost
independent of both the O$_3$Mg$_2$ and Mg$_2$ ratios.

Mg$_2$ in two galaxies, J1127$+$4610 and J1455$+$6107
\citep[Fig.\,\ref{fig8}e, \ref{fig8}n, ][]{I21a}, in Fig.\,\ref{fig10}b
is considerably above the value of 2 in the case of zero optical depth
in Mg~{\sc ii} lines, which is unlikely. However, we note that Mg$_2$s
in these two galaxies are measured with the
largest errors, $\sim$ 2 times higher than typical errors for objects
shown in Fig.\,\ref{fig10}b.
Furthermore, these lines in all galaxies were not corrected for interstellar or
stellar photospheric Mg~{\sc ii} absorption. Equivalent widths of
these absorption lines are somewhat uncertain. \citet{G19}
adopted equal equivalent widths of $\sim$\,0.5\AA\ for each of Mg~{\sc ii}
absorption lines, whereas \citet{PR15} derived 2.33\,\AA\
for both lines, which are consistent with the value of $\sim$\,1\AA\ for
Mg~{\sc ii}\,$\lambda$2796 absorption line in star-forming galaxies with stellar
masses $\la$\,10$^{9.5}$M$_\odot$ \citep{Ma12} and the values adopted by
\citet*{Pr11}. All these values are lower than equivalent
widths of Mg~{\sc ii} emission lines (Table\,\ref{taba2}). Assuming that
equivalent widths of Mg~{\sc ii} $\lambda$2796 and $\lambda$2803 absorption
lines are equal and correcting emission lines by multiplying with
(EW$_{\rm em}$+EW$_{\rm abs}$)/EW$_{\rm em}$
results in a reduction of Mg$_2$ ratio if this ratio is above 1. This
is because the equivalent width of the Mg~{\sc ii} $\lambda$2796 emission line
is greater than that of the Mg~{\sc ii} $\lambda$2803 emission line. The
effect is larger for higher values of Mg$_2$ reducing the number of galaxies
with Mg$_2$ above 2.

Using the analytic work of \citet{Ch20}, a
Mg$_2$ of 1.3 would correspond to an Mg~{\sc ii} 2803\AA\ optical depth of 0.43
(or a 2796 optical depth near 1). For the typical abundances of the sample, that
would lead to H~{\sc i} column densities near 9.4$\times$10$^{16}$ cm$^{-2}$,
which is very close to being optically thin for the LyC emission.
It is notable that Mg$_2$ in all five galaxies with high $f_{\rm esc}$(LyC)
observed with the high SNR at the XShooter by \citet{G20} is very close to 2
(black symbols in Fig.\,\ref{fig10}b), in agreement with 
expectations for the low optical depth \citep[e.g. ][]{Ch20}. 

In Fig.\,\ref{fig10}c and \ref{fig10}d we show the relations
of $f_{\rm esc}$(LyC) with the O$_3$Mg$_2$ and Mg$_2$ flux ratios, respectively.
We note an interesting feature in Figs.\,\ref{fig10}b and \ref{fig10}d that
the LyC leakers have preferentially Mg$_2$ $\ga$ 1.3, as expected because
high values of Mg$_2$ indicate low optical depth \citep{Ch20}.
Similarly, \citet{N22} found that galaxies with low $f_{\rm esc}$(LyC) have
preferentially low Mg$_2$ $\sim$ 0.9.
Possibly, a tendency of increasing $f_{\rm esc}$(LyC) with increasing of 
the O$_3$Mg$_2$ and Mg$_2$ is present albeit scatter of the data is large.

The statistics in Fig.\,\ref{fig10} are small and subject to large errors
of individual mesurements. Therefore, for a comparison we
selected $\sim$\,6000 galaxies with $z$\,$\geq$\,0.3 from the sample of compact
star-forming galaxies by \citet{I21c} in which both the Mg~{\sc ii}
$\lambda$2796 and 2803 emission lines were observed. The errors
of Mg~{\sc} line fluxes in this sample are also large. However, large
statistics in each bin of the O$_3$Mg$_2$ and Mg$_2$ flux ratios
considerably reduces the impact of uncertain individual values. These galaxies
constitute 60 per cent of the total number of galaxies in the catalogue
of \citet{I21c} with $z$\,$\geq$\,0.3. Mg~{\sc ii} in the remaining galaxies
is either in absorption or only one of the two lines is detected.

The distribution of
Mg$_2$ for selected galaxies is shown in Fig.\,\ref{fig11}a. This distribution
is broad and approximately 1/3 galaxies have Mg$_2$ $>$ 2. The scatter
is likely caused not only by errors of measurements. It remains even if only
brightest galaxies with well measured Mg~{\sc ii} fluxes are considered
(compare Fig.\,\ref{fig11}c and Fig.\,\ref{fig11}d).
On the other hand, correction for underlying absorption can make the
distribution narrower together with the decreasing number of galaxies with
Mg$_2$ $>$ 2. We find that nearly
2/3 of the sample is characterised by a Mg$_2$\,$>$\,1.3 implying that
most of selected compact star-forming galaxies could possibly be LyC leakers.

The distribution of ionizing photon production efficiency $\xi_{\rm ion}$ for the
same galaxies is shown in Fig.\,\ref{fig11}b. Here
$\xi_{\rm ion}$ = $N$(LyC)/$L_\nu$, where $N$(LyC) and $L_\nu$ are the
production rate of the LyC radiation in photons s$^{-1}$ and the intrinsic
monochromatic luminosity at the rest-frame wavelength of 1500\AA\ in
erg s$^{-1}$ Hz$^{-1}$. It is seen that log $\xi_{\rm ion}$ in the
sample galaxies is high. In most of galaxies it is above the threshold of 25.2,
adopted in models of reionization \citep[e.g. ][]{R13}. Finally,
we show the relations between log $\xi_{\rm ion}$ and Mg$_2$
for all selected SDSS galaxies (Fig.\,\ref{fig11}c) and brightest SDSS galaxies
in the sense that the H$\beta$ fluxes in these galaxies are above
5$\times$10$^{-16}$ erg s$^{-1}$cm$^{-2}$ and equivalent widths of the
Mg~{\sc ii} $\lambda$2796 emission line are above 10\,\AA\
(Fig.\,\ref{fig11}d).
The unshaded region in Fig.\,\ref{fig11}c and \ref{fig11}d is populated by the
galaxies with Mg$_2$ $\geq$ 1.3
and log $\xi_{\rm ion}$ $\geq$ 25.2, which constitute nearly half of the
total sample and somewhat more for the brightest galaxies.
Most of low-$z$ LyC leakers (black filled circles) are located in this region.
The few galaxies with log $\xi_{\rm ion}$ below 25.2 are only from the LzLCS
  sample by \citet{F22a,F22b}, which contains, in general, lower-excitation
H~{\sc ii} regions compared e.g. with the galaxies from the \citet{I16a,I16b,I18a,I18b,I21a} sample.
Thus, the criterion Mg$_2$ $<$ 1.3 can be a useful cut to selected
LyC leaker candidates at low- and high-redshifts due to the fact that
strong Mg~{\sc ii} emission is present in most LyC leaking galaxies.

\section{Conclusions}\label{sec:summary}

We present new {\sl HST} COS low- and medium-resolution spectra of
seven compact SFG in the redshift range
$z$ = 0.3161 -- 0.4276, with various O$_3$Mg$_2$ =
[O~{\sc iii}]$\lambda$5007/Mg~{\sc ii} $\lambda$2796+2803 and
Mg$_2$ = Mg~{\sc ii} $\lambda$2796/Mg~{\sc ii} $\lambda$2803 emission-line
ratios. We aim to obtain properties of leaking LyC and resolved Ly$\alpha$
emission and to study the dependence
of leaking LyC emission on the characteristics of Mg~{\sc ii} emission
along with other indirect indicators of escaping ionizing radiation.
This study is an extension of the work
reported earlier in \citet{I16a,I16b,I18a,I18b,I21a}. Our main
results are summarised as follows:

1. Emission of Lyman continuum is detected in four out of
the seven galaxies with the escape fraction $f_{\rm esc}$(LyC) in the range
between 3.1 per cent (J1137+3605) and 4.6 per cent (J0844+5312).
Only upper limits $f_{\rm esc}$(LyC) $\sim$ 1 -- 3 per cent are
obtained for the remaining three galaxies.

2. A Ly$\alpha$ emission line with two peaks is observed in the
spectra of five galaxies. The Ly$\alpha$ emission line in two galaxies,
J0130$-$0014 and J1157+5801, is very weak. Our new observations
support a strong
anti-correlation between $f_{\rm esc}$(LyC) and the peak velocity separation
$V_{\rm sep}$ of the Ly$\alpha$ profile, confirming the finding of
\citet{I18b,I21a} and making $V_{\rm sep}$ the most robust indirect
indicator of Lyman continuum radiation leakage.

3. Other characteristics such as O$_{32}$ ratio, escape fraction of
the Ly$\alpha$ emission line $f_{\rm esc}$(Ly$\alpha$) and the stellar mass
$M_\star$ show weak or no correlations with $f_{\rm esc}$(LyC), with a high spread
of values, in agreement with earlier studies by e.g. \citet{I16b,I18a,I21a},
\citet{F22b}.

4. We study the characteristics of Mg~{\sc ii} $\lambda$2796+2803 emission, such
as O$_3$Mg$_2$ and Mg$_2$ ratios, as possible indirect indicators of escaping
LyC emission. We find that galaxies with detected LyC emission have
preferentially Mg$_2$ $\ge$ 1.3, the latter indicating low optical depths.
A high Mg$_2$ ratio of $\geq$ 1.3 can be used to select LyC leaker candidates.
A tendency of an increase of $f_{\rm esc}$(LyC) with increasing of both the
  O$_3$Mg$_2$ and Mg$_2$ is possibly present. However, there is substantial
scatter in these relations due to the low signal-to-noise ratio
in the blue part of the SDSS spectra near the observed Mg~{\sc ii} emission
  not allowing their use for reliable prediction of $f_{\rm esc}$(LyC).

5. We find that galaxies with Mg$_2$ $\ge$ 1.3 and ionizing photon
production efficiency $\xi_{\rm ion}$ greater than the value of
10$^{25.2}$ erg$^{-1}$ Hz used in modelling of the process of reionization of the
Universe \citep[e.g. ][]{R13} constitute $\sim$ 40 per cent of all compact
star-forming galaxies at redshift $z$ $\geq$ 0.3, which were selected by
\citet{I21c} from the Data Release 16 of the Sloan Digital Sky Survey. 

6. A bright compact star-forming region superimposed on a
low-surface-brightness component is seen in the COS near ultraviolet
(NUV) acquisition images of five galaxies (two images are
missing due to technical problems). The surface brightness at the
outskirts of our galaxies can be approximated by an exponential disc, with a
scale length of $\sim$ 0.20 -- 0.63 kpc. This is $\sim$ 4 times lower than the
scale lengths of the LyC leakers observed by \citet{I16b,I18a,I18b},
but is similar to that in low-mass galaxies with $M_\star$ $<$ 10$^8$ M$_\odot$
by \citet{I21a}.
Part of this difference may be explained by acquisition
exposure times that are $\sim$ 2 shorter compared to those used by
\citet{I16b,I18a,I18b}, resulting in less deep images.

7. The star formation rates in the range SFR $\sim$ 4 -- 36 M$_\odot$ yr$^{-1}$
and the metallicities of our new galaxies, ranging from 12 + logO/H
= 7.81 to 8.06, are overlapping with those in the LyC leakers
studied by \citet{I16a,I16b,I18a,I18b,I21a}.

\section*{Acknowledgements}

Based on observations made with the NASA/ESA {\sl Hubble Space Telescope}, 
obtained from the data archive at the Space Telescope Science Institute. 
Support for this work was provided by NASA through grant number HST-GO-15845
from the Space Telescope Science Institute. STScI is operated by the
Association of Universities for Research in Astronomy,
Inc. under NASA contract NAS 5-26555. 
Y.I. and N.G. acknowledge support from the National Academy of Sciences of 
Ukraine by its priority project No. 0122U002259 ``Fundamental properties of 
the matter and its manifistation in micro world, astrophysics and cosmology''.
Funding for SDSS-III has been provided by the Alfred P. Sloan Foundation, 
the Participating Institutions, the National Science Foundation, and the U.S. 
Department of Energy Office of Science. The SDSS-III web site is 
http://www.sdss3.org/. SDSS-III is managed by the Astrophysical Research 
Consortium for the Participating Institutions of the SDSS-III Collaboration. 
{\sl GALEX} is a NASA mission  managed  by  the  Jet  Propulsion  Laboratory.
This research has made use of the NASA/IPAC Extragalactic Database (NED) which 
is operated by the Jet  Propulsion  Laboratory,  California  Institute  of  
Technology,  under  contract with the National Aeronautics and Space 
Administration.

\section*{Data availability}

The data underlying this article will be shared on reasonable request to the 
corresponding author.





\begin{thebibliography}{}

\bibitem[Ade et al.(2014)]{P14} Ade P. A. R. et al.,
2014, \aap, 571, A16

\bibitem[Aller(1984)]{A84} Aller L. H., 1984, Physics of Thermal
Gaseous Nebulae. Dordrecht: Reidel

\bibitem[Bian et al.(2017)]{B17} Bian F., Fan X., McGreer I., Cai Z., Jiang L.,
2017, \apj, 837, 12

\bibitem[Borthakur et al.(2014)]{B14} Borthakur S., Heckman T. M., 
Leitherer C., Overzier R. A., 2014, \sci, 346, 216

\bibitem[Bouwens et al.(2015)]{B15a} Bouwens R. J., Illingworth G. D., 
Oesch P. A., Caruana J., Holwerda B., Smit R.,  Wilkins S., 
2015, \apj, 811, 140

\bibitem[Bouwens et al.(2017)]{Bo17}  Bouwens R. J., Illingworth G. D., 
Oesch P. A., Atek H, Lam D, Stefanon M., 2017, \apj, 843, 41

\bibitem[Calzetti et al.(2000)]{C00} Calzetti D., Armus L., Bohlin R. C.,
Kinney A. L., Koornneef J., Storchi-Bergmann T., 2000, \apj, 533, 682

\bibitem[Cardamone et al.(2009)]{Ca09} Cardamone C. et al.,
2009, \mnras, 399, 1191

\bibitem[Cardelli et al.(1989)Cardelli, Clayton \& Mathis]{C89} 
Cardelli J. A., Clayton G. C., Mathis J. S., 1989, \apj, 345, 245

\bibitem[Caruana et al.(2018)]{Ca18} Caruana J. et al.,
2018, \mnras, 473, 30

\bibitem[Chisholm et al.(2017)]{C17} Chisholm J., Orlitov\'a I., 
Schaerer D., Verhamme A., Worseck G., Izotov Y. I., Thuan T. X., Guseva N. G.,
2017, \aap, 605, A67

\bibitem[Chisholm et al.(2018)]{Ch18} Chisholm J. et al.,
2018, \aap, 616, 30

\bibitem[Chisholm et al.(2020)]{Ch20} Chisholm J., Prochaska J. X., 
Schaerer D., Gazagnes S., Henry A., 2020, \mnras, 498, 2554

\bibitem[Cowie et al.(2009)Cowie, Barger \& Trouille]{C09} 
Cowie L. L., Barger A. J.,  Trouille L., 2009, \apj, 692, 1476

\bibitem[Curtis-Lake et al.(2016)]{CL16} Curtis-Lake E. et al., 
2016, \mnras, 457, 440

\bibitem[de Barros et al.(2016)]{B16} de Barros S. et al., 2016, \aap, 585, A51

\bibitem[de Barros et al.(2019)]{B19} de Barros S., Oesch P. A., Labb\'e I., 
Stefanon M., Gonz\'alez V., Smit R., Bouwens R. J., Illingworth G. D., 
2019, \mnras, 489, 2355

\bibitem[Dijkstra et al.(2016)Dijkstra, Gronke \& Venkatesan]{Di16}
Dijkstra M., Gronke M., Venkatesan A., 2016, \apj, 828, 71

\bibitem[Dressler et al.(2015)]{D15} Dressler A., Henry A., Martin C. L., 
Sawicki M., McCarthy P., Villaneuva E., 2015, \apj, 806, 19

\bibitem[Endsley et al.(2021)]{E20} Endsley R., Stark D. P., Chevallard J.,
Charlot S., 2021, \mnras, 500, 5229

\bibitem[Erb et al.(2012)]{E12} Erb D. K., Quider A. M., Henry A. L.,
Martin C. L., 2012, \apj, 759, 26

\bibitem[Faisst(2016)]{F16} Faisst A. L., 2016, \apj, 829, 99

\bibitem[Finkelstein et al.(2019)]{Fi19} Finkelstein S. L. et al., 
2019, \apj, 879, 36

\bibitem[Finley et al.(2017)]{F17} Finley H. et al. 2017, \aap, 608, A7

\bibitem[Fletcher et al.(2019)]{Fl19} Fletcher T. J., Tang M., Robertson B. E., 
Nakajima K., Ellis R. S., Stark D. P., Inoue A., 2019, \apj, 878, 87

\bibitem[Flury et al.(2022a)]{F22a} Flury S. R. et al. 2022a, \apjs, 260, 1

\bibitem[Flury et al.(2022b)]{F22b} Flury S. R. et al. 2022b, \apj, 930, 126

\bibitem[Gazagnes et al.(2018)]{Ga18} Gazagnes S., Chisholm J., Schaerer D., 
Verhamme A., Rigby J. R., Bayliss M., 2018, \aap, 616, 29

\bibitem[Gazagnes et al.(2020)]{Ga20} Gazagnes S., Chisholm J., Schaerer D., 
Verhamme A., Izotov Y., 2020, \aap, 639, 85 

\bibitem[Girardi et al.(2000)]{G00} Girardi L., Bressan A., Bertelli G., 
Chiosi C., 2000, \aaps, 141, 371

\bibitem[Grazian et al.(2016)]{Gr16} Grazian A. et al., 2016, \aap, 585, A48 

\bibitem[Gronke et al.(2021)]{Gr21} Gronke M. et al., 2021, \mnras, 508, 3697

\bibitem[Guseva et al.(2013)]{G13} Guseva N. G., Izotov Y. I., Fricke K. J.,
Henkel C., 2013, \aap, 555, A90

\bibitem[Guseva et al.(2019)]{G19} Guseva N. G., Izotov Y. I., Fricke K. J.,
Henkel C., 2019, \aap, 624, A21

\bibitem[Guseva et al.(2020)]{G20} Guseva N. G. et al., 2020, \mnras, 497, 4293

\bibitem[Henry et al.(2015)]{H15} Henry A., Scarlata C., Martin C. S., Erb D.,
2015, \apj, 809, 19

\bibitem[Henry et al.(2018)]{Hen18} Henry A., Berg D. A., Scarlata C., 
Verhamme A., Erb D., 2018, \apj, 855, 96

\bibitem[Inoue et al.(2014)]{Inoue14} Inoue A.~K., Shimizu 
I., Iwata I., Tanaka M., 2014, \mnras, 442, 1805 

\bibitem[Izotov et al.(1994)Izotov, Thuan \& Lipovetsky]{ITL94} Izotov Y. I.,
Thuan T. X., Lipovetsky V. A., 1994, \apj, 435, 647

\bibitem[Izotov et al.(2006)]{I06} Izotov Y. I., Stasi\'nska G., Meynet G.,
Guseva N. G., Thuan T. X., 2006, \aap, 448, 955

\bibitem[Izotov et al.(2011)Izotov, Guseva \& Thuan]{I11} Izotov Y. I., 
Guseva N. G., Thuan T. X., 2011, \apj, 728, 161

\bibitem[Izotov et al.(2015)]{I15} Izotov Y. I., Guseva N. G., 
Fricke K. J.,  Henkel C., 2015, \mnras, 451, 2251

\bibitem[Izotov et al.(2016a)]{I16a} Izotov Y. I., Orlitov\'a I., Schaerer D.,
Thuan T. X., Verhamme A., Guseva N. G.,  Worseck G., 2016a, \nat, 529, 178

\bibitem[Izotov et al.(2016b)]{I16b} Izotov Y. I., Schaerer D., Thuan, T. X., 
Worseck G., Guseva N. G., Orlitov\'a I., Verhamme A., 2016b, \mnras, 461, 3683

\bibitem[Izotov et al.(2017)]{I17} Izotov Y. I., Guseva N. G., 
Fricke K. J.,  Henkel C., Schaerer D., 2017, \mnras, 467, 4718

\bibitem[Izotov et al.(2018a)]{I18a} Izotov Y. I., Schaerer D.,
Worseck G., Guseva N. G., Thuan, T. X., Verhamme A., Orlitov\'a I., Fricke K. J,
 2018a, \mnras, 474, 4514

\bibitem[Izotov et al.(2018b)]{I18b} Izotov Y. I., Worseck G., Schaerer D.,
Guseva N. G., Thuan, T. X., Fricke K. J, Verhamme A., Orlitov\'a I.,
2018b, \mnras, 478, 4851

\bibitem[Izotov et al.(2018c)]{I18c} Izotov Y. I., Thuan T. X., Guseva N. G., 
Liss S. E., 2018c, \mnras, 473, 1956

\bibitem[Izotov et al.(2020)]{I20} Izotov Y. I., Schaerer D., Worseck G., 
Verhamme A., Guseva N. G., Thuan T. X., Orlitov\'a I., Fricke K. J.,  
2020, \mnras, 491, 468

\bibitem[Izotov et al.(2021a)]{I21a} Izotov Y. I., Worseck, G., Schaerer D., 
Guseva N. G., Chisholm J., Thuan T. X., Fricke K. J., Verhamme A., 2021a,
\mnras, 503, 1734

\bibitem[Izotov et al.(2021b)Izotov, Thuan \& Guseva]{I21b}
  Izotov Y. I., Thuan T. X., Guseva N. G., 2021b, \mnras, 504, 3996

\bibitem[Izotov et al.(2021c)]{I21c} Izotov Y. I., Guseva N. G., Fricke K. J.,
  Henkel C., Schaerer D., Thuan T. X., 2021c, \aap, 646, A138

\bibitem[Jaskot \& Oey(2013)]{JO13} Jaskot A. E., Oey M. S.,
2013, \apj, 766, 91

\bibitem[Jaskot \& Oey(2014)]{JO14} Jaskot A. E., Oey M. S.,
2014, \apj, 791, L19

\bibitem[Katz et al.(2022)]{Ka22} Katz H. et al., 2022, \mnras,
  https://doi.org/10.1093/mnras/stac1437

\bibitem[Kennicutt(1998)]{K98} Kennicutt R. C., Jr.,
1998, \araa, 36, 189

\bibitem[Khaire et al.(2016)]{K16} Khaire V., Srianand R., Choudhury T. R.,
Gaikwad P., 2016, \mnras, 457, 4051

\bibitem[Kim et al.(2020)]{Ki20} Kim K., Malhotra S., Rhoads J. E., 
Butler N. R., Yang H., 2020, \apj, 893, 134

\bibitem[Kornei et al.(2013)]{K13}Kornei K. A., Shapley A. E., Martin C. L.,
  Coil A. L., Lotz J. M., Weiner B. J., 2013, \apj, 774, 50

\bibitem[Kroupa(2001)]{K01} Kroupa P., 2001, \mnras, 322, 231

\bibitem[Labb\'e et al.(2013)]{La13} Labb\'e I. et al., 2013, \apj, 777, L19

\bibitem[Leitet et al.(2013)]{L13} Leitet E., Bergvall N., Hayes M., 
Linn\'e S., Zackrisson E., 2013, \aap, 553, A106

\bibitem[Leitherer et al.(2016)]{L16} Leitherer C., Hernandez S., 
Lee J. C., Oey M. S., 2016, \apj, 823, L64

\bibitem[Lejeune et al.(1997)Lejeune, Buser \& Cuisiner]{L97} Lejeune T., 
Buser R., Cuisinier F., 1997, \aaps, 125, 229

\bibitem[Lewis et al.(2020)]{Le20} Lewis J. S. W. et al., 
2020, \mnras, 496, 4342
\bibitem[Makan et al.(2021)]{Ma21} Makan K., Worseck G., Davies F. B.,
Hennawi J. F., Prochaska J. X., Richter P., 2021, \apj, 912, 38

\bibitem[Marchi et al.(2017)]{Ma17} Marchi F. et al., 2017, \aap, 601, 73

\bibitem[Marchi et al.(2018)]{Ma18} Marchi F. et al., 2018, \aap, 614, 11

\bibitem[Martin et al.(2012)]{Ma12} Martin C. L., Shapley A. E., Coil A. L.,
  Kornei K. A., Bundi K., Weiner B. J., Noeske K. G., Schiminovich D.,
  2012, \apj, 760, 127

\bibitem[Mathis(1990)]{Ma90} Mathis J. S., 1990, \araa, 28, 10
  
\bibitem[Matsuoka et al.(2018)]{Mats18} Matsuoka Y. et al., 2018, \apj, 869, 150

\bibitem[Me\v{s}tric et al.(2020)]{Mes20} Me\v{s}tric U. et al., 2020,
\mnras, 494, 4986

\bibitem[Meyer et al.(2020)]{Me20} Meyer R. A. et al., 2020, 494, 1560

\bibitem[Mitra et al.(2013)Mitra, Ferrara \& Choudhury]{M13} 
Mitra S., Ferrara A., Choudhury T. R., 2013, \mnras, 428, L1

\bibitem[Naidu et al.(2020)]{N20} Naidu R. P., Tacchella S., Mason C. A., 
Bose S., Oesch P. A., Conroy C., 2020, \apj, 892, 109

\bibitem[Naidu et al.(2022)]{N22} Naidu R. P. et al., 2022, \mnras, 510, 4582

\bibitem[Nakajima \& Ouchi(2014)]{NO14} Nakajima K., Ouchi M.,
2014, \mnras, 442, 900

\bibitem[Nakajima et al.(2018)]{Na18} Nakajima K., Fletcher T., Ellis R. S.,
Robertson B. E., Iwata I., 2018, \mnras, 477, 2098

\bibitem[Nakajima et al.(2020)]{Na20} Nakajima K., Ellis R. S., Robertson B. E.,
 Tang M., Stark D. P., 2020, \apj, 889, 161 

\bibitem[Ouchi et al.(2009)]{O09} Ouchi M. et al., 2009, \apj, 706, 1136

\bibitem[Paulino-Afonso et al.(2018)]{PA18} Paulino-Afonso A. et al., 2018,
\mnras, 476, 5479

\bibitem[P\'erez-R\`afols et al.(2015)]{PR15} P\'erez-R\`afols I.,
  Miralda-Escud\'e J., Lundgren B., Ge J., Petitjean P., Schneider
D. P., York D. G., Weaver B. A., 2015, \mnras, 447, 2784

\bibitem[Prochaska et al.(2011)Prochaska, Kasen \& Rubin]{Pr11}
  Prochaska J. X., Kasen D., Rubin K., 2011, \apj, 734, 24

\bibitem[Rivera-Thorsen et al.(2017)]{RT17} Rivera-Thorsen T. E. et al., 
2017, \aap, 608, L4

\bibitem[Rivera-Thorsen et al.(2019)]{RT19} Rivera-Thorsen T. E. et al., 
2019, \sci, 366, 738

\bibitem[Robertson et al.(2013)]{R13} Robertson B. E. et al.,
2013, \apj, 768, 71

\bibitem[Robertson et al.(2015)]{Robertson15} Robertson B.~E., 
Ellis R.~S., Furlanetto S.~R., Dunlop J.~S., 2015, \apjl, 802, L19 

\bibitem[Saha et al.(2020)]{Sa20} Saha K. et al., 2020, \nata, 4, 1185

\bibitem[Saldana-Lopez et al.(2022)]{SL22} Saldana-Lopez A. et al., 2022,
  \aap, in press; preprint arXiv:2201.11800

\bibitem[Salpeter(1955)]{S55} Salpeter E. E., 1955, ApJ, 121, 161

\bibitem[Schmutz et al.(1992)Schmutz, Leitherer \& Gruenwald]{S92} 
Schmutz W., Leitherer C.,  Gruenwald R., 1992, \pasp, 104, 1164

\bibitem[Shapley et al.(2016)]{Sh16} Shapley A. E., Steidel C. C., 
Strom A. L., Bogosavljevi\'c M., Reddy N. A., Siana B. Mostardi R. E., 
Rudie G. C., 2016, \apj, 826, L24

\bibitem[Smit et al.(2014)]{Sm14} Smit R. et al., 2014, \apj, 784, 58

\bibitem[Stasi\'nska et al.(2015)]{S15} Stasi\'nska G., Izotov Y., 
Morisset C., Guseva N., 2015, \aap, 576, A83

\bibitem[Steidel et al.(2018)]{St18} Steidel C. C., Bogosavljevi\'c M., 
Shapley A.E., Reddy N. A., Rudie G. C., Pettini M., Trainor R. F., 
Strom A. L., 2018, \apj, 869, 123

\bibitem[Thuan \& Martin(1981)]{TM81} Thuan T. X., Martin G. E., 1981, \apj,
247, 823
 
\bibitem[Trebitsch et al.(2017)]{T17} Trebitsch M., Blaizot J., Rosdahl J.,
Devriendt J., Slyz A., 2017, \mnras, 470, 224

\bibitem[Vanzella et al.(2010)]{V10} Vanzella E. et al., 
2010, \apj, 725, 1011

\bibitem[Vanzella et al.(2012)]{V12} Vanzella E. et al., 2012, \apj, 751, 70 

\bibitem[Vanzella et al.(2015)]{Va15} Vanzella E. et al., 2015, \aap, 576, A116 

\bibitem[Vanzella et al.(2018)]{Va18} Vanzella E. et al., 2018, \mnras, 476, 
L15 

\bibitem[Vanzella et al.(2020)]{Va20} Vanzella E. et al., 2020, \mnras, 491, 
1093 

\bibitem[Verhamme et al.(2015)]{V15}  Verhamme A., Orlitov\'a I., 
Schaerer D., Hayes M., 2015, \aap, 578, A7

\bibitem[Verhamme et al.(2017)]{V17} Verhamme A., Orlitov\'a I., Schaerer D., 
Izotov Y., Worseck G., Thuan T. X., Guseva N., 2017, \aap, 597, A13

\bibitem[Vielfaure et al.(2020)]{Vi20} Vielfaure J.-B., et al. 2020, \aap,
640, 30

\bibitem[Wang et al.(2021)]{W21} Wang B. et al., 2021, \apj, 916, 3

\bibitem[Weiner et al.(2009)]{W09} Weiner B. J. et al., 2009, \apj, 692, 187

\bibitem[Witstok et al.(2021)]{Wi21} Witstok J., Smit R., Maiolino R., Curti M.,
 Laporte N., Massey R., Richard J., Swinbank M., 2021, \mnras, 508, 1686

\bibitem[Wise \& Chen(2009)]{WC09} Wise J. H., Cen R.,
2009, \apj, 693, 984

\bibitem[Wise et al.(2014)]{W14} Wise J.~H., Demchenko 
V.~G., Halicek M.~T., Norman M. L., Turk M. J., Abel T., Smith B. D., 
2014, \mnras, 442, 2560 

\bibitem[Worseck et al.(2016)]{W16} Worseck G., Prochaska J. X., Hennawi J. F.,
McQuinn M., 2016, \apj, 825, 144

\bibitem[Wright(2006)]{W06} Wright E. L., 2006, \pasp, 118, 1711

\bibitem[Xu et al.(2022)]{Xu22} Xu X. et al., 2022, \apj, in press;
  preprint arXiv:2205.11317

\bibitem[Yajima et al.(2011)Yajima, Choi \& Nagamine]{Y11} 
Yajima H., Choi J.-H., Nagamine K., 2011, \mnras, 412, 411

\bibitem[Yang et al.(2017a)]{Y17} Yang H. et al., 2017a, \apj, 844, 171

\bibitem[Yang et al.(2017b)]{Ya17} Yang H., Malhotra S., Rhoads J. E., 
Leitherer C., Wofford A., Jiang T., Wang J., 2017b, \apj, 847, 38

\end{thebibliography}


\appendix

\section{Apparent magnitudes}

  \begin{table*}
  \caption{Apparent AB magnitudes with errors in parentheses compiled
from the SDSS and {\sl GALEX} databases and apparent Vega magnitudes from
the {\sl WISE} database
\label{taba1}}
\begin{tabular}{lccccccccccccc} \hline
Name&\multicolumn{5}{c}{SDSS}
&&\multicolumn{2}{c}{\sl GALEX}&&\multicolumn{4}{c}{{\sl WISE}} \\ 
    &\multicolumn{1}{c}{$u$}&\multicolumn{1}{c}{$g$}&\multicolumn{1}{c}{$r$}&\multicolumn{1}{c}{$i$}&\multicolumn{1}{c}{$z$}&&FUV&NUV&&\multicolumn{1}{c}{$W1$}&\multicolumn{1}{c}{$W2$}&\multicolumn{1}{c}{$W3$}&\multicolumn{1}{c}{$W4$}
\\
    &\multicolumn{1}{c}{(err)}&\multicolumn{1}{c}{(err)}&\multicolumn{1}{c}{(err)}&\multicolumn{1}{c}{(err)}&\multicolumn{1}{c}{(err)}&&(err)&(err)&&\multicolumn{1}{c}{(err)}&\multicolumn{1}{c}{(err)}&\multicolumn{1}{c}{(err)}&\multicolumn{1}{c}{(err)} \\
\hline
J0130$-$0014& 22.09& 21.95& 21.09& 22.18& 21.80&& 22.42& 22.15&&\multicolumn{1}{c}{16.81}&\multicolumn{1}{c}{16.10}&\multicolumn{1}{c}{ ... }&\multicolumn{1}{c}{ ... } \\
            &(0.18)&(0.08)&(0.05)&(0.17)&(0.42)&&(0.16)&(0.14)&&\multicolumn{1}{c}{(0.09)}&\multicolumn{1}{c}{(0.17)}&\multicolumn{1}{c}{(...)}&\multicolumn{1}{c}{(...)} \\
J0141$-$0304& 21.45& 21.25& 21.22& 20.88& 20.83&& 21.35& 22.04&&\multicolumn{1}{c}{ ... }&\multicolumn{1}{c}{ ... }&\multicolumn{1}{c}{ ... }&\multicolumn{1}{c}{ ... } \\
            &(0.12)&(0.04)&(0.05)&(0.05)&(0.19)&&(0.35)&(0.44)&&\multicolumn{1}{c}{(...)}&\multicolumn{1}{c}{(...)}&\multicolumn{1}{c}{(...)}&\multicolumn{1}{c}{(...)} \\
J0844$+$5312& 21.50& 21.44& 21.57& 20.35& 22.59&& 21.99& 21.61&&\multicolumn{1}{c}{ ... }&\multicolumn{1}{c}{ ... }&\multicolumn{1}{c}{ ... }&\multicolumn{1}{c}{ ... } \\
            &(0.15)&(0.06)&(0.09)&(0.04)&(0.73)&&(0.53)&(0.49)&&\multicolumn{1}{c}{(...)}&\multicolumn{1}{c}{(...)}&\multicolumn{1}{c}{(...)}&\multicolumn{1}{c}{(...)} \\
J1014$+$5501& 21.94& 21.73& 21.66& 21.58& 21.45&& 21.88& 22.29&&\multicolumn{1}{c}{17.54}&\multicolumn{1}{c}{15.70}&\multicolumn{1}{c}{12.51}&\multicolumn{1}{c}{ ... } \\
            &(0.16)&(0.06)&(0.09)&(0.11)&(0.30)&&(0.59)&(0.70)&&\multicolumn{1}{c}{(0.16)}&\multicolumn{1}{c}{(0.11)}&\multicolumn{1}{c}{(0.43)}&\multicolumn{1}{c}{(...)} \\
J1137$+$3605& 22.23& 21.79& 20.96& 22.09& 20.56&&  ... & 22.46&&\multicolumn{1}{c}{17.50}&\multicolumn{1}{c}{15.89}&\multicolumn{1}{c}{12.37}&\multicolumn{1}{c}{ ... } \\
            &(0.23)&(0.07)&(0.05)&(0.18)&(0.17)&& (...)&(0.20)&&\multicolumn{1}{c}{(0.17)}&\multicolumn{1}{c}{(0.15)}&\multicolumn{1}{c}{(0.42)}&\multicolumn{1}{c}{(...)} \\
J1157$+$5801& 24.03& 22.32& 21.33& 22.59& 21.73&&  ... & 22.94&&\multicolumn{1}{c}{ ... }&\multicolumn{1}{c}{ ... }&\multicolumn{1}{c}{ ... }&\multicolumn{1}{c}{ ... } \\
            &(0.82)&(0.10)&(0.07)&(0.34)&(0.48)&& (...)&(0.37)&&\multicolumn{1}{c}{(...)}&\multicolumn{1}{c}{(...)}&\multicolumn{1}{c}{(...)}&\multicolumn{1}{c}{(...)} \\
J1352$+$5617& 22.17& 21.73& 21.54& 21.10& 21.55&& 21.83& 21.81&&\multicolumn{1}{c}{ ... }&\multicolumn{1}{c}{ ... }&\multicolumn{1}{c}{ ... }&\multicolumn{1}{c}{ ... } \\
            &(0.23)&(0.06)&(0.07)&(0.08)&(0.41)&&(0.17)&(0.13)&&\multicolumn{1}{c}{(...)}&\multicolumn{1}{c}{(...)}&\multicolumn{1}{c}{(...)}&\multicolumn{1}{c}{(...)} \\
\hline
\end{tabular}

  \end{table*}

\section{Emission line fluxes and chemical composition}

  \begin{table*}
  \caption{Extinction-corrected fluxes and rest-frame equivalent widths of
the emission lines in SDSS spectra
\label{taba2}}
\begin{tabular}{lcrrrrrrrr} \hline
 & &\multicolumn{8}{c}{Galaxy}\\
Line &\multicolumn{1}{c}{$\lambda$}&\multicolumn{2}{c}{J0130$-$0014}& \multicolumn{2}{c}{J0141$-$0304}& \multicolumn{2}{c}{J0844$+$5312}& \multicolumn{2}{c}{J1014$+$5501}\\
     &&\multicolumn{1}{c}{$I$$^{\rm a}$}&\multicolumn{1}{c}{EW$^{\rm b}$}&\multicolumn{1}{c}{$I$$^{\rm a}$}&\multicolumn{1}{c}{EW$^{\rm b}$}&\multicolumn{1}{c}{$I$$^{\rm a}$}&\multicolumn{1}{c}{EW$^{\rm b}$}&\multicolumn{1}{c}{$I$$^{\rm a}$}&\multicolumn{1}{c}{EW$^{\rm b}$}\\
\hline
Mg~{\sc ii}          &2796&\multicolumn{1}{c}{...}& ...& 26.3$\pm$3.2&  11& 43.4$\pm$4.5&  15& 24.0$\pm$4.5          &   8\\
Mg~{\sc ii}          &2803&\multicolumn{1}{c}{...}& ...& 16.2$\pm$2.8&   7& 18.2$\pm$3.5&   5& 20.1$\pm$4.3          &   7\\
$[$O~{\sc ii}$]$         &3727&  92.3$\pm$7.7         & 109&118.0$\pm$5.5  & 142&137.7$\pm$7.0&  98& 91.8$\pm$6.9          & 107\\
H12                  &3750&\multicolumn{1}{c}{...}& ...&  3.3$\pm$2.0&   3&\multicolumn{1}{c}{...}& ...&\multicolumn{1}{c}{...}& ...\\
H11                  &3771&\multicolumn{1}{c}{...}& ...&  5.2$\pm$1.9&   6&\multicolumn{1}{c}{...}& ...&\multicolumn{1}{c}{...}& ...\\
H10                  &3798&   6.0$\pm$2.8         &   9&  7.7$\pm$2.0&   9&  7.1$\pm$2.3          &   6&\multicolumn{1}{c}{...}& ...\\
H9                   &3836&  10.2$\pm$3.0         &  21&  9.1$\pm$1.9&  11& 14.0$\pm$2.6          &  14&  9.7$\pm$2.9          &  12\\
$[$Ne~{\sc iii}$]$   &3869&  47.1$\pm$5.4         &  78& 54.1$\pm$3.5&  65& 58.5$\pm$4.4          &  47& 51.8$\pm$5.2          &  64\\
H8+He~{\sc i}        &3889&  18.6$\pm$3.6         &  36& 21.4$\pm$2.4&  28& 22.8$\pm$3.0          &  21& 20.6$\pm$3.6          &  25\\
H7+$[$Ne~{\sc iii}$]$&3969&  29.3$\pm$4.4         &  47& 35.6$\pm$2.9&  49& 39.2$\pm$3.7          &  37& 37.9$\pm$4.4          &  59\\
H$\delta$            &4101&  24.8$\pm$4.0         &  52& 30.0$\pm$2.6&  47& 30.6$\pm$3.3          &  30& 28.1$\pm$4.0          &  33\\
H$\gamma$            &4340&  47.0$\pm$5.3         &  96& 47.9$\pm$3.1&  85& 48.6$\pm$3.8          &  55& 48.1$\pm$4.9          &  66\\
$[$O~{\sc iii}$]$    &4363&  12.3$\pm$3.0         &  30&  9.7$\pm$1.6&  18& 11.3$\pm$2.2          &  16& 10.9$\pm$2.7          &  16\\
He~{\sc i}           &4471&\multicolumn{1}{c}{...}& ...&  4.8$\pm$1.3&  10&  6.4$\pm$1.9          &   8&  4.2$\pm$2.3          &   5\\
H$\beta$             &4861& 100.0$\pm$7.8         & 200&100.0$\pm$4.4& 220&100.0$\pm$5.2          & 196&100.0$\pm$6.7          & 240\\
$[$O~{\sc iii}$]$    &4959& 228.2$\pm$12.         & 507&222.9$\pm$7.2& 501&223.7$\pm$8.4          & 485&217.8$\pm$10.          & 619\\
$[$O~{\sc iii}$]$    &5007& 685.9$\pm$25.         &1697&656.0$\pm$17.&1419&671.4$\pm$18.          &1495&625.9$\pm$20.          &1654\\
He~{\sc i}           &5876&   8.7$\pm$2.6         &  24& 12.6$\pm$1.5&  32& 10.7$\pm$1.8          &  27&\multicolumn{1}{c}{...}& ...\\
H$\alpha$            &6563& 280.0$\pm$15.         &2306&289.3$\pm$8.9& 712&285.5$\pm$10.          & 723&282.7$\pm$13.          & 901\\
$[$N~{\sc ii}$]$     &6583&   4.2$\pm$1.8         &  27& 15.3$\pm$1.6&  39& 10.2$\pm$1.7          &  35&  4.9$\pm$1.6          &  32\\
$[$S~{\sc ii}$]$     &6717&   5.5$\pm$1.9         &  20&  9.0$\pm$1.2&  28& 10.7$\pm$1.7          &  35&\multicolumn{1}{c}{...}& ...\\
$[$S~{\sc ii}$]$     &6731&   6.3$\pm$2.1         &  25&  8.0$\pm$1.2&  23&  6.2$\pm$1.4          &  18&\multicolumn{1}{c}{...}& ...\\
$C$(H$\beta$)$_{\rm int}$$^{\rm c}$  &&\multicolumn{2}{c}{0.000$\pm$0.065}&\multicolumn{2}{c}{0.268$\pm$0.038}&\multicolumn{2}{c}{0.172$\pm$0.043}&\multicolumn{2}{c}{0.084$\pm$0.054}\\
$C$(H$\beta$)$_{\rm MW}$$^{\rm d}$   &&\multicolumn{2}{c}{0.046}&\multicolumn{2}{c}{0.030}&\multicolumn{2}{c}{0.036}&\multicolumn{2}{c}{0.019}\\
$I$(H$\beta$)$^{\rm e}$     &&\multicolumn{2}{c}{5.0$\pm$0.5}&\multicolumn{2}{c}{32.6$\pm$1.2}&\multicolumn{2}{c}{17.3$\pm$0.9}&\multicolumn{2}{c}{8.1$\pm$0.6}\\
\hline

 & &\multicolumn{6}{c}{Galaxy}\\
Line &\multicolumn{1}{c}{$\lambda$}&\multicolumn{2}{c}{J1137$+$3605}& \multicolumn{2}{c}{J1157$+$5801}& \multicolumn{2}{c}{J1352$+$5617} \\
     &&\multicolumn{1}{c}{$I$$^{\rm a}$}&\multicolumn{1}{c}{EW$^{\rm b}$}&\multicolumn{1}{c}{$I$$^{\rm a}$}&\multicolumn{1}{c}{EW$^{\rm b}$}&\multicolumn{1}{c}{$I$$^{\rm a}$}&\multicolumn{1}{c}{EW$^{\rm b}$}\\
\cline{1-8}
Mg~{\sc ii}          &2796&  16.7$\pm$3.1         &  20&  9.2$\pm$4.1&   8& 37.2$\pm$4.7&  16\\
Mg~{\sc ii}          &2803&  10.9$\pm$2.8         &  13&  9.5$\pm$4.1&   8& 24.8$\pm$4.0&  11\\
$[$O~{\sc ii}$]$     &3727&  81.5$\pm$5.2         & 132& 72.0$\pm$5.8& 111&159.9$\pm$8.6& 153\\
H12                  &3750&\multicolumn{1}{c}{...}& ...&  6.7$\pm$2.8&  21&\multicolumn{1}{c}{...}& ...\\
H11                  &3771&\multicolumn{1}{c}{...}& ...&  6.4$\pm$2.7&  22&\multicolumn{1}{c}{...}& ...\\
H10                  &3798&   5.7$\pm$2.0         &  11&  9.5$\pm$2.9&  28&  4.8$\pm$2.3&   5\\
H9                   &3836&  11.1$\pm$2.3         &  23& 10.6$\pm$2.9&  31&  8.0$\pm$2.4&   9\\
$[$Ne~{\sc iii}$]$   &3869&  51.5$\pm$4.0         & 103& 59.6$\pm$5.3&  70& 50.3$\pm$4.7&  47\\
H8+He~{\sc i}        &3889&  19.4$\pm$2.7         &  43& 22.6$\pm$3.9&  39& 18.2$\pm$3.2&  17\\
H7+$[$Ne~{\sc iii}$]$&3969&  35.5$\pm$3.4         &  81& 44.5$\pm$4.7&  50& 28.7$\pm$3.7&  27\\
H$\delta$            &4101&  26.6$\pm$2.9         &  61& 31.2$\pm$3.9&  40& 25.3$\pm$3.5&  23\\
H$\gamma$            &4340&  47.6$\pm$3.6         & 104& 48.8$\pm$4.5& 101& 51.2$\pm$4.4&  74\\
$[$O~{\sc iii}$]$    &4363&  13.9$\pm$2.1         &  42& 15.4$\pm$2.8&  33&  9.3$\pm$2.4&  11\\
He~{\sc i}           &4471&   3.7$\pm$1.5         &   8&  5.0$\pm$2.0&  14&  5.3$\pm$2.0&   8\\
H$\beta$             &4861& 100.0$\pm$5.1         & 280&100.0$\pm$6.0& 263&100.0$\pm$6.0& 172\\
$[$O~{\sc iii}$]$    &4959& 210.8$\pm$7.9         & 546&226.2$\pm$9.5& 582&212.6$\pm$9.2& 464\\
$[$O~{\sc iii}$]$    &5007& 602.2$\pm$15.         &1524&647.8$\pm$16.&1796&610.5$\pm$19.&1136\\
He~{\sc i}           &5876&  11.8$\pm$1.6         &  33& 11.6$\pm$2.1&  50& 12.6$\pm$2.2&  36\\
H$\alpha$            &6563& 286.0$\pm$10.         &2127&283.2$\pm$11.&1109&284.6$\pm$12.& 884\\
$[$N~{\sc ii}$]$     &6583&  16.3$\pm$1.8         &  80&  5.0$\pm$1.5&  26& 19.8$\pm$2.6&  63\\
$[$S~{\sc ii}$]$     &6717&   5.9$\pm$1.1         &  42&  5.5$\pm$1.5&  30& \multicolumn{1}{c}{...}& ...\\
$[$S~{\sc ii}$]$     &6731&   5.4$\pm$1.1         &  40&  4.4$\pm$1.4&  24& \multicolumn{1}{c}{...}& ...\\
$C$(H$\beta$)$_{\rm int}$$^{\rm c}$  &&\multicolumn{2}{c}{0.292$\pm$0.041}&\multicolumn{2}{c}{0.204$\pm$0.0.047}&\multicolumn{2}{c}{0.108$\pm$0.049}\\
$C$(H$\beta$)$_{\rm MW}$$^{\rm d}$   &&\multicolumn{2}{c}{0.023}&\multicolumn{2}{c}{0.030}&\multicolumn{2}{c}{0.011}\\
$I$(H$\beta$)$^{\rm e}$     &&\multicolumn{2}{c}{24.2$\pm$0.7}&\multicolumn{2}{c}{13.9$\pm$1.0}&\multicolumn{2}{c}{10.8$\pm$0.8}\\
\cline{1-8}
  \end{tabular}

\hbox{$^{\rm a}$$I$=100$\times$$I$($\lambda$)/$I$(H$\beta$), where $I$($\lambda$) 
and $I$(H$\beta$) are fluxes of emission lines, corrected for both the 
Milky Way and internal extinction.}

\hbox{$^{\rm b}$Rest-frame equivalent width in \AA.}


\hbox{$^{\rm c}$Internal galaxy extinction coefficient.}

\hbox{$^{\rm d}$Milky Way extinction coefficient from the NED.}


\hbox{$^{\rm e}$Extinction-corrected flux but not corrected for $f_{\rm esc}$(LyC),
in 10$^{-16}$ erg s$^{-1}$ cm$^{-2}$.}

  \end{table*}

  \begin{table*}
  \caption{Electron temperatures, electron number densities and 
element abundances in H~{\sc ii} regions \label{taba3}}
  \begin{tabular}{lcccc} \hline
Galaxy &J0130$-$0014 &J0142$-$0304 &J0844$+$5312 &J1014$+$5501 \\ \hline
$T_{\rm e}$ ($[$O {\sc iii}$]$), K      & 14530$\pm$1570       & 13370$\pm$920       & 14120$\pm$1170     & 14260$\pm$1560\\
$T_{\rm e}$ ($[$O {\sc ii}$]$), K       & 13720$\pm$1380       & 12940$\pm$830       & 13460$\pm$1040     & 13550$\pm$1390 \\
$N_{\rm e}$ ($[$S {\sc ii}$]$), cm$^{-3}$& 1070$\pm$1070       &  353$\pm$353  &       10$\pm$10   &    10$\pm$10    \\ \\
O$^+$/H$^+$$\times$10$^{5}$             &1.25$\pm$0.17        &1.81$\pm$0.18        &1.77$\pm$0.20       &1.16$\pm$0.18 \\
O$^{2+}$/H$^+$$\times$10$^{5}$          &8.16$\pm$0.42        &9.80$\pm$0.43        &8.62$\pm$0.37      &7.93$\pm$0.43  \\
O/H$\times$10$^{5}$                    &9.41$\pm$0.46        &11.61$\pm$0.46        &10.39$\pm$0.42      &9.08$\pm$0.46  \\
12+log O/H                             &7.97$\pm$0.02        &8.06$\pm$0.02        &8.02$\pm$0.02      &7.96$\pm$0.02  \\ \\
N$^+$/H$^+$$\times$10$^{6}$             &0.38$\pm$0.16       &1.54$\pm$0.17        &0.93$\pm$0.16      &0.44$\pm$0.15  \\
ICF(N)$^{\rm a}$                        &    6.89            & 5.98                 &5.55               &    7.17      \\
N/H$\times$10$^{6}$                     &2.59$\pm$1.23       &9.19$\pm$1.07        &5.19$\pm$0.96      &3.15$\pm$1.17 \\
log N/O                                &~$-$1.56$\pm$0.21~~\, &~$-$1.10$\pm$0.05~~\, &~$-$1.30$\pm$0.08~~\,&~$-$1.46$\pm$0.16~~\, \\ \\
Ne$^{2+}$/H$^+$$\times$10$^{5}$          &1.38$\pm$0.23        &2.06$\pm$0.23        &1.87$\pm$0.23       &1.61$\pm$0.26 \\
ICF(Ne)$^{\rm a}$                       &1.05                 &1.07                 &1.08                &1.05           \\
Ne/H$\times$10$^{5}$                    &1.45$\pm$0.24        &2.20$\pm$0.25        &2.03$\pm$0.25       &1.69$\pm$0.27 \\
log Ne/O                               &~$-$0.81$\pm$0.07~~\, &~$-$0.72$\pm$0.05~~\, &~$-$0.71$\pm$0.06~~\,&~$-$0.73$\pm$0.07~~\, \\ \\
Mg$^{+}$/H$^+$$\times$10$^{6}$          &        ...          & 0.55$\pm$0.09    & 0.68$\pm$0.12      &0.47$\pm$0.11 \\
ICF(Mg)$^{\rm a}$                        &       ...          &      12.14         &      12.26       &14.31        \\
Mg/H$\times$10$^{6}$                    &        ...          & 6.55$\pm$1.08     & 8.38$\pm$1.52     &6.67$\pm$1.58 \\
log Mg/O                               &         ...          &~$-$1.24$\pm$0.07~~\, &~$-$1.09$\pm$0.08~~\, &~$-$1.13$\pm$0.11~~\, \\
\hline
Galaxy &J1137$+$3605 &J1157$+$5801 &J1352$+$5617 \\ \cline{1-4}
$T_{\rm e}$ ($[$O {\sc iii}$]$), K      & 16190$\pm$1240       & 16420$\pm$1530       & 13490$\pm$1440 \\
$T_{\rm e}$ ($[$O {\sc ii}$]$), K       & 14610$\pm$1040       & 14710$\pm$1280       & 13030$\pm$1300 \\
$N_{\rm e}$ ($[$S {\sc ii}$]$), cm$^{-3}$&  450$\pm$450       &  172$\pm$172  &        10$\pm$10    \\ \\
O$^+$/H$^+$$\times$10$^{5}$             &0.84$\pm$0.08        &0.71$\pm$0.08        &2.30$\pm$0.35   \\
O$^{2+}$/H$^+$$\times$10$^{5}$          &5.56$\pm$0.10        &5.78$\pm$1.49        &8.96$\pm$0.57   \\
O/H$\times$10$^{5}$                    &6.40$\pm$0.13        &6.49$\pm$0.17        &11.26$\pm$0.67   \\
12+log O/H                             &7.81$\pm$0.01        &7.81$\pm$0.01        &8.05$\pm$0.03    \\ \\
N$^+$/H$^+$$\times$10$^{6}$             &1.27$\pm$0.15        &0.38$\pm$0.12        &1.96$\pm$0.27    \\
ICF(N)$^{\rm a}$                        &7.03               &8.34                   &4.78             \\
N/H$\times$10$^{6}$                     &8.91$\pm$1.10       &3.16$\pm$1.10        &9.35$\pm$1.33     \\
log N/O                                &~$-$0.86$\pm$0.05~~\, &~$-$1.31$\pm$0.15~~\, &~$-$1.08$\pm$0.07~~\, \\ \\
Ne$^{2+}$/H$^+$$\times$10$^{5}$          &1.11$\pm$0.11        &1.24$\pm$0.14        &1.86$\pm$0.30   \\
ICF(Ne)$^{\rm a}$                       &1.05                 &1.04                 &1.12             \\
Ne/H$\times$10$^{5}$                    &1.17$\pm$0.12        &1.28$\pm$0.15        &2.08$\pm$0.35    \\
log Ne/O                               &~$-$0.74$\pm$0.04~~\, &~$-$0.70$\pm$0.05~~\, &~$-$0.73$\pm$0.08~~\, \\ \\
Mg$^{+}$/H$^+$$\times$10$^{6}$          &0.21$\pm$0.04         &  0.12$\pm$0.04     &0.78$\pm$0.18       \\
ICF(Mg)$^{\rm a}$                        &     13.94           &     15.33          &     9.50           \\
Mg/H$\times$10$^{6}$                    &2.91$\pm$0.54        & 1.80$\pm$0.58      &7.39$\pm$1.68       \\
log Mg/O                               &~$-$1.34$\pm$0.08~~\, &~$-$1.48$\pm$0.14~~\, &~$-$1.18$\pm$0.10~~\,  \\
\cline{1-4}
\end{tabular}

\hbox{$^{\rm a}$Ionization correction factor.}
  \end{table*}



\bsp	
\label{lastpage}
\end{document}